\PassOptionsToPackage{table}{xcolor}
\documentclass[acmsmall, screen]{acmart}

\usepackage{comment}
\usepackage{pgfplots}
\usepackage{pgfplotstable}
\pgfplotsset{compat=1.13}
\usepackage{tikz}
\usetikzlibrary{pgfplots.groupplots}
\usetikzlibrary{pgfplots.statistics}
\usetikzlibrary{calc,tikzmark}

\usepackage{fancyvrb}
\usepackage[ruled,vlined,linesnumbered]{algorithm2e}
\usepackage{xspace}
\usepackage{listings}
\usepackage{balance}
\usepackage{array}
\usepackage[normalem]{ulem}
\usepackage{booktabs}
\usepackage{longtable}
\usepackage{pdflscape}
\usepackage{subcaption}
\usepackage{caption}
\usepackage{framed}
\usepackage{algorithmic}
\usepackage{graphicx}
\usepackage{textcomp}
\usepackage{soul}
\usepackage{multirow}
\usepackage{pifont}
\usepackage{makecell}
\usepackage{bbm}
\usepackage{MnSymbol}
\usepackage{epstopdf}
\usepackage{colortbl}
\usepackage{tabularx}
\usepackage{expl3}
\usepackage[most]{tcolorbox}
\usepackage{wasysym}

\usepackage{enumitem}
\newlist{rqs}{enumerate}{1}
\setlist[rqs,1]{label={\bfseries RQ\arabic*},align=left}

\usepackage{cleveref}
\crefname{section}{Section}{Sections}
\crefname{table}{Table}{Tables}
\crefname{figure}{Figure}{Figures}
\crefname{subfigure}{Figure}{Figures}
\crefname{definition}{Definition}{Definitions}
\crefname{equation}{Equation}{Equations}
\crefname{example}{Ex.}{Examples}
\crefname{algorithm}{Algorithm}{Algorithms}
\crefname{line}{Line}{Lines}

\newcommand{\cmark}{\ding{51}}
\newcommand{\xmark}{\ding{55}}

\newcommand{\eg}{e.g.,\xspace}
\newcommand{\ie}{i.e.,\xspace}
\newcommand{\etal}{et al.\xspace}
\newcommand{\etc}{etc.\xspace}

\newcolumntype{Z}[1]{>{\hbox to #1\bgroup}l<{\hss\egroup}}

\ExplSyntaxOn
\newcommand{\calculate}[1]{\fp_eval:n { \expanded{#1} }}
\ExplSyntaxOff
\newcommand{\ratio}[2]{$\calculate{round({#1} /{#2} *100, 1)}\%$}

\newcommand{\tablefontsize}{\scriptsize}

\definecolor{javared}{rgb}{0.6,0,0}
\definecolor{javagreen}{rgb}{0.25,0.5,0.35}
\definecolor{javapurple}{rgb}{0.5,0,0.35}
\definecolor{javadocblue}{rgb}{0.25,0.35,0.75}
\newcommand{\numtotalpaper}{15,777\xspace}
\newcommand{\numkeywordpaper}{6,778\xspace}
\newcommand{\numvenuepaper}{11,872\xspace}
\newcommand{\numempiricalpaper}{4,502\xspace}
\newcommand{\numkeyphrasepaper}{11,280\xspace}
\newcommand{\numselectedpaper}{150\xspace}
\newcommand{\numdataset}{151\xspace}

\newcommand{\java}{Java\xspace}
\newcommand{\csharp}{C\#\xspace}
\newcommand{\qsharp}{Q\#\xspace}
\newcommand{\fsharp}{F\#\xspace}
\newcommand{\scala}{Scala\xspace}
\newcommand{\cncpp}{C/C++\xspace}
\newcommand{\python}{Python\xspace}
\newcommand{\javascript}{JavaScript\xspace}
\newcommand{\php}{PHP\xspace}
\newcommand{\ruby}{Ruby\xspace}
\newcommand{\perl}{Perl\xspace}
\newcommand{\go}{Go\xspace}
\newcommand{\swift}{Swift\xspace}
\newcommand{\kotlin}{Kotlin\xspace}
\newcommand{\typescript}{TypeScript\xspace}
\newcommand{\rust}{Rust\xspace}
\newcommand{\objectiveC}{Objective-C\xspace}
\newcommand{\sql}{SQL\xspace}
\newcommand{\shell}{Shell\xspace}
\newcommand{\matlab}{MATLAB\xspace}
\newcommand{\groovy}{Groovy\xspace}
\newcommand{\lua}{Lua\xspace}
\newcommand{\haskell}{Haskell\xspace}
\newcommand{\solidity}{Solidity\xspace}
\newcommand{\julia}{Julia\xspace}
\newcommand{\cpp}{C++\xspace}
\newcommand{\fortran}{Fortran\xspace}
\newcommand{\unknownpl}{unspecified\xspace}

\newcommand{\pytorch}{PyTorch\xspace}
\newcommand{\tensorflow}{TensorFlow\xspace}
\newcommand{\keras}{Keras\xspace}
\newcommand{\android}{Android\xspace}
\newcommand{\ethereum}{Ethereum\xspace}
\newcommand{\qiskit}{Qiskit\xspace}
\newcommand{\github}{GitHub\xspace}
\newcommand{\stackoverflow}{Stack Overflow\xspace}
\newcommand{\stackexchange}{Stack Exchange\xspace}
\newcommand{\mapreduce}{MapReduce\xspace}
\newcommand{\hadoop}{Hadoop\xspace}
\newcommand{\hdfs}{HDFS\xspace}
\newcommand{\hbase}{HBase\xspace}
\newcommand{\cassandra}{Cassandra\xspace}
\newcommand{\zookeeper}{ZooKeeper\xspace}
\newcommand{\flume}{Flume\xspace}
\newcommand{\chrome}{Chrome\xspace}
\newcommand{\chromium}{Chromium\xspace}
\newcommand{\firefox}{Firefox\xspace}
\newcommand{\webassembly}{WebAssembly\xspace}
\newcommand{\iot}{IoT\xspace}
\newcommand{\npm}{npm\xspace}
\newcommand{\jupyternotebook}{Jupyter Notebook\xspace}
\newcommand{\api}{API\xspace}
\newcommand{\apache}{Apache\xspace}
\newcommand{\jira}{Jira\xspace}
\newcommand{\bugzilla}{Bugzilla\xspace}
\newcommand{\cicd}{CI/CD\xspace}
\newcommand{\qa}{Q\&A\xspace}
\newcommand{\gdrive}{Google Drive\xspace}
\newcommand{\figsshare}{figshare\xspace}
\newcommand{\dfj}{Defects4J\xspace}
\newcommand{\git}{Git\xspace}

\newif\ifshowwebcites
\showwebcitesfalse

\newcommand{\webcite}[1]{
  \ifshowwebcites
    \unskip\nobreakspace\cite{#1}
  \else
    \unskip\ignorespaces
  \fi
}

\definecolor{revisionbg}{RGB}{233,248,245}

\sethlcolor{revisionbg}

\soulregister\cite7
\soulregister\citep7
\soulregister\citet7
\soulregister\citealt7
\soulregister\citealp7
\soulregister\Cite7
\soulregister\Citep7
\soulregister\Citet7
\soulregister\Citealt7
\soulregister\Citealp7
\soulregister\ref7
\soulregister\pageref7
\soulregister\cref7
\soulregister\Cref7
\soulregister\autoref7
\soulregister\eqref7
\soulregister\url7
\soulregister\path7
\soulregister\href7
\soulregister\numdataset7
\soulregister\numtotalpaper7
\soulregister\numkeywordpaper7
\soulregister\numvenuepaper7
\soulregister\numempiricalpaper7
\soulregister\numkeyphrasepaper7
\soulregister\numselectedpaper7
\soulregister\xspace7
\soulregister\eg7
\soulregister\ie7
\soulregister\etal7
\soulregister\etc7
\soulregister\java7
\soulregister\csharp7
\soulregister\qsharp7
\soulregister\fsharp7
\soulregister\scala7
\soulregister\cncpp7
\soulregister\python7
\soulregister\javascript7
\soulregister\php7
\soulregister\ruby7
\soulregister\perl7
\soulregister\go7
\soulregister\swift7
\soulregister\kotlin7
\soulregister\typescript7
\soulregister\rust7
\soulregister\objectiveC7
\soulregister\sql7
\soulregister\shell7
\soulregister\matlab7
\soulregister\groovy7
\soulregister\lua7
\soulregister\haskell7
\soulregister\solidity7
\soulregister\julia7
\soulregister\cpp7
\soulregister\fortran7
\soulregister\unknownpl7
\soulregister\pytorch7
\soulregister\tensorflow7
\soulregister\keras7
\soulregister\android7
\soulregister\ethereum7
\soulregister\qiskit7
\soulregister\github7
\soulregister\stackoverflow7
\soulregister\stackexchange7
\soulregister\mapreduce7
\soulregister\hadoop7
\soulregister\hdfs7
\soulregister\hbase7
\soulregister\cassandra7
\soulregister\zookeeper7
\soulregister\flume7
\soulregister\chrome7
\soulregister\chromium7
\soulregister\firefox7
\soulregister\webassembly7
\soulregister\iot7
\soulregister\npm7
\soulregister\jupyternotebook7
\soulregister\api7
\soulregister\apache7
\soulregister\jira7
\soulregister\cicd7
\soulregister\qa7
\soulregister\gdrive7
\soulregister\figsshare7
\soulregister\dfj7
\soulregister\git7
\soulregister\cmark7
\soulregister\xmark7
\soulregister\itemize7
\soulregister\item7
\soulregister\bugzilla7
\soulregister\footnote7

\newtcolorbox{revisionblock}{
  enhanced,
  breakable,
  colback=revisionbg,
  colframe=revisionbg,
  boxrule=0pt,
  sharp corners,
  left=6pt, right=6pt, top=6pt, bottom=6pt,
  before skip=8pt, after skip=8pt,
}

\setcopyright{acmlicensed}
\copyrightyear{2026}
\acmYear{2026}
\acmDOI{XXXXXXX.XXXXXXX}

\setcopyright{none}
\renewcommand\footnotetextcopyrightpermission[1]{}

\acmJournal{CSUR}

\begin{document}

\title[A Comprehensive Survey of Software Defect Datasets]{From Bugs to Benchmarks: \\A Comprehensive Survey of Software Defect Datasets}

\author{Hao-Nan Zhu}
\email{hnzhu@ucdavis.edu}
\affiliation{
  \institution{University of California, Davis}
  \country{United States}
}

\author{Robert M. Furth}
\email{rmfurth@ucdavis.edu}
\affiliation{
  \institution{University of California, Davis}
  \country{United States}
}

\author{Michael Pradel}
\email{michael@binaervarianz.de}
\affiliation{
  \institution{University of Stuttgart}
  \country{Germany}
}

\author{Cindy Rubio-Gonz\'alez}
\email{crubio@ucdavis.edu}
\affiliation{
  \institution{University of California, Davis}
  \country{United States}
}

\thanks{This work was supported by the National Science Foundation (awards CNS-2016735, CNS-2346396, and CCF-2119348), by the European Research Council (ERC, grant agreements 851895 and 101155832), and by the German Research Foundation through the ConcSys, DeMoCo, and QPTest projects.}

\renewcommand{\shortauthors}{Zhu et al.}

\begin{abstract}

Software defect datasets, which are collections of software bugs, are essential resources to facilitate empirical research and enable standardized benchmarking for a wide range of software engineering techniques, including emerging areas like agentic AI-based software development. Over the years, numerous software defect datasets have been developed, providing rich resources for the community, yet making it increasingly difficult to navigate the landscape. This article provides a comprehensive survey of \numdataset software defect datasets, covering their scope, construction, availability, usability, and practical uses. We also suggest potential opportunities for future research based on our findings, such as addressing underrepresented kinds of defects. A complete catalog of all surveyed software defect datasets is available at \url{https://defect-datasets.github.io/}.

\end{abstract}



\maketitle

\section{Introduction}
\label{sec-introduction}

Given the escalating scale and complexity of real-world software systems, software defects are inevitable and have become increasingly challenging to manage~\cite{DBLP:journals/tse/FentonO00,DBLP:conf/sosp/ChouYCHE01}.
Software defects, commonly referred to as bugs, are
unintentional errors in software that can cause software systems to
behave unexpectedly or fail to meet their intended requirements. To
address the challenges posed by software defects, in the last few
decades, a variety of techniques have been proposed to detect,
localize, and fix them. These techniques include, but are
not limited to, static and dynamic
analysis~\cite{DBLP:journals/software/AyewahHMPP08,ball1999concept},
test generation and prioritization~\cite{DBLP:journals/tse/NebutFTJ06,DBLP:journals/tse/ElbaumMR02},
fault detection and localization~\cite{DBLP:journals/tse/WongGLAW16}, and
automated program repair~\cite{DBLP:journals/software/GouesPRC21}. To
evaluate the effectiveness and efficiency of these techniques, the research community  heavily relies on software defect datasets, which serve as foundational resources for benchmarking and
validation.

Software defect datasets are curated collections of real-world software bugs.
These datasets serve as standardized benchmarks, fostering consistent and
reproducible evaluations of novel techniques across diverse
studies. Additionally, the realism, diversity, scale, and accessibility of these
datasets support empirical analyses of defect characteristics and enable the
training and fine-tuning of machine learning models for software quality
assurance.  For instance, the widely adopted \dfj
dataset~\cite{DBLP:conf/issta/JustJE14} has been used for a wide range of
technical evaluations
(\eg~\cite{DBLP:journals/ese/MartinezDSXM17,10.1145/3688834,DBLP:conf/issta/OuyangY024}),
empirical studies
(\eg~\cite{DBLP:conf/wcre/SobreiraDDMM18,DBLP:conf/sigsoft/DurieuxDMA19}), and
to train machine learning models
(\eg~\cite{DBLP:journals/tse/ZhangFSLHHC24}).
As another example, the recently proposed SWE-Bench~\cite{DBLP:conf/iclr/JimenezYWYPPN24} dataset is widely used to evaluate the effectiveness of large language models (LLMs) in automating software engineering~\cite{DBLP:conf/issta/0002RFR24,DBLP:conf/nips/YangJWLYNP24} and serves as the basis for other benchmarks~\cite{DBLP:journals/corr/abs-2410-06992,DBLP:journals/corr/abs-2406-12952}, demonstrating the practical value
of software defect datasets.

With the continuous growth in the complexity and scale of modern software
systems, there is an increasing demand for comprehensive and high-quality defect
datasets. This demand is further amplified by the rising popularity of artificial intelligence (AI)-based software engineering approaches~\cite{DBLP:journals/tosem/HouZLYWLLLGW24,DBLP:journals/csur/YangXLG22}. In response to this escalating demand,
industry challenges, competitions, and
workshops have been organized to encourage the development of software defect
datasets of high quality. As a result, these datasets have experienced a
significant increase in both quantity and diversity in recent years, becoming
essential resources for researchers and practitioners in multiple fields, including software engineering, programming languages, computer security, computer systems, and artificial intelligence.

With the increasing number of software defect datasets, researchers and
practitioners face challenges in navigating the landscape of software defect
datasets for two main reasons. First, different datasets may have different
focuses, such as types of defects or programming languages, requiring additional
effort to identify the most suitable dataset for specific research or practical
needs.  Second, based on how the datasets are constructed, they exhibit
diverse characteristics, such as size and usability, affecting the impact of these datasets. Moreover, creators and maintainers of
software defect datasets also encounter challenges in developing datasets that
address the evolving needs of software engineering research and practice, as
well as in ensuring the long-term sustainability and usability of the created
datasets.

\begin{figure}[t]
    \centering
    \includegraphics[width=\textwidth]{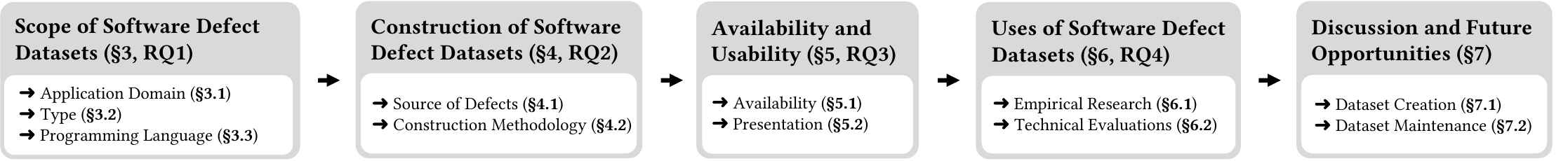}
    \caption{Topics covered in this survey. }
    \label{fig-sections-structure}
\end{figure}

To address these challenges, this paper provides a comprehensive overview of
existing software defect datasets. Specifically, we follow a systematic
methodology (detailed in \cref{sec-methodology}) for collecting and filtering
relevant literature, resulting in the identification of \numdataset software
defect datasets.
Our survey focuses on the following four research questions (RQs), as illustrated in \cref{fig-sections-structure}:

\begin{itemize}[left=0pt]
    \item \textbf{RQ1: What is the scope of existing software defect datasets? (\cref{sec-scope-focus})}
    For each dataset surveyed, we examine the application domain of the software systems from which the defects are collected (\eg machine learning systems or mobile applications), the types of defects (\eg functional, security, or performance), the programming languages (\eg \java, \cncpp, or \python), and other characteristics that define the focus of the dataset.
    We also explore potential interrelationships among these dimensions to provide a holistic understanding of the dataset landscape.
    \item \textbf{RQ2: How are existing software defect datasets constructed? (\cref{sec-construction})}
    For each dataset, we investigate how the dataset is constructed. This includes the sources of defects (\eg issue reports, version control systems, or continuous integration \& delivery) and the methodologies used to obtain the defects (\eg manual inspection, keyword matching, or machine learning techniques).
    \item \textbf{RQ3: To what extent are existing software defect datasets available and usable in practice? (\cref{sec-artifact})}
    For each dataset, we evaluate its availability and usability. Availability refers to whether the dataset is publicly available, and if so, where it is hosted. Usability refers to how the defects are presented in the dataset, such as providing only textual descriptions, code snippets, or executable artifacts to reproduce the defects.

    \item \textbf{RQ4: How are software defect datasets actually used by research community? (\cref{sec-usage})}
    While prior literature reviews~\cite{DBLP:journals/jss/HirschH22,DBLP:conf/worldcist/HolekBC23} have provided high-level summaries of the intended uses of software defect datasets from the perspective of their creators, this research question seeks to offer a more comprehensive and detailed understanding of how these datasets are actually used. By analyzing citations of the most influential datasets, we aim to uncover usage patterns and trends over time, thereby revealing their impact on the research community.

\end{itemize}

\newcommand{\covered}{\ding{52}}
\newcommand{\notcovered}{\ding{55}}
\newcommand{\partiallycovered}{\Circle}

\begin{table}[!t]
\centering
\tablefontsize
\caption{Comparison between surveys of software defect datasets.}
\begin{tabularx}{\textwidth}{l *{3}{>{\centering\arraybackslash}X}}
\toprule
 & \textbf{\citet{DBLP:journals/jss/HirschH22}} & \textbf{\citet{DBLP:conf/worldcist/HolekBC23}} & \textbf{This Survey} \\
\midrule
Number of Datasets & 73 & 14 & \numdataset \\
Cutoff Date For Dataset Selection & November 2021 & April 2022 & January 2025 \\
\midrule
\textbf{Scope of Datasets} & & & \\
- Application Domains & \notcovered & \notcovered & \covered \\
- Types of Defects & \notcovered & \notcovered & \covered \\
- Programming Languages & \covered & \notcovered & \covered \\
\midrule
\textbf{Dataset Construction} & & & \\
- Source of Defects & \notcovered & \covered & \covered \\
- Mining Methodology & \notcovered & \notcovered & \covered \\
\midrule
\textbf{Availability and Usability} & & & \\
- Availability and Hosting Platforms & \covered & \covered & \covered \\
- Presentation & \covered & \covered & \covered \\
\midrule
\textbf{Uses of Datasets} & \notcovered & \notcovered & \covered \\
\bottomrule
\end{tabularx}
\label{tab-survey-comparison}
\end{table}

Finally, based on our analysis of existing datasets, \cref{sec-discussion} identifies potential opportunities for future research, including automating the mining of high-quality datasets, e.g., by leveraging CI pipelines and LLMs. Additionally, we identify the need for specialized datasets targeting specific domains, defect types, and new languages, as well as techniques ensuring dataset reproducibility and continuous updates.

Our work contributes the first comprehensive survey on software defect datasets, providing a detailed discussion of their scope, construction, availability, and real-world usage.
Orthogonal to existing surveys on software defects, which focus on categorizing~\cite{DBLP:journals/ijseke/NagwaniV14,DBLP:journals/chinaf/ZhangWHXZM15}, detecting/localizing~\cite{DBLP:journals/tse/WongGLAW16,DBLP:journals/pieee/LinWHZX20,DBLP:journals/corr/abs-2410-00650,DBLP:journals/csur/ChenPPXZHZ20,DBLP:journals/csur/AndreasenGMPSSS17}, and repairing software defects~\cite{DBLP:journals/csur/Monperrus18,DBLP:journals/corr/abs-2303-18184,DBLP:journals/tosem/ZhangFMSC24}, this survey focuses on datasets that collect software defects prevalent in software codebases.
Previous literature reviews on 73 benchmarks~\cite{DBLP:journals/jss/HirschH22} and 14 bug datasets~\cite{DBLP:conf/worldcist/HolekBC23} provide high-level overviews of existing benchmarking datasets.
However, as summarized in \cref{tab-survey-comparison}, these studies do not cover several important aspects of software defect datasets.
In contrast, this article provides (1) significantly broader coverage of \numdataset{} software defect datasets with a more recent cutoff date, (2) a systematic and in-depth analysis of their scope, construction, availability, and usability, (3) insights into the relationships between different dataset characteristics, and (4) an examination of longitudinal trends in the actual usage of the most influential datasets.

We envision this survey to be useful for both dataset users and creators.
For dataset users, this survey provides a comprehensive overview of
existing software defect datasets, enabling them to identify the most
suitable datasets for their research or practical needs.
For dataset creators, our work highlights the current focuses and methodologies
used in existing datasets, which can guide the development of new datasets that fill the gaps in the current literature.
In addition to this article itself, we provide an interactive website\footnote{\url{https://defect-datasets.github.io/}} that
allows users to find datasets based on specific criteria, such as programming
languages or application domains of interest.

\section{Survey Methodology}
\label{sec-methodology}

\begin{table}[t]
    \caption{Searching strategies to collect papers}
    \tablefontsize
    \begin{tabularx}{\textwidth}{r|X}
        \toprule
        \textbf{Strategy} & \textbf{Search Content}                                                                                                                                                                                                                                                                                                               \\
        \midrule
        Key Phrase     & bug benchmark, bug dataset, defect benchmark, defect dataset, patch benchmark, patch dataset, fail pass, fault benchmark, fault dataset, mining software, software benchmark, software bug, software dataset, software defect, software fault, software patch, software vulnerabilit(y/ies), vulnerabilit(y/ies) benchmark, vulnerabilit(y/ies) dataset, empirical bug, empirical defect, empirical error, empirical failure, empirical fault, empirical issue, empirical patch, study bug, study defect, study error, study failure, study fault, study issue, study patch             \\
        \midrule
         Keyword       & benchmark, bug, dataset, defect, fault, issue, patch, vulnerabilit(y/ies)      \\
         \texttimes     &                                                                                                                                                                                                                                                                                                              \\

         Venue          & ASE, ESE, FSE, ICSE, ISSRE, ISSTA, MSR, OOPSLA, SANER, TOSEM, TSE, ESEC, AAAI, ACL, EMNLP, ICLR, ICML, NeurIPS, USENIX Security, IEEE S\&P, CCS, NDSS                                                                                                                                                              \\
        \bottomrule
    \end{tabularx}
    \label{tab-paper-collection}
\end{table}

\begin{figure}[t]
\centering
\includegraphics[width=0.9\textwidth]{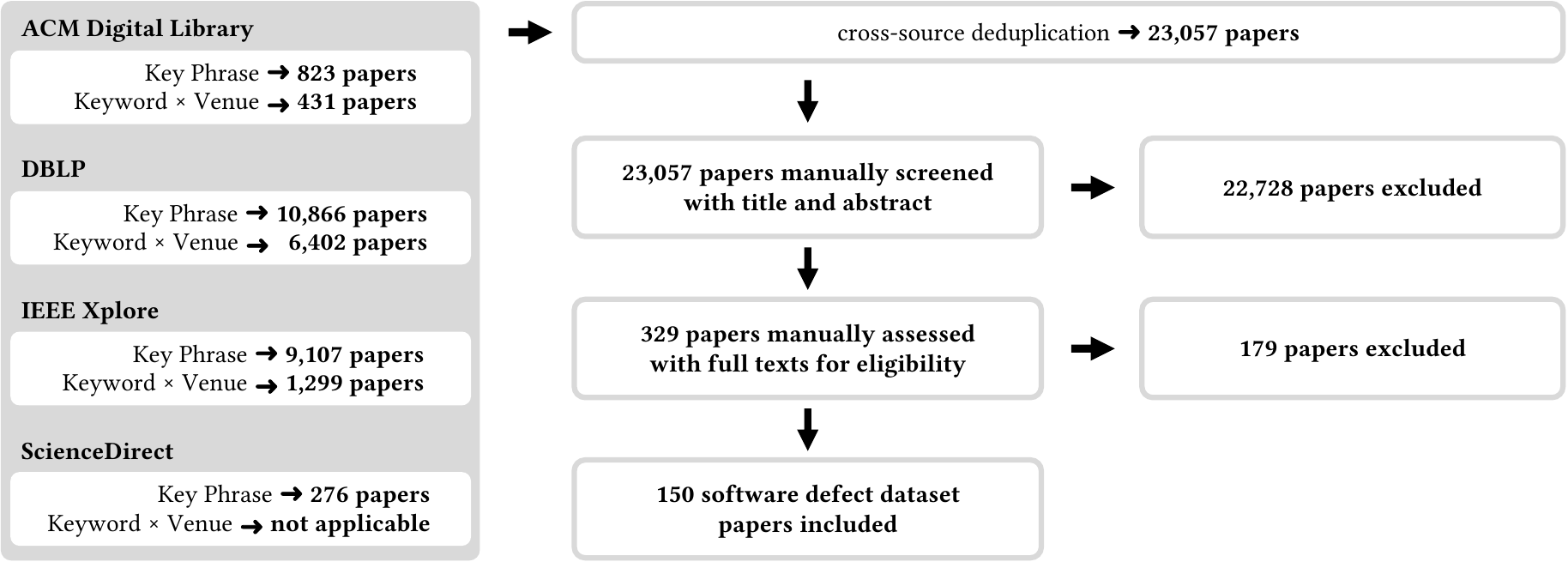}

\caption{Collection and selection process for software defect dataset papers.}
\label{fig-paper-collection-prisma}
\end{figure}

This section describes the methodology used in this survey to identify, select, and analyze existing software defect datasets, guided by research questions described in the introduction.
Our survey covers \numdataset software defect datasets.
To identify these datasets, we follow a systematic process that collects papers from multiple bibliographic databases and selects papers through a rigorous manual review.
\cref{fig-paper-collection-prisma} illustrates the complete paper collection and selection process.
We conduct a comprehensive literature search across four major bibliographic databases: ACM Digital Library, DBLP, IEEE Xplore, and ScienceDirect.
For each source, we execute two complementary search strategies, summarized in \cref{tab-paper-collection}.
We include papers that have been peer-reviewed and published before January 1, 2025, excluding preprints.

\paragraph{(1) Key Phrase-Based Search}
We use specific key phrases to identify papers explicitly related to software defect datasets or empirical studies on software defects.
We search for papers containing all words from a key phrase in their titles, abstracts, and keywords if available.
This strategy yields 823 papers from the ACM Digital Library, 10,866 papers from DBLP, 9,107 papers from IEEE Xplore, and 276 papers from ScienceDirect.

\paragraph{(2) Keyword \texttimes Venue Search}
To complement the key phrase-based search, we use individual keywords combined with venue restrictions.
As individual keywords yield many papers, we restrict this search to high-impact venues in software engineering, machine learning, and software security.
We search for papers containing at least one keyword in their title, abstract, and keywords if available.
This strategy yields 431 papers from the ACM Digital Library, 6,402 papers from DBLP, and 1,299 papers from IEEE Xplore.
Due to API query constraints, this strategy was not applicable to ScienceDirect.

After merging results from both search strategies across all bibliographic databases and performing cross-source deduplication, we obtain 23,057 unique papers potentially related to software defect datasets.
Two authors independently review the titles and abstracts of these papers to identify those potentially introducing a software defect dataset.
The authors then compare their selections and reach consensus through discussion, excluding 22,728 irrelevant papers and identifying 329 papers for full-text assessment.
For the full-text assessment phase, the same two authors independently review each of the 329 papers to determine their eligibility for inclusion.
A paper is included if its main contribution is one or more software defect datasets, or if it includes software defects as byproducts.
Papers are excluded if they only use existing datasets without introducing a new dataset, present datasets for purposes unrelated to software defects (\eg code generation or performance benchmarking), or focus solely on techniques for mining, injecting, or labeling defects.
The inter-rater agreement between the two authors for the full-text assessment is almost perfect, with a Cohen's kappa~\cite{cohen1960coefficient} of 0.871 (95\% CI: 0.818--0.924).
The authors agreed on 308 out of 329 papers (93.6\%), and disagreements on the remaining 21 papers were resolved through discussion and, when necessary, consultation with a third author.
This rigorous assessment ultimately leads to \numselectedpaper papers, from which we identified \numdataset software defect datasets (one paper~\cite{DBLP:journals/tse/GouesHSBDFW15} presents two datasets).

Finally, the same two authors conduct an in-depth manual review of all included datasets to extract and annotate their attributes (\eg application domain, defect type, and programming language for RQ1; sources of defects and construction methodologies for RQ2; and availability and usability for RQ3).
Each dataset is independently cross-checked by both authors to ensure consistency, with ambiguities resolved through discussion.
The annotation process is performed manually over several months.

\begin{figure}[t]
\centering
\begin{minipage}{\textwidth}
\centering
\small
\begin{subfigure}[b]{0.5\textwidth}
    \centering
    \includegraphics[width=0.75\linewidth]{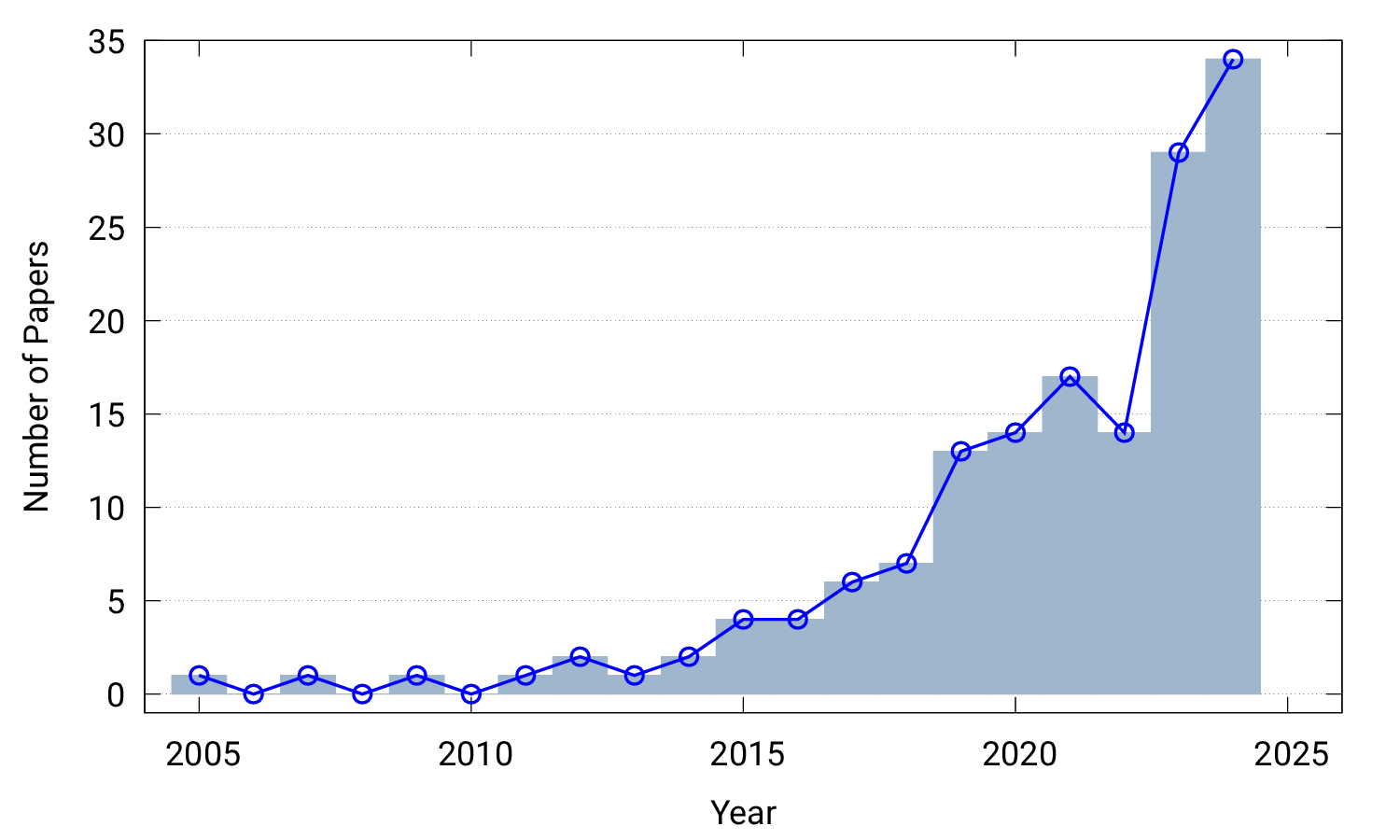}
    \caption{Number of papers by year.}
    \label{fig-paper-count-by-year}
\end{subfigure}\hfill
\begin{subfigure}[b]{0.5\textwidth}
    \centering
    \includegraphics[width=0.75\linewidth]{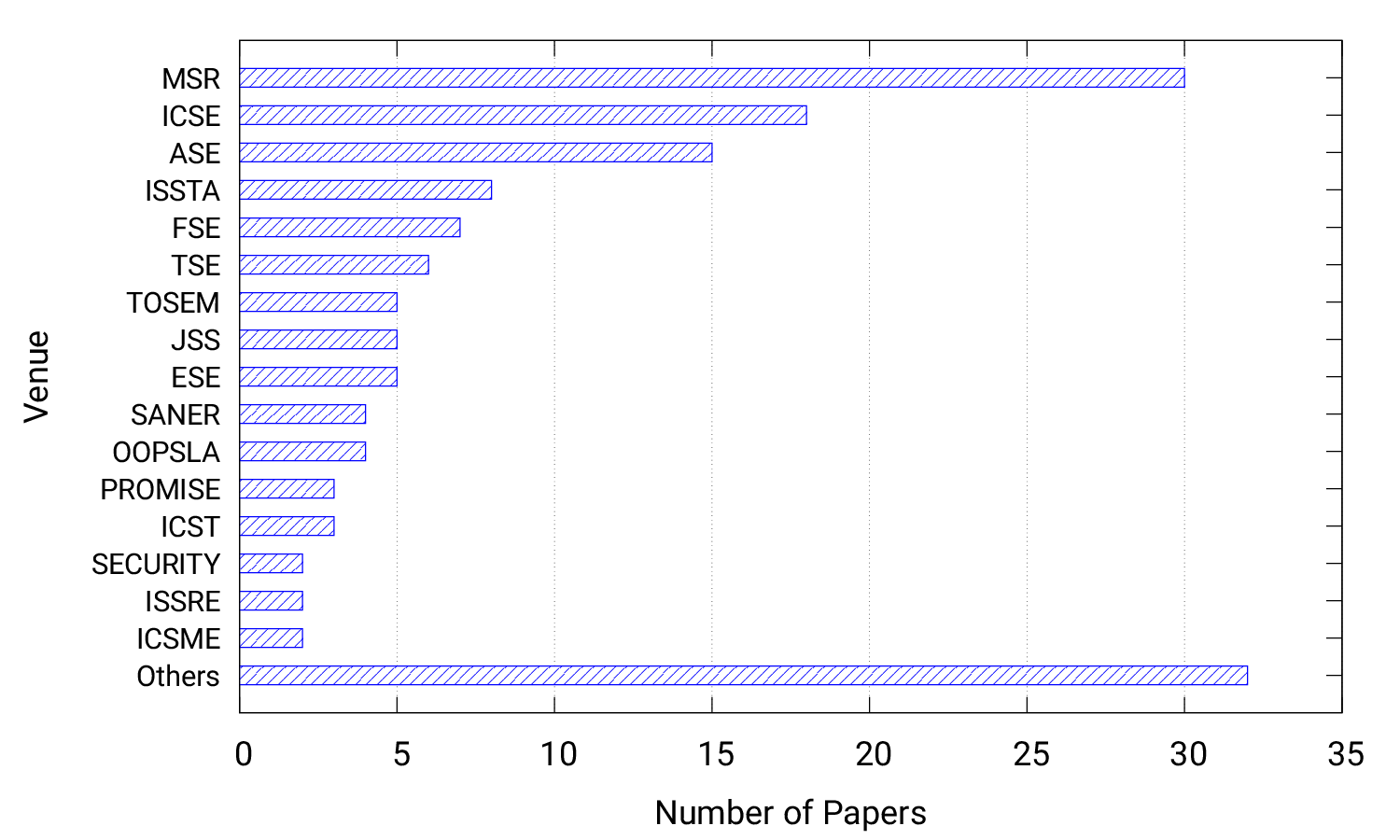}
    \caption{Number of papers by venue.}
    \label{fig-paper-count-by-venue}
\end{subfigure}
\end{minipage}
\caption{General statistics of software defect dataset papers.}
\label{fig-paper-count}
\end{figure}

\cref{fig-paper-count-by-year} plots the number of papers related to software defect datasets published by year, which shows a significant increase in the number of new software defect datasets published in the past eight years. The venues that publish papers related to software defect datasets are shown in \cref{fig-paper-count-by-venue}, where different tracks of a conference are grouped together. The top venues that publish papers related to software defect datasets include MSR, ICSE, and ASE. All venues that have only one paper included in this study are grouped into the \textit{others} category, such as ICLR and SoCC.

\section{Scope of Software Defect Datasets (RQ1)}
\label{sec-scope-focus}

Software defect datasets often have specific purposes. In this section, we
discuss and compare the scope of the various datasets.  In particular, we
consider three dimensions: (1) the application domain of the software that
contains the defects (Section~\ref{sec-domain}), (2) the type of defects that are included in the dataset (Section~\ref{sec-type}),
and (3) the programming languages used in the defective software (Section~\ref{sec-pl}).

\subsection{Application Domains of Defective Software}
\label{sec-domain}

\begin{table}[t]
    \centering
    \caption{Application domains of defective software. One dataset may appear in multiple categories. For clarity, only datasets that explicitly specify the corresponding attribute (\ie  application domain) are listed. The complete list of \numdataset{} datasets is available at \url{https://defect-datasets.github.io/} and in the supplementary materials~\cite{defect_datasets_2025_17402613}. The same convention applies to all tables in \cref{sec-scope-focus,sec-construction}.}
    \tablefontsize
    \begin{tabularx}{\textwidth}{ll|X|r}
        \toprule
        \textbf{Category}            &                          & \textbf{Datasets}                                                                                                                                                                                                                                                                                               & \textbf{Count} \\
        \midrule
        Machine Learning             & \textit{Applications}    & \cite{DBLP:journals/ese/MorovatiNKJ23,DBLP:conf/icse/LiangLSSFD22,DBLP:conf/msr/KimKL21,DBLP:conf/issta/ZhangCCXZ18,DBLP:conf/icse/ChenYLCLWL21,DBLP:conf/kbse/WangWCCY22,DBLP:conf/icse/HumbatovaJBR0T20}                                                                                                      & 7              \\
                                     & \textit{Frameworks}      & \cite{DBLP:conf/icse/GuanXLLB23,DBLP:conf/issre/LiuZDHDM022,DBLP:conf/icsm/HoMI0SKNR23,DBLP:conf/qsw/ZhaoWLLZ23,DBLP:journals/ese/TambonNAKA24}                                                                                                                                                                 & 5              \\
                                     & \textit{Compilers}       & \cite{DBLP:conf/issre/Du0MZ21,DBLP:conf/sigsoft/ShenM0TCC21}                                                                                                                                                                                                                                                    & 2              \\
        \midrule
        Compilers \& Runtime Engines &                          & \cite{DBLP:journals/tosem/ZhangCWCLMMHL24,DBLP:conf/issta/YuXZ0LS24,DBLP:conf/qrs/YuW24,DBLP:conf/icse/PaulTB21a,DBLP:conf/issre/Du0MZ21,DBLP:conf/msr/ZamanAH12,DBLP:conf/uss/XuLDDLWPM23,DBLP:journals/infsof/WangBWYGS23,DBLP:conf/wcre/WangZRLJ23,DBLP:conf/sigsoft/ShenM0TCC21,DBLP:conf/kbse/RomanoLK021} & 11             \\
        \midrule
        Mobile Applications          &                          & \cite{DBLP:conf/pst/TebibAAG24,DBLP:conf/kbse/DasAM24,DBLP:conf/secrypt/SenanayakeKAP023,DBLP:conf/issta/XiongX0SWWP0023,DBLP:conf/msr/WendlandSMMHMRF21,DBLP:conf/msr/RiganelliMMM19,DBLP:conf/promise/MitraR17,DBLP:journals/ese/LiuWWXCWYZ19,DBLP:conf/icse/ChenYLCLWL21,DBLP:conf/icse/TanDGR18}            & 10             \\
        \midrule
        Blockchain                   & \textit{Smart Contracts} & \cite{DBLP:journals/jss/CaiLZZS24,DBLP:journals/tse/ChenXLGLC22,DBLP:conf/icsm/ZhangXL20,DBLP:journals/jss/WangCHZBZ23}                                                                                                                                                                                         & 4              \\
                                     & \textit{Clients}         & \cite{DBLP:conf/msr/KimKL22}                                                                                                                                                                                                                                                                                    & 1              \\
        \midrule
        Cloud Computing              &                          & \cite{DBLP:conf/issta/XuG024,DBLP:conf/cloud/GunawiHLPDAELLM14,DBLP:conf/sigsoft/GaoDQGW0HZW18,DBLP:journals/pacmpl/DrososSAM024}                                                                                                                                                                               & 4              \\
        \midrule
        Autonomous Systems           &                          & \cite{DBLP:journals/ese/TimperleyHSDW24,DBLP:conf/icse/GarciaF0AXC20,DBLP:conf/sigsoft/WangLX0S21}                                                                                                                                                                                                              & 3              \\
        \midrule
        Quantum Computing            & \textit{Applications}    & \cite{DBLP:journals/jss/ZhaoMLZ23}                                                                                                                                                                                                                                                                              & 1              \\
                                     & \textit{Platforms}       & \cite{DBLP:journals/pacmpl/PaltenghiP22,DBLP:conf/qsw/ZhaoWLLZ23}                                                                                                                                                                                                                                               & 2              \\
        \midrule
        Scientific Computing         &                          & \cite{DBLP:conf/hotsos/MurphyBSR20,DBLP:conf/msr/AzadIHR23,DBLP:conf/kbse/FrancoGR17}                                                                                                                                                                                                                           & 3              \\
        \midrule
        Databases                    &                          & \cite{DBLP:journals/tosem/LiuMC24, DBLP:conf/icse/CuiD0WSZWYXH0024}                                                                                                                                                                                                                                             & 2              \\
        \midrule
        Computational Notebooks      &                          & \cite{DBLP:journals/tosem/SantanaNAA24}                                                                                                                                                                                                                                                                         & 1              \\
        \midrule
        Internet of Things           &                          & \cite{DBLP:conf/icse/Makhshari021}                                                                                                                                                                                                                                                                              & 1              \\

        \bottomrule
    \end{tabularx}\\[1mm]
    \label{tab-domain}

\end{table}

The \textit{application domain} of defective software refers to the specific field or area in which the software operates, such as machine learning or cloud computing. Of the \numdataset software defect datasets considered here, the majority (over $60\%$) focus on general software development and are not tied to any specific application domain. Among the remaining datasets that explicitly target specific domains, we identify eleven distinct application domains (\cref{tab-domain}), including machine learning, compilers and runtime engines, mobile applications, and blockchain.\footnote{In this and other classifications of datasets, a single dataset may appear in multiple categories.}
Some domains attract more attention from dataset creators than others. For example, there are 14 datasets related to machine learning defects, which we further categorize into machine learning applications, machine learning frameworks, and machine learning compilers. The rest of this section discusses the application domains focused on by the software defect datasets, highlighting the uniqueness of defects in each domain.

\subsubsection{Defects Related to Machine Learning}
\label{sec-domain-ml}

The reliability and robustness of machine learning applications
and their underlying software platforms are crucial for their successful deployment. Machine
learning-related software defects can cause severe consequences, such
as incorrect predictions, biased decisions, and performance
degradation~\cite{DBLP:conf/issta/ZhangCCXZ18}.
We identify three subdomains of machine learning-related software,
each with several defect datasets: machine learning applications,
machine learning frameworks, and machine learning compilers.

\paragraph{Machine Learning Applications}
Defects in machine learning applications often arise during model
development, training, and deployment, leading to critical issues like
incorrect predictions or silent errors. Datasets such as
defect4ML~\cite{DBLP:journals/ese/MorovatiNKJ23},
gDefects4DL~\cite{DBLP:conf/icse/LiangLSSFD22}, and
Denchmark~\cite{DBLP:conf/msr/KimKL21} focus on common machine
learning defects, including API misuses, performance problems, and
numerical errors. There are also machine learning defect datasets
provided as by-products of empirical studies, such as studies on bugs
from programs relying on
\tensorflow~\cite{DBLP:conf/issta/ZhangCCXZ18}, on numerical defects
in deep learning programs~\cite{DBLP:conf/kbse/WangWCCY22}, and on
deployment-related bugs in machine learning-based mobile
applications~\cite{DBLP:conf/icse/ChenYLCLWL21}.

\paragraph{Machine Learning Frameworks}
Machine learning frameworks, such as PyTorch and TensorFlow, are
crucial for building and optimizing models, but themselves may suffer from defects. Software defect datasets for machine learning
frameworks typically result from empirical research. For example,
MOB-Dataset~\cite{DBLP:conf/icse/GuanXLLB23} collects real-world
machine learning optimization defects from \tensorflow and \pytorch,
and comes with an empirical study of model optimization. Likewise,
empirical studies of machine learning frameworks have produced
defect datasets of common bugs in
\pytorch~\cite{DBLP:conf/icsm/HoMI0SKNR23} and quantum
machine learning frameworks~\cite{DBLP:conf/qsw/ZhaoWLLZ23}.
Other work presents a dataset and a defect taxonomy of aging-related bugs in machine learning
libraries~\cite{DBLP:conf/issre/LiuZDHDM022}. These studies provide critical insights into the
defects that can impact the functionality, efficiency, and stability of machine
learning frameworks.

\paragraph{Machine Learning Compilers}
In addition to applications and frameworks, machine learning compilers
play a vital role in optimizing models,
but they also introduce unique bugs. Based on datasets
of defects in machine learning compilers, \citet{DBLP:conf/issre/Du0MZ21} and
\citet{DBLP:conf/sigsoft/ShenM0TCC21} both examine frequent issues
in these compilers, such as incorrect optimizations, numerical
inaccuracies, and incompatibilities with hardware accelerators.

\subsubsection{Defects in Compilers and Runtime Engines}
\label{sec-domain-compilers}

Compilers and runtime engines are critical components of software systems,
responsible for translating and executing code.
Defects in compilers and runtime
engines can significantly impact software performance, security,
and execution correctness.
We here mean by ``runtime engine'' both stand-alone implementations of a programming language, e.g., the Java Virtual Machine, and engines integrated into a larger software, e.g., a JavaScript engine embedded into a web browser.
Datasets of defects in compilers and runtime engines
often come as by-products of empirical research. For example,
\citet{DBLP:conf/qrs/YuW24} conduct an empirical study on bugs in implementations of the \rust
language\webcite{rustlang_github}, presenting a dataset of 194 bugs from \rust
compilers and \rust infrastructure. \citet{DBLP:conf/uss/XuLDDLWPM23} investigate
security bugs introduced by compilers for \cncpp, highlighting the potential
risks when silent bugs go undetected. In addition, as also mentioned above,
\citet{DBLP:conf/issre/Du0MZ21} and \citet{DBLP:conf/sigsoft/ShenM0TCC21} study
defects in deep-learning compilers, identifying common root causes and symptoms.
For runtime engines, Bugs in Pods~\cite{DBLP:conf/issta/YuXZ0LS24} provides 429
bugs in the container runtime systems containerd\webcite{containerd_github},
cri-o\webcite{crio_github}, gvisor\webcite{gvisor_github}, and
runc\webcite{runc_github}.  Additionally, \citet{DBLP:conf/icse/PaulTB21a} provide
a detailed dataset of security-related changes in the Chromium OS project, and
\citet{DBLP:conf/msr/ZamanAH12} present a dataset of performance-related bugs in
web browsers, focusing on how such bugs degrade system performance.
Besides browsers, datasets constructed from studies of other runtime engines
include defects in \webassembly
engines~\cite{DBLP:conf/wcre/WangZRLJ23,DBLP:journals/tosem/ZhangCWCLMMHL24},
and bugs that impact the correctness and performance of \javascript
engines~\cite{DBLP:journals/infsof/WangBWYGS23}. Altogether, these datasets are
critical for understanding and addressing defects in compilers and runtime
environments.

\subsubsection{Defects in Mobile Applications}
\label{sec-domain-mobile}

Mobile applications are prone to various types of defects,
including functional bugs, security vulnerabilities, and resource
issues.
Several datasets, mostly based on \android applications, have been developed
to represent these defects. For example,
LVDAndro~\cite{DBLP:conf/secrypt/SenanayakeKAP023} and
Ghera~\cite{DBLP:conf/promise/MitraR17} focus on security
vulnerabilities, providing labeled vulnerability datasets and
benchmarks for AI-based detection models and vulnerability analysis.
Functional bugs are collected by
Andror2~\cite{DBLP:conf/msr/WendlandSMMHMRF21} and by
\citet{DBLP:conf/issta/XiongX0SWWP0023}, both of which analyze the
behavior of defective mobile applications by manually reproducing the bugs.
More specialized defects, such as data loss issues and resource
leaks, are captured by
DataLossRepository~\cite{DBLP:conf/msr/RiganelliMMM19} and
DroidLeaks~\cite{DBLP:journals/ese/LiuWWXCWYZ19}, respectively. Finally,
DroixBench~\cite{DBLP:conf/icse/TanDGR18} provides defective APK files
particularly for evaluating automated repair techniques on \android
applications. These datasets play a crucial role in enhancing the
reliability and security of \android applications by supporting bug
detection, repair, and vulnerability research.

\subsubsection{Defects Related to Blockchain}
\label{sec-domain-blockchain}

Blockchain, as a decentralized and distributed ledger technology, has been
widely used in various applications, such as cryptocurrencies~\cite{DBLP:conf/bigdata/ZhengXDCW17}.
In general, blockchain applications are implemented as smart contracts, which
are self-executing contracts with the terms of the agreement between multiple
parties directly written into code.
Defects in blockchain applications can have severe implications, such as
financial loss.
Several datasets focus on identifying and
analyzing defects in smart contracts. From a general perspective,
Jiuzhou~\cite{DBLP:conf/icsm/ZhangXL20} is a dataset of \ethereum smart contract
bugs that are classified into 49 categories. Specifically for security
vulnerabilities, SBP-Dataset~\cite{DBLP:journals/jss/CaiLZZS24} contains smart
contract defects generated by removing guards and turning safe functions into
unsafe ones. In addition, there are two datasets constructed from empirical
studies. One defines 20 categories of smart contract
bugs~\cite{DBLP:journals/tse/ChenXLGLC22}, and the other identifies common
patterns of bug fixes~\cite{DBLP:journals/jss/WangCHZBZ23}.
Unlike the above datasets, \citet{DBLP:conf/msr/KimKL22} focus on energy bugs in
\ethereum client software instead of smart contracts. The dataset contains 507 bugs that may cause energy inefficiency mined from the buggy commits of client software operated in the \ethereum network.

\subsubsection{Defects in Cloud Computing}
Cloud computing systems provide on-demand access to computing resources over the internet and are prone to various software defects that can affect performance, reliability, and security.
Datasets of software defects in cloud computing often arise from empirical studies.
For instance, both~\citet{DBLP:conf/cloud/GunawiHLPDAELLM14} and~\citet{DBLP:conf/sigsoft/GaoDQGW0HZW18} investigate real-world issues in distributed cloud systems, with the latter focusing on crash recovery bugs.
Additionally, \citet{DBLP:conf/issta/XuG024} present a dataset of bugs in cloud orchestration systems, while~\citet{DBLP:journals/pacmpl/DrososSAM024} provide a dataset of misconfigurations in cloud infrastructure-as-code scripts.

\subsubsection{Defects in Autonomous Systems}

Autonomous systems, such as autonomous vehicles, aircraft, or robots, rely on
robust software to operate safely and efficiently. Defects in autonomous systems
can lead to severe consequences, such as accidents, injuries, and fatalities.
\citet{DBLP:conf/sigsoft/WangLX0S21} and \citet{DBLP:conf/icse/GarciaF0AXC20}
both present empirical studies of software defects in autonomous systems. The
former study has collected a dataset of 569 bugs from
PX4\webcite{px4_autopilot_github} and ArduPilot\webcite{ardupilot_github}, which are
well-known open-source autopilot software systems for unmanned aerial vehicles.
The latter study presents a dataset of 499 autonomous vehicle bugs from Apollo\webcite{apollo_github}
and Autoware\webcite{autoware_github}. In addition, ROBUST~\cite{DBLP:journals/ese/TimperleyHSDW24}
offers a dataset of 221 bugs from seven popular robot operating systems. Together,
these datasets highlight the unique challenges caused by defects in autonomous
systems.

\subsubsection{Defects Related to Quantum Computing}

Defects in quantum computing systems, including both quantum applications
and quantum platforms, present unique challenges due to the complexity
and novelty of the technology.
Bugs4Q~\cite{DBLP:journals/jss/ZhaoMLZ23} provides a benchmark of bugs
in quantum programs, particularly focusing on the \qiskit framework\webcite{qiskit_github},
enabling controlled testing and debugging studies. Additionally, there
are defect datasets constructed within empirical studies, such as one
that gathers and categorizes bug patterns across various quantum
platforms~\cite{DBLP:journals/pacmpl/PaltenghiP22}, and another that
focuses on defects within the intersection of quantum computing and
machine learning~\cite{DBLP:conf/qsw/ZhaoWLLZ23}. These datasets are
useful for advancing the reliability of quantum
software systems.

\subsubsection{Defects Related to Scientific Computing}

Scientific computing systems are often complex and require high performance and accuracy. Defects in scientific computing can lead to incorrect results, degraded performance, or even compromised system security. \citet{DBLP:conf/hotsos/MurphyBSR20} collect security defects from open-source scientific software projects written in \julia, a programming language designed for numerical analysis and computational science. Likewise, \citet{DBLP:conf/msr/AzadIHR23} build a dataset by investigating 1,729 performance bugs from 23 real-world high-performance computing (HPC) applications such as molecular simulation and computational fluid dynamics. In addition to scientific computing applications, \citet{DBLP:conf/kbse/FrancoGR17} study numerical bugs across scientific computing libraries and present a dataset of 269 numerical bugs from six libraries, including NumPy\webcite{numpy_github}, SciPy\webcite{scipy_github}, and LAPACK\webcite{lapack_github}. Collectively, these datasets provide valuable insights into the functional, performance, and security challenges faced in scientific computing.

\subsubsection{Other Application Domains}

In addition to the above, software defect datasets have been created
for various other application domains. For example, for database-related
software, datasets are available for both transaction bugs in database
systems~\cite{DBLP:conf/icse/CuiD0WSZWYXH0024} and access bugs in
database applications~\cite{DBLP:journals/tosem/LiuMC24}.
Other application domains include computational
notebooks~\cite{DBLP:journals/tosem/SantanaNAA24} and internet of
things (IoT)~\cite{DBLP:conf/icse/Makhshari021}.
While these domains have only a small number of datasets, these datasets provide insights into the challenges caused by domain-specific bugs.

\subsection{Types of Defects}
\label{sec-type}

\begin{table}[t]
    \centering
    \caption{Types of defects. One dataset may contain multiple types. }
    \tablefontsize
    \begin{tabularx}{\textwidth}{l|X|r}
        \toprule
        \textbf{Type}        & \textbf{Datasets} & \textbf{Count} \\
        \midrule
        Functional  & \cite{DBLP:journals/ese/TimperleyHSDW24,DBLP:conf/edcc/AndradeLV24,DBLP:conf/msr/LiuHLZCSHM24,DBLP:journals/tosem/LiuMC24,DBLP:journals/tosem/ZhangCWCLMMHL24,DBLP:conf/icse/CuiD0WSZWYXH0024,DBLP:conf/issta/YuXZ0LS24,DBLP:conf/issta/XuG024,DBLP:conf/ease/WaseemDA0M24,DBLP:conf/qrs/YuW24,DBLP:conf/acl/TianYQCLPWHL0024,DBLP:conf/iclr/JimenezYWYPPN24,DBLP:conf/msr/PramodSTSW24,DBLP:conf/msr/SilvaSM24,DBLP:conf/kbse/SongWCCLLWP23,DBLP:conf/icst/LeeKYY24,DBLP:conf/kbse/AnKCYY23,DBLP:conf/sigsoft/WuLZ024,DBLP:conf/kbse/AvulaVM23,DBLP:conf/msr/MahbubSR23,DBLP:journals/jss/ZhaoMLZ23,DBLP:conf/icst/KimH23,DBLP:journals/ese/MorovatiNKJ23,DBLP:conf/icse/SaavedraSM24,DBLP:journals/tse/JiangLLZCNZHBZ23,DBLP:conf/msr/ApplisP23,DBLP:conf/icse/GuanXLLB23,DBLP:conf/issta/XiongX0SWWP0023,DBLP:conf/msr/CsuvikV22,DBLP:journals/softx/PachoulyAK22,DBLP:conf/icse/LiangLSSFD22,DBLP:conf/msr/KeshavarzN22,DBLP:conf/wcre/ZhangYYCYZ22,DBLP:journals/tse/ChenXLGLC22,DBLP:conf/issta/Song0NW0DM22,DBLP:journals/pacmpl/PaltenghiP22,DBLP:conf/issre/LiuZDHDM022,DBLP:conf/msr/WendlandSMMHMRF21,DBLP:conf/apsec/AkimovaBDKKMM21,DBLP:journals/stvr/GyimesiVSMBFM21,DBLP:conf/msr/KimKL21,DBLP:conf/splc/NgoNNV21,DBLP:conf/msr/KamienskiPBH21,DBLP:conf/issre/Du0MZ21,DBLP:conf/icse/Makhshari021,DBLP:journals/jss/FerencGGTG20,DBLP:conf/icsm/ZhangXL20,DBLP:conf/msr/KarampatsisS20,DBLP:conf/icst/BuresHA20,DBLP:conf/sigsoft/WidyasariSLQPTT20,DBLP:conf/msr/WangBJS20,DBLP:conf/icse/GarciaF0AXC20,DBLP:conf/promise/VieiraSRG19,DBLP:conf/icse/BentonGZ19,DBLP:conf/icse/DmeiriTWBLDVR19,DBLP:conf/msr/RiganelliMMM19,DBLP:conf/wcre/DelfimUMM19,DBLP:conf/dsa/XuYWA19,DBLP:conf/msr/SahaLLYP18,DBLP:conf/promise/FerencTLSG18,DBLP:conf/issta/ZhangCCXZ18,DBLP:conf/msr/MadeyskiK17,DBLP:conf/icse/TanYYMR17,DBLP:conf/msr/OhiraKYYMLFHIM15,DBLP:journals/tse/GouesHSBDFW15,DBLP:journals/tse/GouesHSBDFW15,DBLP:conf/issta/JustJE14,DBLP:conf/msr/LamkanfiPD13,DBLP:journals/ese/DAmbrosLR12,DBLP:conf/cloud/GunawiHLPDAELLM14,DBLP:conf/icsm/HoMI0SKNR23,DBLP:journals/tosem/SantanaNAA24,DBLP:journals/infsof/WangBWYGS23,DBLP:journals/jss/WangCHZBZ23,DBLP:conf/qsw/ZhaoWLLZ23,DBLP:conf/wcre/WangZRLJ23,DBLP:conf/kbse/WangWCCY22,DBLP:conf/sigsoft/ShenM0TCC21,DBLP:conf/sigsoft/WangLX0S21,DBLP:conf/kbse/EghbaliP20,DBLP:conf/compsac/GuWL0019,DBLP:conf/kbse/FrancoGR17,DBLP:conf/oopsla/LinKCS17,DBLP:conf/kbse/YeCG23,DBLP:conf/icse/TanDGR18,lu2005bugbench,BegBunch,DBLP:conf/icse/HumbatovaJBR0T20,DBLP:conf/kbse/RomanoLK021,DBLP:conf/msr/AmannNNNM16,DBLP:conf/msr/OliverDAH24,DBLP:conf/sigsoft/GaoDQGW0HZW18,DBLP:conf/sigsoft/Zhang0C0Z19,DBLP:conf/wcre/ReyesGSBM24,DBLP:conf/msr/BeyerGKRR24,DBLP:journals/access/AntalVKMHF24,DBLP:journals/ese/TambonNAKA24,DBLP:journals/pacmpl/DrososSAM024,DBLP:journals/tosem/TufanoWBPWP19,DBLP:conf/kbse/XuZL23} & 100$^{*}$      \\
        \midrule
        Security    &      \cite{DBLP:conf/kbse/RuanLZL24,DBLP:conf/msr/NiSYZW24,DBLP:conf/pst/TebibAAG24,DBLP:journals/tosem/JiangJWMLZ24,DBLP:conf/issta/YuXZ0LS24,DBLP:conf/ease/WaseemDA0M24,DBLP:conf/kbse/DasAM24,DBLP:conf/qrs/YuW24,DBLP:conf/msr/PramodSTSW24,DBLP:journals/jss/CaiLZZS24,DBLP:conf/icse/WangHGWC024,DBLP:conf/kbse/ZhongLZGWGL23,DBLP:conf/kbse/AnKCYY23,DBLP:conf/raid/0001DACW23,DBLP:conf/secrypt/SenanayakeKAP023,DBLP:journals/compsec/BrustSG23,DBLP:conf/aina/PonteRM23a,DBLP:conf/icse/BhuiyanPVPS23,DBLP:conf/uss/ZhangP0W22,DBLP:conf/msr/BuiSF22,DBLP:conf/prdc/PereiraAV22,DBLP:journals/tse/ChenXLGLC22,DBLP:journals/stvr/GyimesiVSMBFM21,DBLP:conf/icse/PaulTB21a,DBLP:conf/icse/ZhengPLBEYLMS21,DBLP:conf/sigsoft/NikitopoulosDLM21,DBLP:journals/tse/AfroseXRMY23,DBLP:conf/icse/Makhshari021,DBLP:journals/jss/WuZCWM20,DBLP:conf/icsm/ZhangXL20,DBLP:conf/hotsos/MurphyBSR20,DBLP:conf/msr/FanL0N20,DBLP:conf/icse/GarciaF0AXC20,DBLP:conf/msr/RaduN19,DBLP:conf/msr/PontaPSBD19,DBLP:conf/icics/LinXZX19,DBLP:conf/msr/GkortzisMS18,DBLP:conf/promise/MitraR17,DBLP:conf/msr/OhiraKYYMLFHIM15,DBLP:conf/kbse/KuHCL07,DBLP:conf/cloud/GunawiHLPDAELLM14,DBLP:conf/uss/XuLDDLWPM23,DBLP:journals/jss/WangCHZBZ23,DBLP:conf/fie/SandersWA24,DBLP:conf/icse-apr/YadavW24,DBLP:conf/icse/HuZLYH024,DBLP:conf/issta/WuJPLD0BS23,DBLP:conf/sp/Dolan-GavittHKL16}    &    48   \\
        \midrule
        Performance &   \cite{DBLP:journals/tosem/LiuMC24,DBLP:conf/icse/CuiD0WSZWYXH0024,DBLP:conf/issta/YuXZ0LS24,DBLP:conf/ease/WaseemDA0M24,DBLP:conf/qrs/YuW24,DBLP:conf/msr/PramodSTSW24,DBLP:conf/icst/KimH23,DBLP:conf/icse/GuanXLLB23,DBLP:conf/msr/KimKL22,DBLP:journals/stvr/GyimesiVSMBFM21,DBLP:conf/issre/Du0MZ21,DBLP:conf/icse/Makhshari021,DBLP:journals/access/SanchezDMS20,DBLP:conf/icsm/ZhangXL20,DBLP:conf/msr/WangBJS20,DBLP:conf/msr/RaduN19,DBLP:conf/issta/ZhangCCXZ18,DBLP:conf/msr/OhiraKYYMLFHIM15,DBLP:conf/msr/ZamanAH12,DBLP:conf/se/SelakovicP17,DBLP:conf/cloud/GunawiHLPDAELLM14,DBLP:conf/icsm/HoMI0SKNR23,DBLP:conf/msr/AzadIHR23,DBLP:conf/qsw/ZhaoWLLZ23,DBLP:conf/wcre/WangZRLJ23,DBLP:conf/sigsoft/ShenM0TCC21,DBLP:conf/sigsoft/WangLX0S21,DBLP:conf/kbse/FrancoGR17,lu2005bugbench,DBLP:conf/icse/HumbatovaJBR0T20,DBLP:journals/ese/TambonNAKA24,DBLP:journals/pacmpl/DrososSAM024,DBLP:journals/tse/ZhaoXBCL23}       &   33    \\
        \midrule
        Concurrency &    \cite{DBLP:journals/ese/TimperleyHSDW24,DBLP:conf/asiaccs/LiangYDMHZ23,DBLP:conf/cgo/YuanLLLLX21,DBLP:journals/pacmpl/0001FSK20,DBLP:conf/msr/GaoYJLYZ18,DBLP:conf/kbse/LinMZCZ15,DBLP:conf/hotpar/JalbertPPS11,DBLP:conf/cloud/GunawiHLPDAELLM14,DBLP:conf/sigsoft/ShenM0TCC21,DBLP:conf/kbse/WangDGGQYW17,lu2005bugbench}      &    11   \\
        \midrule
        Others      &  \cite{DBLP:journals/ese/LiuWWXCWYZ19,DBLP:conf/icse/ChenYLCLWL21,DBLP:conf/iwpc/RenL020}    &     3  \\
        \bottomrule
    \end{tabularx}\\[1mm]
    \parbox[t]{\linewidth}{
    \raggedright
    \scriptsize
    $^{*}$The ManyBugs and IntroClass Benchmarks~\cite{DBLP:journals/tse/GouesHSBDFW15} are counted as two distinct datasets. This convention applies to all other tables.
        \par
    }
    \label{tab-type}
\end{table}

This section categorizes datasets based on the types of defects they
contain. Across the \numdataset software defect datasets we survey, we identify four commonly considered defect types: functional bugs, security vulnerabilities, performance bottlenecks, and concurrency issues. In addition, we include a fifth category, \textit{others}, to account for datasets that do not clearly fall into any of the four primary categories.

\begin{figure}[t]
\centering
\includegraphics[width=0.9\textwidth]{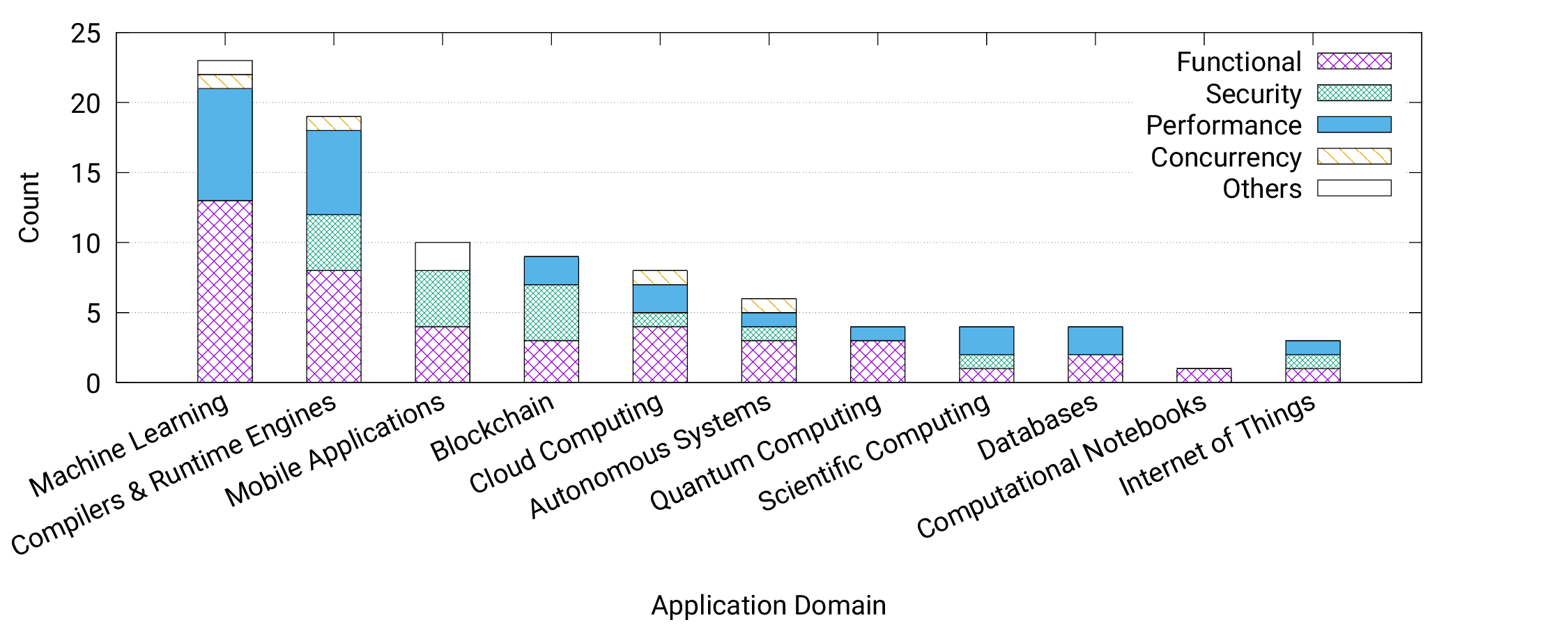}
\caption{Breakdown of types of defects in each application domain.}
\label{fig-domain-type}
\end{figure}

\cref{tab-type} lists the datasets for each category, where a single dataset may
appear under one or more categories. Functional defects are the most common type
of defects in software defect datasets, with 100 (\ratio{100}{151}) datasets in this
category. Security vulnerabilities are the second most common type, which are
present in 48 (\ratio{48}{151}) datasets. Performance defects and concurrency issues are less
common, with 33 (\ratio{33}{151}) and 11 (\ratio{11}{151}) datasets respectively.  \cref{fig-domain-type} shows the
breakdown of types of bugs for each of the application domains discussed in
\cref{sec-domain}.
We see that functional defects dominate most application domains
and other types are also prevalent in specific domains based on the nature of
the applications. The rest of this section further discusses these defect
types and the datasets that target them. For each defect type, we discuss the
characteristics of the defects and highlight representative datasets.

\subsubsection{Functional Defects}
\label{sec-type-functional}

Functional defects, or functional bugs, refer to errors in software
that cause the software to diverge from the intended behavior and to violate its (usually informally provided) specifications.
It is the most significant
type of software defect, not only in the context of software defect
datasets, but also in general software development~\cite{DBLP:conf/ftcs/SullivanC91,DBLP:conf/asplos/LiTWLZZ06}.
Functional defects impact
users in various ways, ranging from minor inconveniences, over
incorrect outputs, to application crashes. \cref{fig-domain-type} shows
that functional defect datasets dominate most application domains,
with the notable exceptions of blockchain and scientific computing.

Functional bugs are the most common focus in software defect datasets,
as they form the backbone of software reliability studies. These
datasets either provide a general collection of functional bugs across
various categories (\eg Defects4J~\cite{DBLP:conf/issta/JustJE14},
Bears~\cite{DBLP:conf/wcre/DelfimUMM19},
Bugs.jar~\cite{DBLP:conf/msr/SahaLLYP18},
BugSwarm~\cite{DBLP:conf/icse/DmeiriTWBLDVR19}) or target a specific
subset of functional bugs to facilitate detailed analysis and testing.
Examples of such specialized datasets include API
misuses~\cite{DBLP:conf/kbse/WangDGGQYW17}, regular expression
defects~\cite{DBLP:journals/ese/WangBJS22}, single-statement
bugs,~\cite{DBLP:conf/msr/KarampatsisS20}, type
errors~\cite{DBLP:conf/wcre/ZhangYYCYZ22}, string related
defects~\cite{DBLP:conf/kbse/EghbaliP20}, and numerical
bugs~\cite{DBLP:conf/kbse/WangWCCY22,DBLP:conf/kbse/FrancoGR17}.
Overall, functional bug datasets cover a wide variety of topics,
making them versatile resources for software defect research.

\subsubsection{Security Defects}
\label{sec-type-security}

Security defects, also known as vulnerabilities, are errors in software that make systems
vulnerable to attacks or unauthorized access. These defects can compromise the
integrity, confidentiality, and availability of a system, leading to severe
consequences, such as data breaches, system crashes, or unauthorized access to
sensitive information~\cite{DBLP:journals/csur/LandwehrBMC94}. Security defects
are particularly critical in software systems where data protection and privacy
are paramount, such as in financial, healthcare, or governmental applications.
While these defects can overlap with other types of bugs, their specific focus
is on potential exploits that can be leveraged by malicious actors. As security
is a major concern in software development, many datasets focus on identifying
and analyzing security vulnerabilities to strengthen defenses.
\cref{fig-domain-type} shows that datasets of security defects cover 7 out of 11
application domains, ranging from mobile applications to the internet of things.
The interest in security defects even surpasses that in functional defects for blockchain-related software,
which we attribute to the direct financial impact that vulnerabilities in this domain may have.

Different vulnerability datasets focus on different kinds of defects. One main type of security software defect dataset, e.g., ReposVul~\cite{DBLP:conf/icse/WangHGWC024},
Big-Vul\cite{DBLP:conf/msr/FanL0N20}, and
CVEjoin~\cite{DBLP:conf/aina/PonteRM23a}, focuses on
well-defined vulnerabilities, such as those cataloged in standardized
databases like CVE (Common Vulnerabilities and Exposures) and
categorized by CWE (Common Weakness Enumeration).
Another kind of dataset focuses on buffer
overflows~\cite{DBLP:conf/kbse/KuHCL07}, which are critical flaws
where programs write memory locations beyond buffer boundaries, and
compiler-introduced security bugs~\cite{DBLP:conf/uss/XuLDDLWPM23},
which arise due to incorrect compiler optimizations. Lastly,
cross-language vulnerabilities, which involve multiple
programming languages, are captured in datasets like
CrossVul~\cite{DBLP:conf/sigsoft/NikitopoulosDLM21}, providing a basis for multi-language defense techniques.

\subsubsection{Performance Defects}
\label{sec-type-performance}

Performance defects are software issues that cause performance
degradation, leading to inefficient use of resources, increased
latency, and longer execution times. These bugs are critical in
high-performance computing, web applications, and real-time systems
where responsiveness and resource management are essential~\cite{DBLP:conf/pldi/JinSSSL12}.
Performance bugs can arise from improper algorithms,
inefficient memory usage, excessive computation, or mismanagement of
concurrency. These defects can significantly impact user experience
and operational costs, making their identification and resolution a
priority in software engineering. As shown in
\cref{fig-domain-type}, datasets of performance issues are prevalent in 9 of 11
application domains, with the exceptions of mobile applications and computational notebooks.

Software defect datasets for performance issues have been constructed for
both general applications and specific domains. For example,
TANDEM~\cite{DBLP:journals/access/SanchezDMS20} offers a taxonomy and
dataset of real-world performance bugs, while
ECench~\cite{DBLP:conf/msr/KimKL22} introduces a benchmark of energy
bugs specifically in Ethereum client software, highlighting issues where
performance inefficiencies lead to excessive energy consumption. There
are also datasets constructed from empirical research of
performance issues, such as studies on performance degradation in
browser environments~\cite{DBLP:conf/msr/ZamanAH12}, in JavaScript
projects~\cite{DBLP:conf/se/SelakovicP17}, in database
systems~\cite{DBLP:conf/icse/CuiD0WSZWYXH0024} and
applications~\cite{DBLP:journals/tosem/LiuMC24}, in high-performance
computing (HPC) systems~\cite{DBLP:conf/msr/AzadIHR23}, in numerical libraries~\cite{DBLP:conf/kbse/FrancoGR17}, and in cloud
systems~\cite{DBLP:conf/cloud/GunawiHLPDAELLM14}. Such
datasets could facilitate future research on pinpointing or mitigating
performance issues.

\begin{table}[t]
    \centering
    \caption{Programming languages targeted in defect datasets. One dataset may cover multiple languages.}
    \tablefontsize

    \renewcommand{\arraystretch}{1.25}
    \begin{tabularx}{\textwidth}{l|X|r}
        \toprule
        \textbf{Language} & \textbf{Datasets}  & \textbf{Count} \\
        \midrule
        \java    & \cite{DBLP:conf/edcc/AndradeLV24,DBLP:conf/msr/LiuHLZCSHM24,DBLP:conf/pst/TebibAAG24,DBLP:journals/tosem/LiuMC24,DBLP:conf/issta/XuG024,DBLP:conf/kbse/DasAM24,DBLP:conf/acl/TianYQCLPWHL0024,DBLP:conf/msr/SilvaSM24,DBLP:conf/icse/WangHGWC024,DBLP:conf/kbse/SongWCCLLWP23,DBLP:conf/kbse/ZhongLZGWGL23,DBLP:conf/icst/LeeKYY24,DBLP:conf/sigsoft/WuLZ024,DBLP:conf/kbse/AvulaVM23,DBLP:conf/secrypt/SenanayakeKAP023,DBLP:journals/tse/JiangLLZCNZHBZ23,DBLP:conf/issta/XiongX0SWWP0023,DBLP:journals/softx/PachoulyAK22,DBLP:conf/msr/KeshavarzN22,DBLP:conf/msr/KimKL22,DBLP:conf/msr/BuiSF22,DBLP:conf/issta/Song0NW0DM22,DBLP:conf/msr/WendlandSMMHMRF21,DBLP:conf/splc/NgoNNV21,DBLP:journals/tse/AfroseXRMY23,DBLP:conf/icse/Makhshari021,DBLP:journals/jss/FerencGGTG20,DBLP:conf/msr/KarampatsisS20,DBLP:conf/icst/BuresHA20,DBLP:journals/pacmpl/0001FSK20,DBLP:conf/msr/WangBJS20,DBLP:conf/icse/DmeiriTWBLDVR19,DBLP:conf/msr/RaduN19,DBLP:conf/msr/RiganelliMMM19,DBLP:conf/wcre/DelfimUMM19,DBLP:conf/msr/SahaLLYP18,DBLP:conf/promise/FerencTLSG18,DBLP:conf/msr/GaoYJLYZ18,DBLP:conf/msr/MadeyskiK17,DBLP:conf/promise/MitraR17,DBLP:journals/ese/LiuWWXCWYZ19,DBLP:conf/kbse/LinMZCZ15,DBLP:conf/issta/JustJE14,DBLP:journals/ese/DAmbrosLR12,DBLP:conf/icse/ChenYLCLWL21,DBLP:conf/cloud/GunawiHLPDAELLM14,DBLP:conf/oopsla/LinKCS17,DBLP:conf/icse/TanDGR18,DBLP:conf/fie/SandersWA24,DBLP:conf/issta/WuJPLD0BS23,DBLP:conf/msr/AmannNNNM16,DBLP:conf/sigsoft/GaoDQGW0HZW18,DBLP:conf/sigsoft/Zhang0C0Z19,DBLP:conf/wcre/ReyesGSBM24,DBLP:journals/tosem/TufanoWBPWP19,DBLP:journals/tse/ZhaoXBCL23,DBLP:conf/kbse/XuZL23} & 57     \\
        \midrule
        \cncpp   & \cite{DBLP:journals/ese/TimperleyHSDW24,DBLP:conf/edcc/AndradeLV24,DBLP:conf/msr/LiuHLZCSHM24,DBLP:conf/pst/TebibAAG24,DBLP:journals/tosem/ZhangCWCLMMHL24,DBLP:journals/tosem/JiangJWMLZ24,DBLP:conf/icse/CuiD0WSZWYXH0024,DBLP:conf/issta/YuXZ0LS24,DBLP:conf/acl/TianYQCLPWHL0024,DBLP:conf/icse/WangHGWC024,DBLP:conf/kbse/ZhongLZGWGL23,DBLP:conf/kbse/AnKCYY23,DBLP:conf/kbse/AvulaVM23,DBLP:conf/icst/KimH23,DBLP:conf/asiaccs/LiangYDMHZ23,DBLP:conf/raid/0001DACW23,DBLP:conf/icse/GuanXLLB23,DBLP:conf/uss/ZhangP0W22,DBLP:conf/prdc/PereiraAV22,DBLP:journals/pacmpl/PaltenghiP22,DBLP:conf/issre/LiuZDHDM022,DBLP:conf/msr/KimKL21,DBLP:conf/icse/PaulTB21a,DBLP:conf/icse/ZhengPLBEYLMS21,DBLP:conf/issre/Du0MZ21,DBLP:conf/icse/Makhshari021,DBLP:conf/msr/FanL0N20,DBLP:conf/icse/GarciaF0AXC20,DBLP:conf/icics/LinXZX19,DBLP:conf/icse/TanYYMR17,DBLP:journals/tse/GouesHSBDFW15,DBLP:journals/tse/GouesHSBDFW15,DBLP:conf/msr/ZamanAH12,DBLP:conf/hotpar/JalbertPPS11,DBLP:conf/kbse/KuHCL07,DBLP:conf/icsm/HoMI0SKNR23,DBLP:conf/uss/XuLDDLWPM23,DBLP:journals/infsof/WangBWYGS23,DBLP:conf/msr/AzadIHR23,DBLP:conf/qsw/ZhaoWLLZ23,DBLP:conf/wcre/WangZRLJ23,DBLP:conf/sigsoft/ShenM0TCC21,DBLP:conf/sigsoft/WangLX0S21,DBLP:conf/compsac/GuWL0019,DBLP:conf/kbse/FrancoGR17,lu2005bugbench,BegBunch,DBLP:conf/icse-apr/YadavW24,DBLP:conf/msr/BeyerGKRR24,DBLP:journals/ese/TambonNAKA24,DBLP:journals/tse/ZhaoXBCL23,DBLP:conf/sp/Dolan-GavittHKL16} & 52     \\
        \midrule
        \python  & \cite{DBLP:journals/ese/TimperleyHSDW24,DBLP:conf/msr/LiuHLZCSHM24,DBLP:conf/acl/TianYQCLPWHL0024,DBLP:conf/iclr/JimenezYWYPPN24,DBLP:conf/icse/WangHGWC024,DBLP:conf/kbse/ZhongLZGWGL23,DBLP:conf/sigsoft/WuLZ024,DBLP:conf/kbse/AvulaVM23,DBLP:conf/msr/MahbubSR23,DBLP:journals/jss/ZhaoMLZ23,DBLP:journals/ese/MorovatiNKJ23,DBLP:conf/icse/GuanXLLB23,DBLP:conf/icse/LiangLSSFD22,DBLP:conf/wcre/ZhangYYCYZ22,DBLP:journals/pacmpl/PaltenghiP22,DBLP:conf/issre/LiuZDHDM022,DBLP:conf/apsec/AkimovaBDKKMM21,DBLP:conf/msr/KimKL21,DBLP:conf/msr/KamienskiPBH21,DBLP:conf/issre/Du0MZ21,DBLP:conf/icse/Makhshari021,DBLP:conf/sigsoft/WidyasariSLQPTT20,DBLP:conf/msr/WangBJS20,DBLP:conf/icse/DmeiriTWBLDVR19,DBLP:conf/msr/RaduN19,DBLP:conf/issta/ZhangCCXZ18,DBLP:conf/icse/ChenYLCLWL21,DBLP:conf/icsm/HoMI0SKNR23,DBLP:journals/tosem/SantanaNAA24,DBLP:conf/qsw/ZhaoWLLZ23,DBLP:conf/kbse/WangWCCY22,DBLP:conf/sigsoft/ShenM0TCC21,DBLP:conf/kbse/FrancoGR17,DBLP:conf/oopsla/LinKCS17,DBLP:conf/icse/HumbatovaJBR0T20,DBLP:journals/access/AntalVKMHF24,DBLP:journals/ese/TambonNAKA24,DBLP:journals/pacmpl/DrososSAM024,DBLP:journals/tse/ZhaoXBCL23} & 39     \\
        \midrule
        \javascript & \cite{DBLP:conf/kbse/ZhongLZGWGL23,DBLP:conf/icse/BhuiyanPVPS23,DBLP:conf/msr/CsuvikV22,DBLP:journals/stvr/GyimesiVSMBFM21,DBLP:conf/icse/Makhshari021,DBLP:conf/msr/WangBJS20,DBLP:conf/se/SelakovicP17,DBLP:journals/infsof/WangBWYGS23,DBLP:conf/wcre/WangZRLJ23,DBLP:conf/kbse/EghbaliP20,DBLP:conf/kbse/WangDGGQYW17,DBLP:conf/msr/OliverDAH24,DBLP:conf/kbse/RomanoLK021} & 13     \\
        \midrule
        \go & \cite{DBLP:conf/icse/CuiD0WSZWYXH0024,DBLP:conf/issta/YuXZ0LS24,DBLP:conf/issta/XuG024,DBLP:conf/icse/SaavedraSM24,DBLP:conf/msr/KimKL22,DBLP:conf/cgo/YuanLLLLX21,DBLP:conf/icse/HuZLYH024} & 7     \\
        \midrule
        \csharp & \cite{DBLP:conf/edcc/AndradeLV24,DBLP:conf/kbse/ZhongLZGWGL23,DBLP:conf/msr/KimKL22,DBLP:journals/pacmpl/PaltenghiP22,DBLP:conf/fie/SandersWA24} & 5     \\
        \midrule
        \rust & \cite{DBLP:journals/tosem/ZhangCWCLMMHL24,DBLP:conf/qrs/YuW24,DBLP:conf/msr/KimKL22,DBLP:conf/wcre/WangZRLJ23,DBLP:conf/kbse/RomanoLK021} & 5     \\
        \midrule
        \solidity & \cite{DBLP:journals/jss/CaiLZZS24,DBLP:journals/tse/ChenXLGLC22,DBLP:conf/icsm/ZhangXL20,DBLP:journals/jss/WangCHZBZ23} & 4     \\
        \midrule
        \kotlin & \cite{DBLP:conf/pst/TebibAAG24,DBLP:conf/issta/XiongX0SWWP0023,DBLP:conf/icse/BentonGZ19} & 3     \\
        \midrule
        \php & \cite{DBLP:conf/edcc/AndradeLV24,DBLP:conf/msr/PramodSTSW24} & 2     \\
        \midrule
        \ruby & \cite{DBLP:conf/msr/MadeyskiK17,DBLP:journals/pacmpl/DrososSAM024} & 2     \\
        \midrule
        Others & \fortran~\cite{DBLP:conf/kbse/FrancoGR17}, \fsharp~\cite{DBLP:journals/pacmpl/PaltenghiP22}, \groovy~\cite{DBLP:conf/icse/BentonGZ19}, \haskell~\cite{DBLP:conf/msr/ApplisP23}, \julia~\cite{DBLP:conf/hotsos/MurphyBSR20}, \qsharp~\cite{DBLP:journals/pacmpl/PaltenghiP22}, \scala~\cite{DBLP:journals/pacmpl/0001FSK20}  & 7     \\

        \bottomrule
    \end{tabularx}
    \label{tab-pl}
\end{table}

\subsubsection{Concurrency Defects}
\label{sec-type-concurrency}

Concurrency defects arise in software when multiple threads or
processes execute simultaneously and interact in unintended ways,
leading to nondeterministic behavior, crashes, or data corruption.
These defects are particularly challenging because they often do not
manifest consistently, making them difficult to detect, reproduce, and
fix. Concurrency bugs include race conditions, deadlocks, and
atomicity violations, which occur when proper synchronization
mechanisms fail, or incorrect access to shared resources occurs. Given
the complexity of multithreaded systems and the rise of concurrent
programming models, identifying and studying concurrency bugs has
become critical in ensuring the stability and reliability of modern
software systems. As shown in \cref{fig-domain-type}, concurrency
defects exist in application domains where parallelism and
real-time execution are critical, e.g., machine learning, compilers and
runtime engines, cloud computing, and autonomous systems.

Several datasets have been developed to address different aspects of
concurrency bugs. For C/C++,
RaceBench~\cite{DBLP:conf/asiaccs/LiangYDMHZ23} provides
triggerable and observable race condition bugs to test race detectors,
and RADBench~\cite{DBLP:conf/hotpar/JalbertPPS11} offers a concurrency
bug benchmark suite focused on race conditions, deadlocks, and
atomicity violations. For Java, Jbench~\cite{DBLP:conf/msr/GaoYJLYZ18}
presents a suite of data race bugs specifically designed for
concurrency testing, and JaConTeBe~\cite{DBLP:conf/kbse/LinMZCZ15}
focuses on concurrency bugs in real-world projects, offering
manually written test cases for verification. Additionally, there are
some datasets of concurrency bugs constructed from empirical studies, e.g.,
on actor-based concurrency
systems~\cite{DBLP:journals/pacmpl/0001FSK20}, cloud computing
systems~\cite{DBLP:conf/cloud/GunawiHLPDAELLM14}, deep learning
compilers~\cite{DBLP:conf/sigsoft/ShenM0TCC21}, and
Node.js~\cite{DBLP:conf/kbse/WangDGGQYW17}, which offer invaluable
insights for future studies.

\subsubsection{Other Types of Defects}
\label{sec-type-others}

Some datasets do not fall neatly into the four categories of
functional, security, performance, or concurrency bugs. For example,
such datasets consider critical bugs that prevent other bugs from
being resolved~\cite{DBLP:conf/iwpc/RenL020}, deployment faults in
deep learning-based mobile applications~\cite{DBLP:conf/icse/ChenYLCLWL21}, or
resource leak bugs in \android
applications~\cite{DBLP:journals/ese/LiuWWXCWYZ19}. These datasets address specialized defect types that fall outside the four main categories.

\subsection{Programming Languages of Defective Software}
\label{sec-pl}

Another way of categorizing datasets is by the programming language in which the defective code is written.
\ratio{134}{151} of our surveyed defect datasets focus on one or multiple programming languages.
These languages include \java, \python, \cncpp, \javascript, \go, and
others. \cref{tab-pl} presents the distribution of programming
languages across software defect datasets. \java stands out as the
most represented language, with a total of 57 (\ratio{57}{151}) datasets, followed by
\cncpp (52 datasets, \ratio{52}{151}) and \python (39 datasets, \ratio{39}{151}). The dominance of these
languages is likely due to their popularity in software development
and due to the focus of the research community on these languages.
Other languages, such as \fsharp, \groovy, \haskell, and \julia, have
fewer datasets available.

\begin{figure}[t]
\centering
\includegraphics[width=0.9\textwidth]{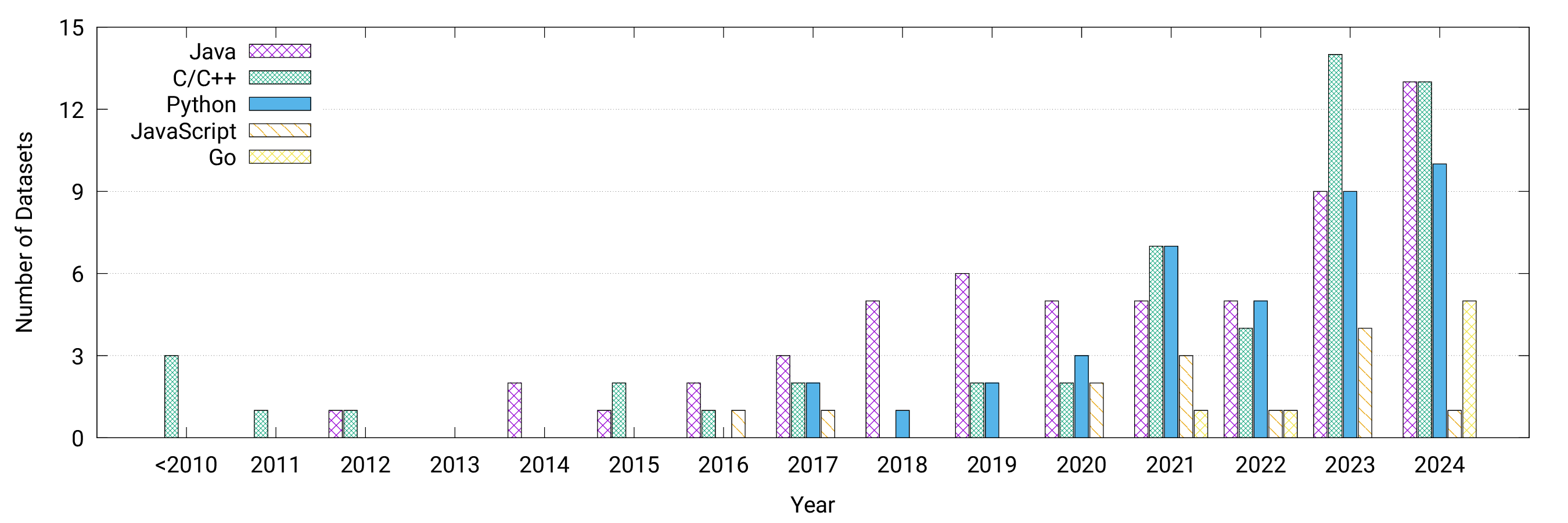}
\caption{Trend of programming languages in software defect datasets.}
\label{fig-pl-trend}
\end{figure}

The distribution of languages in defect datasets has evolved over time.
\cref{fig-pl-trend} shows the trend for the top five programming languages: \java, \cncpp, \python, \javascript, and \go.
These five languages also represent the top five most used languages on \github~\cite{githut}, suggesting that dataset creation reflects real-world development trends.
\java datasets have increased steadily over the past decade, while \python has gained significant popularity since 2017.
\javascript and \go have experienced peaks within the past five years, and \cncpp datasets have appeared consistently over time.
It is important to note the surge in software defect datasets since 2023.
While the general increase in publication volume contributes to this rise, another factor is the growing interest in domain- or type-specific defects.
The surge is particularly pronounced for \cncpp{} datasets, with 14 papers in 2023 (a 250\% increase from 2022), driven by growing interest in performance bugs \cite{DBLP:journals/tse/ZhaoXBCL23, DBLP:conf/msr/AzadIHR23}, security vulnerabilities \cite{DBLP:conf/raid/0001DACW23, DBLP:conf/uss/XuLDDLWPM23}, and bugs in emerging domains such as machine learning frameworks \cite{DBLP:conf/icse/GuanXLLB23, DBLP:conf/icsm/HoMI0SKNR23}, compilers \cite{DBLP:conf/wcre/WangZRLJ23, DBLP:journals/infsof/WangBWYGS23}, and quantum computing \cite{DBLP:conf/qsw/ZhaoWLLZ23}.
The trend continues in 2024 with 13 \cncpp{} datasets, suggesting sustained research momentum in infrastructure-level defect analysis.
Another contributing factor is the rising demand for large-scale datasets to train and evaluate machine learning and large language models, which has spurred automated bug mining and benchmark construction studies in 2023~\cite{DBLP:conf/kbse/ZhongLZGWGL23,DBLP:conf/kbse/AnKCYY23,DBLP:journals/tse/JiangLLZCNZHBZ23,DBLP:conf/kbse/SongWCCLLWP23,DBLP:conf/aina/PonteRM23a,DBLP:conf/msr/MahbubSR23,DBLP:conf/raid/0001DACW23,DBLP:conf/icst/KimH23,DBLP:conf/kbse/AvulaVM23,DBLP:conf/kbse/YeCG23}.
This trend will be further discussed in \cref{sec-usage}.

\begin{figure}[t]
\centering
\begin{minipage}[t]{0.48\textwidth}
  \centering
  \includegraphics[width=\textwidth]{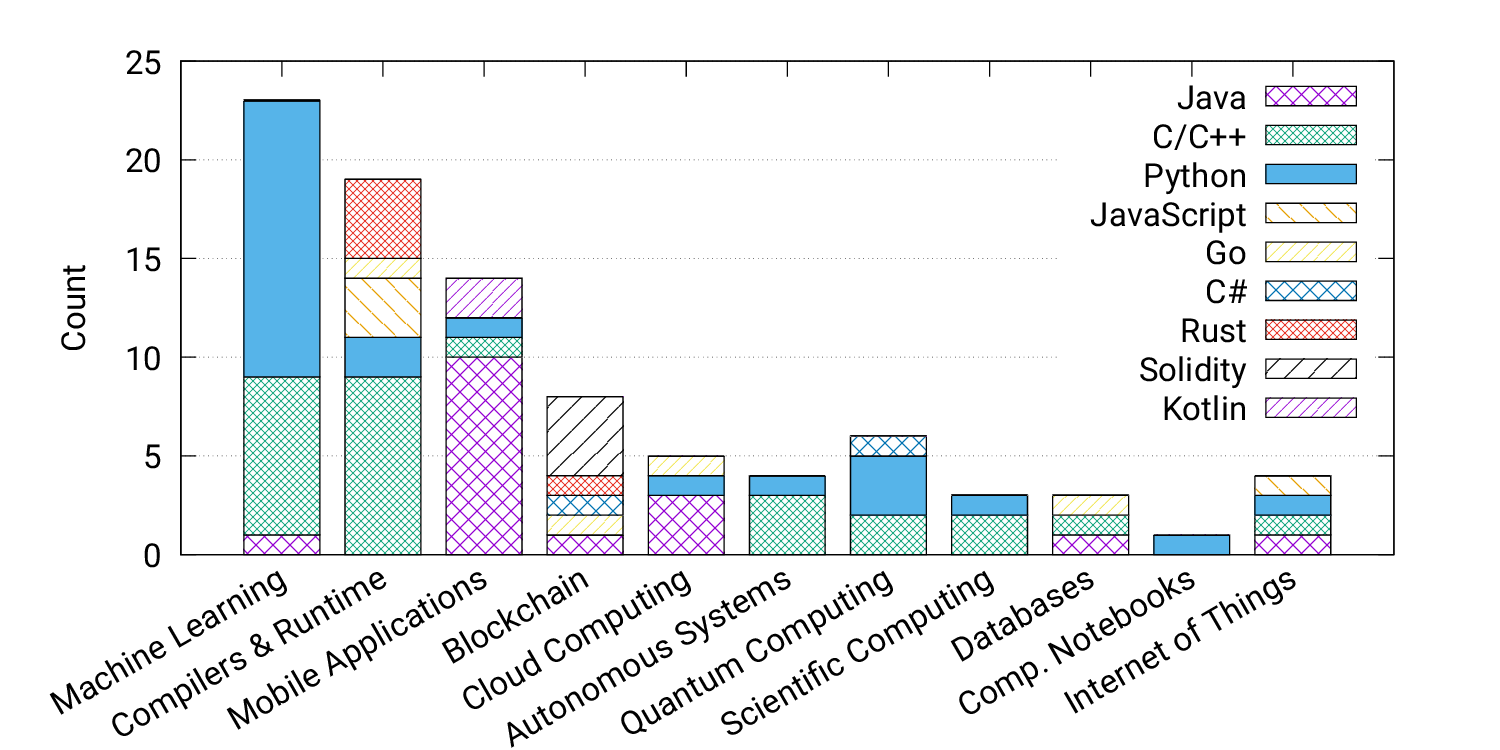}
  \caption{Programming languages across domains.}
  \label{fig-domain-pl}
\end{minipage}\hfill
\begin{minipage}[t]{0.48\textwidth}
  \centering
  \includegraphics[width=\textwidth]{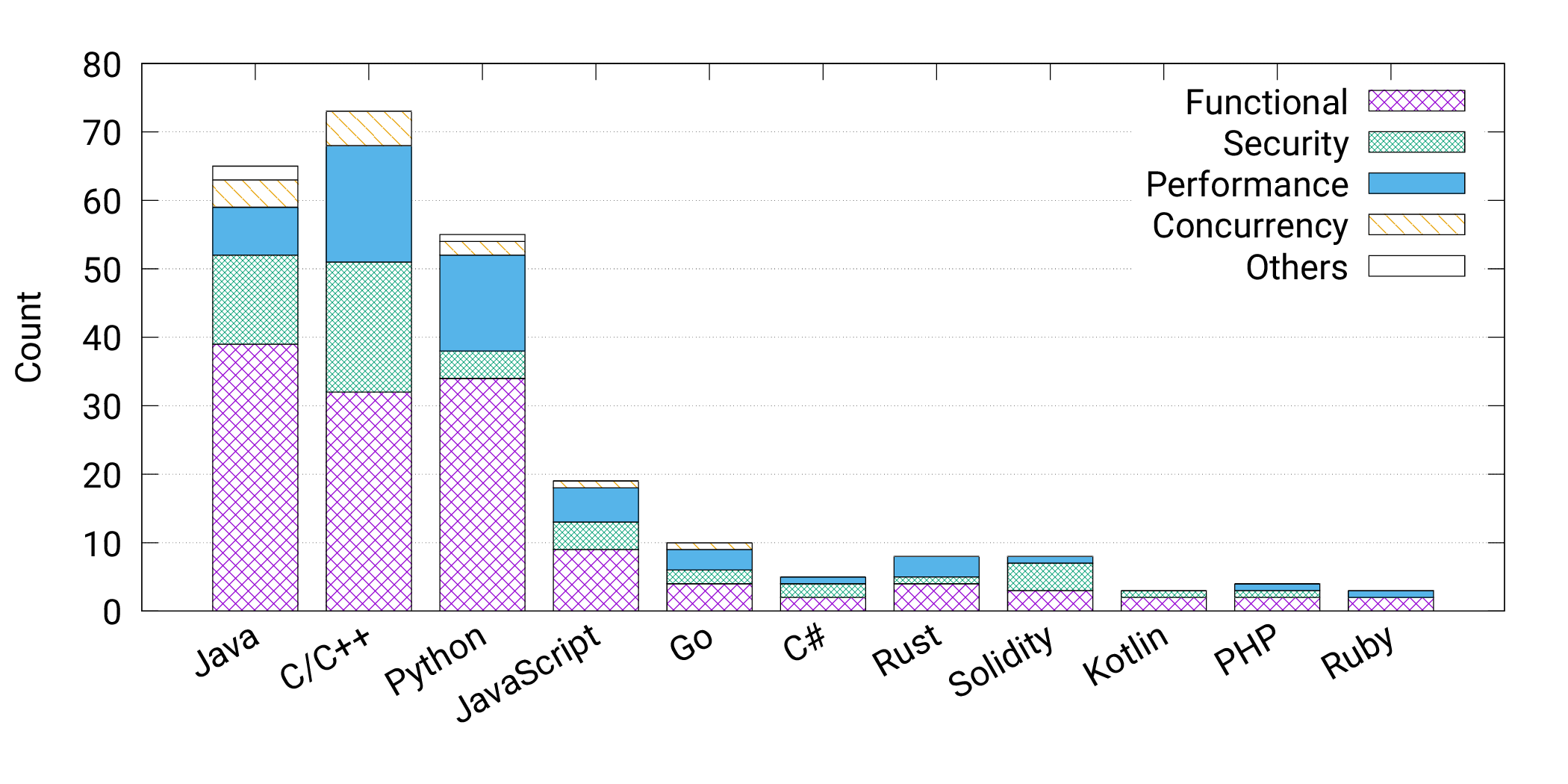}
  \caption{Defect types by programming languages.}
  \label{fig-pl-type-bar}
\end{minipage}
\end{figure}

\cref{fig-domain-pl} shows the distribution of the most popular
programming languages across different application domains. \python
and \cncpp dominate machine learning-related defect datasets, and
\java and \kotlin cover the major datasets for mobile applications.
Compilers and runtime engines have a variety of programming languages,
including \cncpp, \rust, \python, \javascript, and \go.
Most blockchain-related defects are from \solidity smart contracts, except for one dataset that collects
defects from blockchain clients~\cite{DBLP:conf/msr/KimKL22}, which are implemented in \java, \go, \csharp, and \rust. In other domains, such as
autonomous systems, quantum computing, scientific computing, and
computational notebooks, \cncpp and \python are commonly used, whereas datasets for cloud computing, databases, and the internet of things involve \java,
\go, \cncpp, \python, and \javascript.

Defect types are not equally distributed across programming languages.
\cref{fig-pl-type-bar} shows the distribution of software defect datasets across different programming languages and defect types.
\java, \cncpp, \python, \javascript, and \go are common across all defect types, while \csharp, \rust, \solidity, and \php cover functional, security, and performance defects only.
\kotlin and \ruby have the least coverage, with \kotlin covering only functional and security defects, and \ruby covering only functional and performance defects.

\begin{table}[t]
    \centering
    \caption{Source of defects. One dataset may be based on multiple sources.}
    \tablefontsize
    \begin{tabularx}{\textwidth}{l|X|r}
        \toprule
        \textbf{Source}        & \textbf{Datasets} & \textbf{Count} \\
        \midrule
        Issue Reports & \cite{DBLP:journals/ese/TimperleyHSDW24,DBLP:conf/edcc/AndradeLV24,DBLP:journals/tosem/LiuMC24,DBLP:journals/tosem/ZhangCWCLMMHL24,DBLP:conf/icse/CuiD0WSZWYXH0024,DBLP:conf/issta/YuXZ0LS24,DBLP:conf/issta/XuG024,DBLP:conf/ease/WaseemDA0M24,DBLP:conf/kbse/DasAM24,DBLP:conf/qrs/YuW24,DBLP:conf/iclr/JimenezYWYPPN24,DBLP:journals/jss/ZhaoMLZ23,DBLP:journals/tse/JiangLLZCNZHBZ23,DBLP:conf/msr/ApplisP23,DBLP:conf/icse/BhuiyanPVPS23,DBLP:conf/icse/GuanXLLB23,DBLP:conf/issta/XiongX0SWWP0023,DBLP:journals/softx/PachoulyAK22,DBLP:conf/icse/LiangLSSFD22,DBLP:conf/msr/KeshavarzN22,DBLP:conf/issre/LiuZDHDM022,DBLP:conf/msr/WendlandSMMHMRF21,DBLP:conf/msr/KimKL21,DBLP:conf/icse/PaulTB21a,DBLP:conf/issre/Du0MZ21,DBLP:conf/icse/Makhshari021,DBLP:journals/jss/FerencGGTG20,DBLP:journals/jss/WuZCWM20,DBLP:conf/icse/GarciaF0AXC20,DBLP:conf/promise/VieiraSRG19,DBLP:conf/msr/RiganelliMMM19,DBLP:conf/msr/PontaPSBD19,DBLP:conf/msr/SahaLLYP18,DBLP:conf/issta/ZhangCCXZ18,DBLP:conf/msr/OhiraKYYMLFHIM15,DBLP:conf/kbse/LinMZCZ15,DBLP:journals/tse/GouesHSBDFW15,DBLP:conf/issta/JustJE14,DBLP:conf/msr/LamkanfiPD13,DBLP:journals/ese/DAmbrosLR12,DBLP:conf/msr/ZamanAH12,DBLP:conf/hotpar/JalbertPPS11,DBLP:conf/icse/ChenYLCLWL21,DBLP:conf/se/SelakovicP17,DBLP:conf/cloud/GunawiHLPDAELLM14,DBLP:conf/icsm/HoMI0SKNR23,DBLP:conf/uss/XuLDDLWPM23,DBLP:journals/infsof/WangBWYGS23,DBLP:conf/qsw/ZhaoWLLZ23,DBLP:conf/wcre/WangZRLJ23,DBLP:conf/kbse/WangWCCY22,DBLP:conf/sigsoft/ShenM0TCC21,DBLP:conf/sigsoft/WangLX0S21,DBLP:conf/iwpc/RenL020,DBLP:conf/kbse/FrancoGR17,DBLP:conf/kbse/WangDGGQYW17,DBLP:conf/icse/TanDGR18,DBLP:conf/icse/HumbatovaJBR0T20,DBLP:conf/kbse/RomanoLK021,DBLP:conf/msr/OliverDAH24,DBLP:conf/sigsoft/GaoDQGW0HZW18,DBLP:conf/msr/BeyerGKRR24,DBLP:journals/access/AntalVKMHF24,DBLP:journals/ese/TambonNAKA24,DBLP:journals/pacmpl/DrososSAM024,DBLP:journals/tse/ZhaoXBCL23} & 66 \\
        \midrule
        Version Control Systems & \cite{DBLP:journals/ese/TimperleyHSDW24,DBLP:conf/pst/TebibAAG24,DBLP:conf/issta/YuXZ0LS24,DBLP:conf/msr/PramodSTSW24,DBLP:conf/icse/WangHGWC024,DBLP:conf/kbse/SongWCCLLWP23,DBLP:conf/kbse/ZhongLZGWGL23,DBLP:conf/icst/LeeKYY24,DBLP:conf/kbse/AnKCYY23,DBLP:conf/kbse/AvulaVM23,DBLP:conf/msr/MahbubSR23,DBLP:journals/ese/MorovatiNKJ23,DBLP:conf/raid/0001DACW23,DBLP:journals/tse/JiangLLZCNZHBZ23,DBLP:conf/msr/ApplisP23,DBLP:conf/msr/CsuvikV22,DBLP:conf/msr/KimKL22,DBLP:conf/wcre/ZhangYYCYZ22,DBLP:conf/msr/BuiSF22,DBLP:conf/issta/Song0NW0DM22,DBLP:journals/pacmpl/PaltenghiP22,DBLP:conf/apsec/AkimovaBDKKMM21,DBLP:journals/stvr/GyimesiVSMBFM21,DBLP:conf/cgo/YuanLLLLX21,DBLP:conf/icse/ZhengPLBEYLMS21,DBLP:conf/sigsoft/NikitopoulosDLM21,DBLP:conf/msr/KamienskiPBH21,DBLP:conf/msr/KarampatsisS20,DBLP:conf/sigsoft/WidyasariSLQPTT20,DBLP:conf/msr/WangBJS20,DBLP:conf/icse/BentonGZ19,DBLP:conf/msr/RaduN19,DBLP:conf/dsa/XuYWA19,DBLP:journals/ese/LiuWWXCWYZ19,DBLP:conf/issta/JustJE14,DBLP:journals/tosem/SantanaNAA24,DBLP:journals/jss/WangCHZBZ23,DBLP:conf/msr/AzadIHR23,DBLP:conf/kbse/EghbaliP20,DBLP:conf/compsac/GuWL0019,DBLP:conf/icse/HumbatovaJBR0T20,DBLP:conf/issta/WuJPLD0BS23,DBLP:conf/kbse/RomanoLK021,DBLP:conf/msr/AmannNNNM16,DBLP:conf/msr/BeyerGKRR24,DBLP:journals/access/AntalVKMHF24,DBLP:journals/pacmpl/DrososSAM024,DBLP:journals/tosem/TufanoWBPWP19,DBLP:journals/tse/ZhaoXBCL23,DBLP:conf/kbse/XuZL23} & 50 \\
        \midrule
        Vulnerability Databases & \cite{DBLP:conf/kbse/RuanLZL24,DBLP:conf/msr/NiSYZW24,DBLP:journals/tosem/JiangJWMLZ24,DBLP:conf/icse/WangHGWC024,DBLP:conf/aina/PonteRM23a,DBLP:conf/prdc/PereiraAV22,DBLP:conf/sigsoft/NikitopoulosDLM21,DBLP:conf/msr/FanL0N20,DBLP:conf/icics/LinXZX19,DBLP:conf/msr/GkortzisMS18,DBLP:conf/kbse/KuHCL07,DBLP:conf/kbse/YeCG23,BegBunch,DBLP:conf/icse/HuZLYH024,DBLP:conf/issta/WuJPLD0BS23} & 15 \\
        Other Pre-Existing Datasets & \cite{DBLP:conf/icst/KimH23,DBLP:conf/msr/BuiSF22,DBLP:conf/msr/KamienskiPBH21,DBLP:journals/access/SanchezDMS20,DBLP:conf/icsm/ZhangXL20,DBLP:conf/promise/FerencTLSG18,DBLP:conf/kbse/YeCG23,DBLP:conf/msr/AmannNNNM16,DBLP:conf/msr/OliverDAH24,DBLP:conf/sigsoft/GaoDQGW0HZW18} & 10 \\
        \midrule
        Developer \qa Forums & \cite{DBLP:journals/tosem/ZhangCWCLMMHL24,DBLP:journals/jss/ZhaoMLZ23,DBLP:journals/ese/MorovatiNKJ23,DBLP:journals/tse/ChenXLGLC22,DBLP:journals/pacmpl/0001FSK20,DBLP:conf/issta/ZhangCCXZ18,DBLP:conf/icse/ChenYLCLWL21,DBLP:journals/tosem/SantanaNAA24,DBLP:conf/kbse/WangWCCY22,DBLP:conf/icse/HumbatovaJBR0T20} & 10 \\
        \midrule
        \cicd & \cite{DBLP:conf/msr/SilvaSM24,DBLP:conf/icse/SaavedraSM24,DBLP:conf/hotsos/MurphyBSR20,DBLP:conf/icse/DmeiriTWBLDVR19,DBLP:conf/wcre/DelfimUMM19,DBLP:conf/msr/MadeyskiK17,DBLP:conf/sigsoft/Zhang0C0Z19,DBLP:conf/wcre/ReyesGSBM24} & 8 \\
        \midrule
        Defect Synthesis & \cite{DBLP:conf/acl/TianYQCLPWHL0024,DBLP:journals/jss/CaiLZZS24,DBLP:conf/asiaccs/LiangYDMHZ23,DBLP:conf/uss/ZhangP0W22,DBLP:conf/splc/NgoNNV21,DBLP:conf/icst/BuresHA20,DBLP:conf/sp/Dolan-GavittHKL16,DBLP:conf/icse-apr/YadavW24} & 8 \\
        \midrule
        Coding Assignments & \cite{DBLP:conf/msr/LiuHLZCSHM24,DBLP:conf/acl/TianYQCLPWHL0024,DBLP:conf/sigsoft/WuLZ024,DBLP:conf/icse/TanYYMR17,DBLP:journals/tse/GouesHSBDFW15,DBLP:conf/oopsla/LinKCS17,DBLP:conf/fie/SandersWA24} & 7 \\
        \bottomrule
    \end{tabularx}
    \label{tab-source}
\end{table}

\begin{tcolorbox}[
  colback=white,
  colframe=gray!140,
  boxsep=2pt,
  left=2pt,
  right=2pt,
  top=1pt,
  bottom=1pt,
  title=\textbf{RQ1 Summary},
  fonttitle=\bfseries
]
\vspace{0.25em}
\begin{itemize}[leftmargin=10pt, topsep=2pt, itemsep=2pt]

    \item Over 60\% of the datasets target general-purpose software, while the rest cover eleven specific domains including machine learning, compilers \& runtime engines, mobile applications, and blockchain.
    Machine learning is the most popular specific domain with 14 datasets.

    \item Software defect datasets cover functional, security, performance, and concurrency defects.
    Functional bugs dominate (\ratio{100}{151}), followed by security vulnerabilities (\ratio{48}{151}), performance issues (\ratio{33}{151}), and concurrency defects (\ratio{11}{151}).
    Security and performance datasets are more common in domain-specific contexts such as compilers \& runtime engines or blockchain.

    \item Java, C/C++, and Python appear in over 75\% of all datasets, with JavaScript and Go showing increasing presence recently.
    Python and C/C++ predominate in machine learning and compilers \& runtime engines, while Java and Kotlin dominate mobile applications.
    Language trends partially align with defect types: \java, \cncpp, \python, \javascript, and \go span all defect types, while other languages associate with specific types.

\end{itemize}
\end{tcolorbox}

\section{Dataset Construction (RQ2)}
\label{sec-construction}

Creators of new defect datasets must take various decisions that will impact the quality, quantity, representativeness, and usability of the dataset.
In particular, the process of constructing a defect dataset involves selecting an
appropriate source and a suitable methodology.
There are various sources of defects, such
as bug reports from \github or \apache \jira, continuous integration
systems like Travis-CI or \github Actions, and other
pre-existing datasets.
To extract defects from these sources, existing methodologies range from fully automated, rule-based approaches to
semi-automated techniques that incorporate human validation or
machine learning models, striving to ensure that the defects collected
are representative of real-world scenarios.
This section discusses the sources of defects (Section~\ref{sec-source}) and
the construction methodologies (Section~\ref{sec-mining-methodology}) employed by
the datasets considered in our study.

\subsection{Source of Defects}
\label{sec-source}

\begin{figure}[t]
    \centering
    \includegraphics[width=0.75\textwidth]{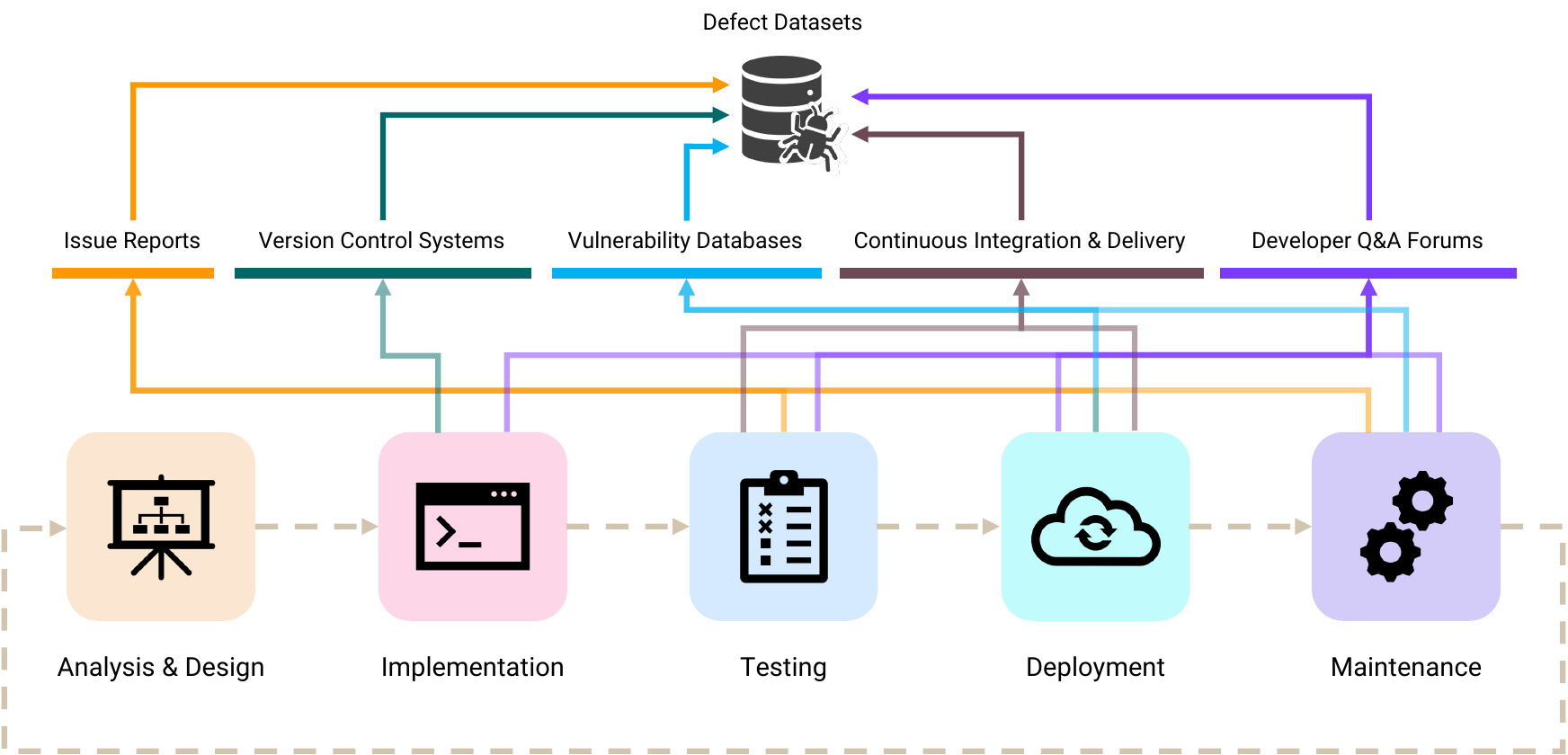}
    \caption{Methods for collecting defects (top) from stages in the software development processes (bottom).}
    \label{fig-source-se-process}
\end{figure}

The \textit{source} of a defect refers to the origin or platform from
which the defect data is collected. As listed in \cref{tab-source},
there are diverse sources for dataset creators to mine defects from,
such as issue reports, version control systems, vulnerability
databases, developer \qa forums, and others. Most of these sources are closely tied to the software development process,
as illustrated in \cref{fig-source-se-process}, especially in software
implementation, testing, deployment, and maintenance stages. Other
sources, such as coding assignments and defect synthesis, offer a more
controlled environment for collecting or creating defects. Among these sources, issue reports and version control systems are the most popular, accounting for 66 (\ratio{66}{151}) and 50 (\ratio{50}{151}) software defect datasets, respectively. Coding assignments, \cicd, and defect synthesis are less frequently used, with no more than 8 (\ratio{8}{151}) datasets utilizing each. The rest of this section provides a detailed discussion of the primary sources of defects in software defect datasets, highlighting their unique characteristics.

\subsubsection{Issue Reports}
\label{sec-source-issue}
A bug issue report is a documented record of a defect or problem
encountered in a software system, typically submitted by developers or
users through a bug tracking system, such as \github issues or \apache
\jira. Issue reports are valuable sources of real-world software defects,
providing insight into bugs as documented during the software lifecycle.

Several datasets collect defects directly from \github
issues. These include gDefects4DL~\cite{DBLP:conf/icse/LiangLSSFD22},
HasBugs~\cite{DBLP:conf/msr/ApplisP23},
Andror2~\cite{DBLP:conf/msr/WendlandSMMHMRF21}, and
Bugs4Q~\cite{DBLP:journals/jss/ZhaoMLZ23}.
In addition to \github, \apache \jira serves as another popular
bug-tracking system, specifically for the large-scale software
projects hosted by the \apache Software Foundation. Datasets like \dfj~\cite{DBLP:conf/issta/JustJE14},
ApacheJIT~\cite{DBLP:conf/msr/KeshavarzN22},
Bugs.jar~\cite{DBLP:conf/msr/SahaLLYP18}, and
\citet{DBLP:conf/msr/OhiraKYYMLFHIM15} source
defects from \apache \jira, where developers log comprehensive bug
reports, including detailed reproduction steps, resolutions, and in
many cases, links to the relevant commits that introduce and fix the bugs.
\bugzilla{} is another prominent bug-tracking system that has been leveraged as a defect source by several software defect datasets, including the Eclipse and Mozilla Defect Tracking Dataset~\cite{DBLP:conf/msr/LamkanfiPD13}, BugHub~\cite{DBLP:conf/edcc/AndradeLV24}, and CISB~\cite{DBLP:conf/uss/XuLDDLWPM23}.

These datasets leverage the wealth of defect information available in issue reports, providing robust resources for research on bug prediction, fixing strategies, and software maintenance practices.
However, relying exclusively on issue reports introduces several limitations.
First, many issue reports lack explicit links to the commits or code regions that caused or fixed the defect, making it difficult to establish accurate traceability between natural-language descriptions and source-code changes~\cite{DBLP:conf/icse/HerzigJZ13,DBLP:conf/icse/ZhouZL12}.
Second, issue reports are often incomplete or poorly described, missing key environmental or configuration information, which complicates defect reproduction and automated analysis~\cite{DBLP:conf/sigsoft/BettenburgJSWPZ08,DBLP:journals/tse/ZimmermannPBJSW10}.
Consequently, datasets built from issue reports typically require extensive manual curation to ensure correctness and reproducibility.

\subsubsection{Version Control Systems}

The commit histories of version control systems offer a
detailed view of software evolution, providing a rich source for
collecting software defects. By analyzing commit messages and code
changes, researchers can identify bug-fixing commits and their
corresponding buggy versions. Several datasets leverage \git commit
histories
as their primary source of defects. For instance,
BugsPHP~\cite{DBLP:conf/msr/PramodSTSW24} mines commit pairs from
GitHub to capture bug fixes in \php projects, while
Defectors~\cite{DBLP:conf/msr/MahbubSR23} focuses on Python,
extracting bug-inducing commits through a combination of rule-based
and keyword-based approaches.
FixJS~\cite{DBLP:conf/msr/CsuvikV22} sources bug-fixing in \javascript
commits, grouping them into function-level changes, and
Minecraft~\cite{DBLP:conf/kbse/AvulaVM23} offers a dataset with
precise bug location information mined from \git commit histories.
Similarly, D2A~\cite{DBLP:conf/icse/ZhengPLBEYLMS21} conducts
differential analysis on commit histories, using static analysis tools to
identify bug-fixing commits.

Besides \github commits, other datasets gather commits from GitLab, such as BugsC++~\cite{DBLP:conf/kbse/AnKCYY23}, or utilize the GitHub archive for large-scale multilingual dataset construction, like MBF~\cite{DBLP:conf/kbse/ZhongLZGWGL23}.
These varied sources ensure a broad and diverse collection of bug data across different platforms and programming languages.
Collecting defects from version control systems allows dataset creators to capture real-world bugs systematically, but this approach faces challenges.
For example, identifying which commits actually introduce or fix defects is difficult, as approaches such as SZZ~\cite{DBLP:journals/sigsoft/SliwerskiZZ05,DBLP:conf/kbse/KimZPW06} often misclassify refactorings or multi-commit fixes~\cite{DBLP:conf/wcre/NetoCK18}.
Additionally, commit messages vary widely in quality, making keyword-based mining unreliable~\cite{DBLP:conf/icse/TianLL12} and often requiring additional context to fully understand the defect.

\subsubsection{Vulnerability Databases \& Other Datasets}
\label{sec-source-cve}

Another prominent source of software defects is pre-existing datasets,
particularly those focused on security vulnerabilities, such as the
Common Vulnerabilities and Exposures (CVE) dataset and the National Vulnerability
Database (NVD). These datasets provide standardized, well-documented
vulnerabilities, which are invaluable for creating benchmarks and
conducting empirical studies. For example,
CrossVul~\cite{DBLP:conf/sigsoft/NikitopoulosDLM21} leverages CVE data
to build comprehensive datasets of Java and cross-language
vulnerabilities, aimed at studying program repair and vulnerability
detection. Similarly, CVEjoin~\cite{DBLP:conf/aina/PonteRM23a}
consolidates information from CVE and other threat intelligence
sources to create a unified dataset for security research.
VulinOSS~\cite{DBLP:conf/msr/GkortzisMS18} mines vulnerabilities from
NVD and CVE to study security issues in open-source projects.

Besides the vulnerability databases mentioned above, some datasets rely
on other well-established sources of defects.
For example, both PreciseBugCollector~\cite{DBLP:conf/kbse/YeCG23} and
BugOss~\cite{DBLP:conf/icst/KimH23} build defect datasets from
OSS-Fuzz~\cite{203944}, a continuous fuzzing platform by Google.
These datasets, built upon existing structured vulnerability
repositories, provide a solid foundation for evaluating and improving
security practices, as well as for developing more robust vulnerability
detection and repair techniques. However, these datasets are limited
to previously mined defects, preventing the inclusion of newly
emerging ones. Furthermore, their quality inherently depends on the
underlying databases, which can vary significantly across different
sources.

\subsubsection{Developer \qa Forums}

\qa platforms, \eg \stackexchange and \stackoverflow, provide valuable real-world insights into common bugs faced by developers.
These platforms contain a wealth of developer-reported issues, which can be mined and categorized to capture bug symptoms, root causes, and their resolutions.
For example, \citet{DBLP:journals/ese/MorovatiNKJ23} and \citet{DBLP:journals/tse/ChenXLGLC22} extract bugs from \qa platforms to study defects in machine learning and blockchain systems, respectively.
These datasets concentrate on practical, frequently encountered defects accompanied by detailed descriptions and resolutions, making them particularly valuable for empirical research.
However, the unstructured nature of data on these platforms poses challenges related to data quality~\cite{DBLP:conf/qsic/PonzanelliMBL14}, readability~\cite{DBLP:conf/msr/DuijnKB15}, and human-factors~\cite{DBLP:journals/tse/WangCH20}, while LLM-based approaches show promise in automating the extraction and structuring of defects from such sources~\cite{DBLP:journals/jss/SilvaSK25}.

\subsubsection{Continuous Integration \& Delivery}
\label{sec-source-cicd}

A continuous integration and continuous delivery (\cicd) system records
histories of the outcome (\eg \textit{failed} or \textit{passed}) of automated
builds and tests after each code change, offering a valuable source of
real-world failures and their resolutions. \cicd is closely tied to version
control systems, as it usually monitors code repositories to trigger builds and
tests on each commit or pull request. By utilizing fail-pass patterns, several
software defect datasets have been built to capture defects that arise during
regular development cycles, ensuring their relevance for both research and
practice. For instance, BugSwarm~\cite{DBLP:conf/icse/DmeiriTWBLDVR19} and
Bears~\cite{DBLP:conf/wcre/DelfimUMM19} mine CI pipelines from Travis-CI,
leveraging fail-pass transitions in the test outcomes to identify and record bug
fixes. Similarly, with this model, BugSwarm~\cite{DBLP:conf/icse/DmeiriTWBLDVR19,DBLP:conf/icse/ZhuGFR23}, GitBug-Java~\cite{DBLP:conf/msr/SilvaSM24}, and GitBug-Actions~\cite{DBLP:conf/icse/SaavedraSM24} also
collect a wide range of defects across different software systems from \github
Actions. These datasets emphasize reproducibility by providing detailed
environments and scripts for setting up and replaying failures, making them
valuable for execution-based benchmarking and debugging. Moreover, by utilizing the fail-pass
pattern from CI/CD systems, these datasets capture real-world defects in an
automated, reproducible manner, making them particularly valuable for studies
that require large-scale, real-world, and reproducible defect datasets.
Nevertheless, collecting defects from \cicd pipelines inherently excludes
defects that do not manifest within these automated processes,
potentially overlooking issues that are caught before committing code into the repository or that manifest only in a production environment.

\subsubsection{Synthesis}
\label{sec-source-synthesis}

Some datasets generate defects through automated code mutation or bug
injection. While this methodology cannot guarantee that the resulting
defects are representative of real-world bugs, it provides a
controlled and repeatable process for creating benchmarks for
research. Specifically, it allows creators to control the location and
types of defects, which is not feasible with naturally occurring bugs.
\citet{DBLP:conf/splc/NgoNNV21} and
RaceBench~\cite{DBLP:conf/asiaccs/LiangYDMHZ23} introduce synthetic
defects, such as variability and concurrency bugs, respectively, to
evaluate bug localization and testing tools.
FixReverter~\cite{DBLP:conf/uss/ZhangP0W22} focuses on bug injection
by reverting CVE-related fixes to recreate real-world vulnerabilities
for fuzz testing. Similarly, \citet{DBLP:conf/icst/BuresHA20} and
\citet{DBLP:journals/jss/CaiLZZS24} inject faults into open-source
projects and smart contracts, respectively, to assess security
vulnerability detection. Additionally, with LLMs, DebugBench~\cite{DBLP:conf/acl/TianYQCLPWHL0024} injects faulty code into correct LeetCode solutions to create a defect dataset for evaluating the debugging ability of LLMs. These datasets provide a systematic way to study the effects of specific types of bugs and the efficacy of tools designed to detect or repair them.

\subsubsection{Coding Assignments}
Another source of defects is submissions to coding assignments,
programming competitions, and online judging platforms. While these
defects may not reflect bugs in complex, real-world software, they
often come with detailed descriptions and test cases, making them
useful for evaluating bug detection and repair techniques.
ConDefects~\cite{DBLP:conf/sigsoft/WuLZ024} draws from coding
platforms like AtCoder to collect bugs observed in a competitive
programming environment.
IntroClass~\cite{DBLP:journals/tse/GouesHSBDFW15} consists of bugs in
student coding assignments and small C programs, providing benchmarks for automated program repair.
Codeflaws~\cite{DBLP:conf/icse/TanYYMR17} collects defects from
competitive programming contests, while
QuixBugs~\cite{DBLP:conf/oopsla/LinKCS17} is based on the Quixey
programming competition, providing a multi-language benchmark for
evaluating program repair techniques.

\subsection{Construction Methodology}
\label{sec-mining-methodology}

The \textit{construction methodology} of a software defect dataset refers to the specific process by which defects are collected, and in some cases, validated and categorized.
While different datasets can be constructed in radically different ways, some strategies
appear repeatedly. Many datasets integrate
several distinct strategies at different stages of their construction
process, such as combining automated keyword-based search with
subsequent manual validation of defects.
This means that the construction methodologies outlined in this section are
not rigid categories, but instead composable strategies that can be
mix-and-matched.

In the following, we separate the construction process into three stages:
defect collection, defect validation, and defect categorization.
Defect collection involves gathering defects from the sources
discussed in \cref{sec-source}. Defect validation is the process of
verifying that the collected defects genuinely represent software
faults and defect categorization refers to assigning
additional classification information to each defect within the
dataset.
Datasets with fully synthetic defects, as discussed in
\cref{sec-source-synthesis}, are excluded from this section, since
their construction does not involve defect mining methodologies. However,
datasets that obtain code from external sources and subsequently
mutate it, such as SBP~\cite{DBLP:journals/jss/CaiLZZS24}, remain
within the scope of this discussion.

\subsubsection{Defect Collection}
\begin{table}[t]
  \tablefontsize
  \centering
  \caption{Collection methodologies of defects. One dataset may use multiple methodologies.}
  \label{tab:mine-collection}
  \renewcommand{\arraystretch}{1.25}
  \begin{tabularx}{\textwidth}{l|X|r}
    \toprule
    \textbf{Methodology} & \textbf{Datasets} & \textbf{Count} \\
    \midrule
    Keyword Matching & \cite{DBLP:journals/ese/TimperleyHSDW24, DBLP:journals/tosem/LiuMC24, DBLP:journals/tosem/JiangJWMLZ24, DBLP:conf/icse/CuiD0WSZWYXH0024, DBLP:conf/issta/YuXZ0LS24, DBLP:conf/issta/XuG024, DBLP:conf/ease/WaseemDA0M24, DBLP:conf/kbse/DasAM24, DBLP:conf/msr/PramodSTSW24, DBLP:conf/kbse/SongWCCLLWP23, DBLP:conf/kbse/ZhongLZGWGL23, DBLP:conf/kbse/AvulaVM23, DBLP:conf/msr/MahbubSR23, DBLP:journals/ese/MorovatiNKJ23, DBLP:conf/raid/0001DACW23, DBLP:conf/icse/GuanXLLB23, DBLP:conf/issta/XiongX0SWWP0023, DBLP:conf/msr/CsuvikV22, DBLP:conf/msr/KimKL22, DBLP:conf/wcre/ZhangYYCYZ22, DBLP:journals/pacmpl/PaltenghiP22, DBLP:conf/msr/WendlandSMMHMRF21, DBLP:journals/stvr/GyimesiVSMBFM21, DBLP:conf/cgo/YuanLLLLX21, DBLP:conf/msr/KimKL21, DBLP:conf/icse/PaulTB21a, DBLP:conf/msr/KamienskiPBH21, DBLP:conf/icse/Makhshari021, DBLP:journals/access/SanchezDMS20, DBLP:conf/icsm/ZhangXL20, DBLP:conf/msr/KarampatsisS20, DBLP:conf/hotsos/MurphyBSR20, DBLP:journals/pacmpl/0001FSK20, DBLP:conf/msr/WangBJS20, DBLP:conf/icse/GarciaF0AXC20, DBLP:conf/icse/BentonGZ19, DBLP:conf/msr/RaduN19, DBLP:conf/msr/RiganelliMMM19, DBLP:conf/dsa/XuYWA19, DBLP:conf/issta/ZhangCCXZ18, DBLP:journals/ese/LiuWWXCWYZ19, DBLP:conf/kbse/LinMZCZ15, DBLP:journals/ese/DAmbrosLR12, DBLP:conf/msr/ZamanAH12, DBLP:conf/icse/ChenYLCLWL21, DBLP:conf/se/SelakovicP17, DBLP:conf/uss/XuLDDLWPM23, DBLP:journals/tosem/SantanaNAA24, DBLP:journals/jss/WangCHZBZ23, DBLP:conf/msr/AzadIHR23, DBLP:conf/kbse/WangWCCY22, DBLP:conf/sigsoft/ShenM0TCC21, DBLP:conf/compsac/GuWL0019, DBLP:conf/kbse/FrancoGR17, DBLP:conf/kbse/WangDGGQYW17, DBLP:conf/icse/TanDGR18, DBLP:conf/icse/HumbatovaJBR0T20, DBLP:conf/kbse/RomanoLK021, DBLP:conf/sigsoft/GaoDQGW0HZW18, DBLP:journals/access/AntalVKMHF24, DBLP:journals/ese/TambonNAKA24, DBLP:journals/pacmpl/DrososSAM024, DBLP:journals/tosem/TufanoWBPWP19, DBLP:journals/tse/ZhaoXBCL23} & 64 \\
    \midrule
    Mapping & \cite{DBLP:conf/kbse/RuanLZL24, DBLP:conf/msr/NiSYZW24, DBLP:journals/tosem/JiangJWMLZ24, DBLP:conf/qrs/YuW24, DBLP:conf/icse/WangHGWC024, DBLP:conf/kbse/SongWCCLLWP23, DBLP:conf/kbse/ZhongLZGWGL23, DBLP:conf/icst/LeeKYY24, DBLP:conf/msr/MahbubSR23, DBLP:conf/raid/0001DACW23, DBLP:journals/tse/JiangLLZCNZHBZ23, DBLP:conf/icse/BhuiyanPVPS23, DBLP:conf/icse/GuanXLLB23, DBLP:conf/issta/XiongX0SWWP0023, DBLP:conf/icse/LiangLSSFD22, DBLP:conf/msr/KeshavarzN22, DBLP:conf/msr/BuiSF22, DBLP:conf/prdc/PereiraAV22, DBLP:conf/apsec/AkimovaBDKKMM21, DBLP:conf/msr/KimKL21, DBLP:conf/icse/PaulTB21a, DBLP:journals/jss/FerencGGTG20, DBLP:journals/jss/WuZCWM20, DBLP:conf/promise/VieiraSRG19, DBLP:conf/msr/RiganelliMMM19, DBLP:conf/msr/PontaPSBD19, DBLP:conf/icics/LinXZX19, DBLP:conf/msr/SahaLLYP18, DBLP:conf/msr/GkortzisMS18, DBLP:conf/kbse/LinMZCZ15, DBLP:conf/issta/JustJE14, DBLP:journals/ese/DAmbrosLR12, DBLP:conf/hotpar/JalbertPPS11, DBLP:conf/icsm/HoMI0SKNR23, DBLP:journals/infsof/WangBWYGS23, DBLP:conf/wcre/WangZRLJ23, DBLP:conf/sigsoft/WangLX0S21, DBLP:conf/iwpc/RenL020, DBLP:conf/kbse/FrancoGR17, DBLP:conf/kbse/WangDGGQYW17, DBLP:conf/icse/TanDGR18, DBLP:conf/issta/WuJPLD0BS23} & 42 \\
    \midrule
    Manual Collection & \cite{DBLP:journals/ese/TimperleyHSDW24, DBLP:journals/tosem/LiuMC24, DBLP:journals/jss/ZhaoMLZ23, DBLP:conf/msr/ApplisP23, DBLP:conf/issre/LiuZDHDM022, DBLP:conf/msr/PontaPSBD19, DBLP:conf/icics/LinXZX19, DBLP:conf/msr/OhiraKYYMLFHIM15, DBLP:conf/hotpar/JalbertPPS11, DBLP:conf/kbse/KuHCL07, DBLP:conf/cloud/GunawiHLPDAELLM14, DBLP:conf/icse/HumbatovaJBR0T20, DBLP:conf/msr/AmannNNNM16, DBLP:conf/msr/OliverDAH24, DBLP:journals/ese/TambonNAKA24, DBLP:journals/tse/ZhaoXBCL23} & 16 \\
    \midrule
    External Oracle & \cite{DBLP:conf/msr/LiuHLZCSHM24, DBLP:conf/acl/TianYQCLPWHL0024, DBLP:conf/msr/SilvaSM24, DBLP:conf/sigsoft/WuLZ024, DBLP:conf/icst/KimH23, DBLP:conf/icse/SaavedraSM24, DBLP:conf/icse/DmeiriTWBLDVR19, DBLP:conf/wcre/DelfimUMM19, DBLP:conf/icse/TanYYMR17, DBLP:journals/tse/GouesHSBDFW15, DBLP:conf/sigsoft/Zhang0C0Z19, DBLP:conf/wcre/ReyesGSBM24, DBLP:conf/kbse/XuZL23} & 13 \\
    \midrule
    SZZ Algorithm~\cite{DBLP:journals/sigsoft/SliwerskiZZ05} & \cite{DBLP:conf/kbse/AvulaVM23, DBLP:conf/msr/MahbubSR23, DBLP:conf/msr/KeshavarzN22, DBLP:conf/icse/PaulTB21a, DBLP:journals/jss/FerencGGTG20, DBLP:conf/msr/KarampatsisS20, DBLP:conf/iwpc/RenL020} & 7 \\
    \midrule
    Static Analysis & \cite{DBLP:conf/icse/WangHGWC024, DBLP:conf/secrypt/SenanayakeKAP023, DBLP:conf/prdc/PereiraAV22, DBLP:conf/icse/ZhengPLBEYLMS21, DBLP:journals/jss/FerencGGTG20, DBLP:conf/msr/KarampatsisS20} & 6 \\
    \midrule
    Machine Learning & \cite{DBLP:conf/icse/WangHGWC024, DBLP:conf/icse/ZhengPLBEYLMS21, DBLP:journals/jss/WuZCWM20} & 3 \\
    \midrule
    Others & \cite{DBLP:journals/tosem/ZhangCWCLMMHL24, DBLP:conf/iclr/JimenezYWYPPN24, DBLP:journals/jss/CaiLZZS24, DBLP:journals/compsec/BrustSG23, DBLP:conf/aina/PonteRM23a, DBLP:journals/tse/ChenXLGLC22, DBLP:conf/issta/Song0NW0DM22, DBLP:journals/tse/AfroseXRMY23, DBLP:conf/issre/Du0MZ21, DBLP:conf/sigsoft/WidyasariSLQPTT20, DBLP:conf/msr/GaoYJLYZ18, DBLP:conf/msr/MadeyskiK17, DBLP:journals/tse/GouesHSBDFW15, DBLP:conf/msr/LamkanfiPD13, DBLP:conf/qsw/ZhaoWLLZ23, DBLP:conf/kbse/EghbaliP20, DBLP:conf/msr/BeyerGKRR24} & 17 \\
    \bottomrule
  \end{tabularx}
\end{table}

Defect collection refers to the process of gathering defects from one
or more mining sources. The particular strategies that dataset creators employ
depend on their choice of defect sources and on the intended scope of the dataset.
For instance, if a dataset is sourced from CI/CD systems, then collecting defects based on historical test failures is a natural choice.
\cref{tab:mine-collection} lists several common defect collection
strategies, showing the diversity of approaches used to construct
software defect datasets. Keyword matching and mapping are the most
common strategies, with 64 and 42 datasets using them, respectively.
Other strategies, such as manual collection, external oracle use, and
specialized algorithms~\cite{DBLP:journals/sigsoft/SliwerskiZZ05}, are less common, but still used by a number of
datasets.

\paragraph{Keyword-Based Collection}
\label{sec-mine-collect-keyword}
Many datasets attempt to find defects by scanning for specific
keywords (\eg ``bug'' or ``fix'') in commit messages, issue reports,
and other relevant text fields. \citet{DBLP:conf/kbse/ZhongLZGWGL23}
make for a prototypical example: they examine all \github push and
pull request events in their target repositories, and consider a
bug-fixing event to be any commit or pull request that contains one of
the terms ``fix,'' ``repair,'' ``patch,'' ``bug,'' ``issue,''
``problem,'' or ``fault'' in the commit message or pull request description.
Those keywords are all typical for this kind of approach, built off
the assumption that a commit that mentions bugs, repairs, and so on is
likely attempting to fix a defect.
Keyword-based defect collection has limited
precision~\cite{DBLP:conf/icse/TianLL12}, but it is still one of the
most common collection approaches observed in this survey, presumably
due to its simplicity and its abundance (64 datasets, \ratio{64}{151}) in previous literature.

\paragraph{Mapping}
\label{sec-mine-collect-mapping}
Another common approach in defect collection is to map defects from
one source to objects from another source, such as mapping commits
from a project's \git history to related bug reports documented in its
issue tracker. This strategy is necessary for datasets that (1) source
defects from bug reports or vulnerability databases (Sections~\ref{sec-source-issue} and~\ref{sec-source-cve}) and (2) wish to
include source-code-level information (\cref{sec-presentation}). Accordingly, mapping is common, with 42
datasets (\ratio{42}{151}) utilizing it. The difficulty of this process
can vary based on the mapping source and destination. For instance,
mapping \github issues to \github commits is quite easy, since \github
tracks commits that reference issues and exposes that relationship in
its website and API. In contrast, mapping CVE entries to fixing
commits can be more challenging, since many CVE datasets do not
reference the commits that fix a vulnerability, and likewise many
vulnerability-fixing commits do not reference their associated CVEs.
\citet{DBLP:journals/tosem/JiangJWMLZ24} outline their process for
mapping CVEs to bug-fixing commits in detail, using six different
strategies to construct these relationships.

\paragraph{Manual Collection}
\label{sec-mine-collect-manual}
Some dataset creators forego automatic defect collection, but instead manually collect defects.
For instance, \citet{DBLP:conf/hotpar/JalbertPPS11} create RADBench by
searching through the issue trackers of several projects to find concurrency
bugs. An advantage of this approach is
that it allows researchers to make qualitative decisions about what
defects to include in their datasets, such as only including ``vital''
issues, which can depend on a variety of factors that are not easy to
judge automatically. Conversely, a notable disadvantage is the
substantial increase in manual effort imposed on dataset creators.
Consequently, it is unsurprising
that manually collected datasets typically exhibit a smaller median
size compared to automatically collected datasets (298 versus 1158
entries, respectively). Also, while qualitative judgments about a
defect can be useful, they are also inherently subjective, which may
deter potential dataset users who prioritize objectivity. These
disadvantages are probably why this strategy is, with only 16 datasets, relatively uncommon among
the surveyed datasets.

\paragraph{External Oracle}
Some datasets use historical test failures or continuous integration
(CI) failures as external oracles to distinguish buggy from non-buggy
software. This should not be confused with datasets that
\textit{reproduce} test failures, which are described in
\cref{sec-mine-validate-reproduce}. For example,
BugSwarm~\cite{DBLP:conf/icse/DmeiriTWBLDVR19,DBLP:conf/icse/ZhuGFR23} is created by scanning
through the commit history of repositories that use \github Actions or Travis-CI, looking for cases where one commit's CI run fails and
a subsequent commit's run passes. Using historical failures to
distinguish bugs from non-bugs allows for better confidence in a
dataset's accuracy, given that failing tests are a tell-tale sign of a
bug. However, since this strategy relies on historical test or CI
results, it inherently limits the bugs that can appear in a dataset to those that have this historical information available. This naturally
limits the defect source of a dataset, which can be an issue if the
dataset creators are concerned about dataset size. Moreover, in the
case of CI failures, care must be taken to ensure that a given failure
was actually caused by a failing test (as opposed to, say, a code
style checker).

\paragraph{SZZ-Based Collection}
\label{sec-mine-collect-szz}
The SZZ algorithm, introduced by \citet{DBLP:journals/sigsoft/SliwerskiZZ05} and refined multiple
times since~\cite{DBLP:conf/kbse/KimZPW06, DBLP:conf/issta/WilliamsS08}, attempts to identify the
software change that introduced a bug given the change that fixes the
bug. This addresses a weakness in some other collection strategies
(\eg keyword search), which is that while they can readily find
software changes that \textit{fix} bugs, they have no way to identify
the change(s) that \textit{caused} those bugs. There are seven datasets in
our survey using SZZ or a similar algorithm to mine bug-introducing
commits. \citet{DBLP:conf/msr/MahbubSR23} demonstrate a typical
approach: after finding bug-fixing pull requests in their target
repositories, they use SZZ to associate one or more bug-introducing commits
to each bug fix.

\paragraph{Static Analysis}
There are six datasets in the survey that use static analysis to collect
defects.
To this end, the dataset creators run existing or newly created static analyses on
candidate code. As an example of the former,
\citet{DBLP:conf/icse/ZhengPLBEYLMS21} create D2A by running the
static analyzer Infer\webcite{infer_github} on pairs of commits, looking for
detected issues that are present in one commit but absent in the next.
As an example of the latter,
ManySStuBs4J~\cite{DBLP:conf/msr/KarampatsisS20} uses a custom static
analysis approach that compares the ASTs of candidate buggy and fixed
programs and searches for a set of specific bug-fix patterns. An
advantage of using static analysis is that, depending on the specific
tools or techniques used, one can achieve better precision than
strategies such as keyword matching (i.e., fewer non-bugs classified
as bugs). The main disadvantage is that relying on static analysis to
separate bugs from non-bugs will limit a dataset's scope to the abilities of the static analysis.
Many static analyzers only detect specific classes of bugs (\eg null
pointer bugs), and even more generalist analyzers like Cppcheck\webcite{cppcheck_github} will necessarily fail to catch some bugs~\cite{DBLP:conf/kbse/HabibP18a}.

\paragraph{Machine Learning-Based Collection}
There are three datasets in the survey that use machine learning
techniques as part of their data collection processes.
\citet{DBLP:conf/icse/WangHGWC024} input candidate patches into LLMs, asking the models to
evaluate whether those patches fix software vulnerabilities. Their
dataset also illustrates that extra care must be taken to ensure that
LLM output is sound, which they achieve by using LLMs in
conjunction with static analysis tools and by including only defects
where the two agree. Similarly, \citet{DBLP:conf/icse/ZhengPLBEYLMS21} and \citet{DBLP:journals/jss/WuZCWM20} use machine learning models to distinguish between defects and non-defects based on semantic information, such as bug reports or commit messages.

\subsubsection{Validation}
\label{sec-mine-validate}
\begin{table}[t]
  \tablefontsize
  \centering
  \caption{Validation methodologies of defects. One dataset may use multiple methodologies.}
  \label{tab:mine-validate}
  \begin{tabularx}{\textwidth}{ll|X|r}
    \toprule
    \textbf{Methodology} & & \textbf{Datasets} & \textbf{Count} \\
    \midrule
    Manual Inspection & \textit{Full} & \cite{DBLP:journals/ese/TimperleyHSDW24, DBLP:journals/tosem/LiuMC24, DBLP:journals/tosem/ZhangCWCLMMHL24, DBLP:conf/icse/CuiD0WSZWYXH0024, DBLP:conf/issta/YuXZ0LS24, DBLP:conf/issta/XuG024, DBLP:conf/ease/WaseemDA0M24, DBLP:conf/kbse/DasAM24, DBLP:conf/acl/TianYQCLPWHL0024, DBLP:conf/msr/SilvaSM24, DBLP:conf/icst/LeeKYY24, DBLP:journals/jss/ZhaoMLZ23, DBLP:conf/icst/KimH23, DBLP:journals/ese/MorovatiNKJ23, DBLP:conf/msr/ApplisP23, DBLP:conf/icse/GuanXLLB23, DBLP:conf/issta/XiongX0SWWP0023, DBLP:conf/icse/LiangLSSFD22, DBLP:conf/msr/KimKL22, DBLP:conf/wcre/ZhangYYCYZ22, DBLP:journals/pacmpl/PaltenghiP22, DBLP:conf/issre/LiuZDHDM022, DBLP:conf/msr/WendlandSMMHMRF21, DBLP:journals/stvr/GyimesiVSMBFM21, DBLP:conf/cgo/YuanLLLLX21, DBLP:conf/icse/PaulTB21a, DBLP:conf/issre/Du0MZ21, DBLP:conf/icsm/ZhangXL20, DBLP:conf/hotsos/MurphyBSR20, DBLP:conf/sigsoft/WidyasariSLQPTT20, DBLP:journals/pacmpl/0001FSK20, DBLP:conf/msr/WangBJS20, DBLP:conf/icse/BentonGZ19, DBLP:conf/msr/RaduN19, DBLP:conf/msr/RiganelliMMM19, DBLP:conf/wcre/DelfimUMM19, DBLP:conf/msr/PontaPSBD19, DBLP:conf/dsa/XuYWA19, DBLP:conf/msr/SahaLLYP18, DBLP:conf/issta/ZhangCCXZ18, DBLP:journals/ese/LiuWWXCWYZ19, DBLP:conf/kbse/LinMZCZ15, DBLP:conf/issta/JustJE14, DBLP:conf/kbse/KuHCL07, DBLP:conf/icse/ChenYLCLWL21, DBLP:conf/cloud/GunawiHLPDAELLM14, DBLP:conf/icsm/HoMI0SKNR23, DBLP:conf/uss/XuLDDLWPM23, DBLP:journals/tosem/SantanaNAA24, DBLP:conf/msr/AzadIHR23, DBLP:conf/qsw/ZhaoWLLZ23, DBLP:conf/kbse/WangWCCY22, DBLP:conf/sigsoft/ShenM0TCC21, DBLP:conf/kbse/EghbaliP20, DBLP:conf/kbse/FrancoGR17, DBLP:conf/kbse/WangDGGQYW17, DBLP:conf/icse/TanDGR18, DBLP:conf/icse/HumbatovaJBR0T20, DBLP:conf/sigsoft/GaoDQGW0HZW18, DBLP:conf/msr/BeyerGKRR24, DBLP:journals/ese/TambonNAKA24, DBLP:journals/tse/ZhaoXBCL23} & 62 \\
     & \textit{Partial} & \cite{DBLP:conf/icse/WangHGWC024, DBLP:conf/kbse/ZhongLZGWGL23, DBLP:conf/kbse/AvulaVM23, DBLP:conf/issta/Song0NW0DM22, DBLP:conf/apsec/AkimovaBDKKMM21, DBLP:conf/msr/KimKL21, DBLP:conf/icse/ZhengPLBEYLMS21, DBLP:conf/icse/Makhshari021, DBLP:journals/jss/FerencGGTG20, DBLP:journals/jss/WuZCWM20, DBLP:journals/ese/DAmbrosLR12, DBLP:conf/msr/ZamanAH12, DBLP:journals/infsof/WangBWYGS23, DBLP:journals/jss/WangCHZBZ23, DBLP:conf/compsac/GuWL0019, DBLP:conf/kbse/RomanoLK021, DBLP:conf/msr/OliverDAH24, DBLP:journals/pacmpl/DrososSAM024, DBLP:journals/tosem/TufanoWBPWP19} & 19 \\
    \midrule
    Defect Reproduction & & \cite{DBLP:journals/ese/TimperleyHSDW24, DBLP:conf/issta/YuXZ0LS24, DBLP:conf/acl/TianYQCLPWHL0024, DBLP:conf/iclr/JimenezYWYPPN24, DBLP:conf/msr/PramodSTSW24, DBLP:conf/msr/SilvaSM24, DBLP:conf/icst/LeeKYY24, DBLP:journals/jss/ZhaoMLZ23, DBLP:conf/icst/KimH23, DBLP:conf/icse/SaavedraSM24, DBLP:journals/compsec/BrustSG23, DBLP:journals/tse/JiangLLZCNZHBZ23, DBLP:conf/msr/ApplisP23, DBLP:conf/icse/BhuiyanPVPS23, DBLP:conf/issta/XiongX0SWWP0023, DBLP:conf/wcre/ZhangYYCYZ22, DBLP:conf/msr/BuiSF22, DBLP:conf/issta/Song0NW0DM22, DBLP:conf/msr/WendlandSMMHMRF21, DBLP:journals/stvr/GyimesiVSMBFM21, DBLP:conf/cgo/YuanLLLLX21, DBLP:conf/sigsoft/WidyasariSLQPTT20, DBLP:conf/icse/BentonGZ19, DBLP:conf/icse/DmeiriTWBLDVR19, DBLP:conf/msr/RiganelliMMM19, DBLP:conf/wcre/DelfimUMM19, DBLP:conf/msr/SahaLLYP18, DBLP:conf/issta/ZhangCCXZ18, DBLP:conf/msr/GaoYJLYZ18, DBLP:journals/tse/GouesHSBDFW15, DBLP:conf/issta/JustJE14, DBLP:conf/hotpar/JalbertPPS11, DBLP:conf/se/SelakovicP17, DBLP:conf/uss/XuLDDLWPM23, DBLP:conf/issta/WuJPLD0BS23, DBLP:conf/wcre/ReyesGSBM24, DBLP:journals/access/AntalVKMHF24, DBLP:journals/ese/TambonNAKA24, DBLP:conf/kbse/XuZL23} & 39 \\
    \bottomrule
  \end{tabularx}
\end{table}

Dataset validation is about ensuring that the defects included in a
dataset are actual defects. This is an important step for making
a useful dataset, since a high false positive rate (i.e., many
non-bugs classified as bugs) can taint further research that uses a
dataset. Even so, not all dataset creators take this step, and there
are legitimate reasons for doing so. For instance, a dataset might be
so large that validation becomes impractical (\eg \cite{DBLP:conf/edcc/AndradeLV24,DBLP:conf/msr/CsuvikV22,DBLP:conf/msr/MahbubSR23,DBLP:conf/raid/0001DACW23,DBLP:conf/aina/PonteRM23a}), or a dataset's data
collection approach might already have a low false positive
rate (\eg \cite{DBLP:conf/msr/LiuHLZCSHM24,DBLP:journals/jss/CaiLZZS24,DBLP:conf/kbse/SongWCCLLWP23,DBLP:journals/tse/ChenXLGLC22,DBLP:conf/icse/TanYYMR17,DBLP:conf/sp/Dolan-GavittHKL16}). \cref{tab:mine-validate} shows common validation methodologies and datasets using them.

\paragraph{Manual Inspection}
\label{sec-mine-validate-manual}
Some dataset creators, such as \citet{DBLP:conf/icst/LeeKYY24},
\citet{DBLP:conf/msr/WangBJS20}, and \citet{DBLP:conf/msr/SilvaSM24},
opt to manually inspect the code and description of every bug in their
datasets (\ie full manual inspection). This allows researchers to
construct a dataset with a low false positive rate and to make
qualitative judgments about what bugs should be included, without
spending the time and resources needed for manual defect collection.
Full manual inspection is quite popular among the datasets surveyed,
with 62 (\ratio{62}{151}) datasets employing it. The
largest~\cite{DBLP:conf/acl/TianYQCLPWHL0024} and
smallest~\cite{DBLP:conf/icst/KimH23} datasets that employ this
strategy have 4,253 and 18 entries respectively, with a median of 260
entries.

Conversely, some dataset creators choose to only manually inspect a
subset of the defects in their dataset (\ie partial manual
inspection). This is less common than full inspection, with only 19 (\ratio{19}{151})
datasets employing it. However, it is attractive for particularly
large datasets where full inspection is infeasible, such as
D2A~\cite{DBLP:conf/icse/ZhengPLBEYLMS21}, the largest dataset that
uses this approach, which has over 1.3 million entries. The smallest
such dataset~\cite{DBLP:conf/msr/ZamanAH12} still contains 191
entries, with the median size being 5,565 entries.

Typically, partial inspection has no effect on the actual construction
of the dataset, but is performed to demonstrate
the accuracy of an automated methodology~\cite{DBLP:conf/icse/ZhengPLBEYLMS21}. However, there is at least one
case where partial inspection is used to filter out defects from a
dataset. \citet{DBLP:conf/kbse/ZhongLZGWGL23} automatically classify
each bug in their dataset into one of multiple fault categories. They
then manually analyze 50 bugs from each category to ascertain that
category's accuracy, and only include bugs from categories with over
90\% accuracy in the final dataset.

\paragraph{Defect Reproduction}
\label{sec-mine-validate-reproduce}

Some dataset creators aim to ensure the accuracy of their datasets by reproducing each defect they collect. Reproduction typically involves running the buggy program in a controlled environment and observing the faulty behavior, often through failing test cases or other verifiable symptoms. While the exact meaning of \textit{reproducibility} can vary depending on the goals and assumptions of different datasets~\cite{DBLP:conf/icse/ZhuR23}, it generally requires the ability to build and run the faulty version of the software.

Among our surveyed datasets, 39 (\ratio{39}{151}) involve defect reproduction, and only one dataset attempts to reproduce only a subset of its defects, while the rest aim for full reproduction. This process is often facilitated by sources that provide a rich environment and execution context, such as continuous integration (CI) systems. For example, BugSwarm~\cite{DBLP:conf/icse/DmeiriTWBLDVR19} leverages CI configurations and historical build logs to automatically reproduce bugs in Docker-based environments, thus making the automation of reproduction feasible.

However, automation has limitations. Many defects are not covered by existing tests, or their reproduction may require complex setup steps beyond test execution (\eg user interface-related bugs in AndroR2~\cite{DBLP:conf/msr/WendlandSMMHMRF21}). Still, automated reproduction enables the construction of much larger datasets than manual validation strategies. The largest dataset in this category~\cite{DBLP:journals/compsec/BrustSG23} contains approximately 168,000 entries. The median size, 225 entries, is comparable to that of datasets which use full manual inspection.

\subsubsection{Categorization}
\begin{table}[t]
  \tablefontsize
  \centering
  \caption{Categorization methodologies of defects. One dataset may use multiple methodologies.}
  \label{tab:mine-categorization}
  \begin{tabularx}{\textwidth}{l|X|r}
    \toprule
    \textbf{Methodology} & \textbf{Datasets} & \textbf{Count} \\
    \midrule
    Manual Categorization & \cite{DBLP:journals/ese/TimperleyHSDW24, DBLP:journals/tosem/LiuMC24, DBLP:journals/tosem/ZhangCWCLMMHL24, DBLP:journals/tosem/JiangJWMLZ24, DBLP:conf/icse/CuiD0WSZWYXH0024, DBLP:conf/issta/YuXZ0LS24, DBLP:conf/issta/XuG024, DBLP:conf/ease/WaseemDA0M24, DBLP:conf/kbse/DasAM24, DBLP:conf/qrs/YuW24, DBLP:conf/kbse/SongWCCLLWP23, DBLP:journals/jss/ZhaoMLZ23, DBLP:conf/icst/KimH23, DBLP:journals/ese/MorovatiNKJ23, DBLP:conf/msr/ApplisP23, DBLP:conf/icse/GuanXLLB23, DBLP:conf/issta/XiongX0SWWP0023, DBLP:conf/icse/LiangLSSFD22, DBLP:conf/msr/KimKL22, DBLP:conf/wcre/ZhangYYCYZ22, DBLP:journals/tse/ChenXLGLC22, DBLP:journals/pacmpl/PaltenghiP22, DBLP:conf/issre/LiuZDHDM022, DBLP:conf/cgo/YuanLLLLX21, DBLP:conf/icse/PaulTB21a, DBLP:conf/msr/KamienskiPBH21, DBLP:conf/issre/Du0MZ21, DBLP:conf/icse/Makhshari021, DBLP:journals/access/SanchezDMS20, DBLP:conf/icsm/ZhangXL20, DBLP:conf/hotsos/MurphyBSR20, DBLP:journals/pacmpl/0001FSK20, DBLP:conf/msr/WangBJS20, DBLP:conf/icse/GarciaF0AXC20, DBLP:conf/msr/RaduN19, DBLP:conf/issta/ZhangCCXZ18, DBLP:conf/msr/GaoYJLYZ18, DBLP:journals/ese/LiuWWXCWYZ19, DBLP:conf/msr/OhiraKYYMLFHIM15, DBLP:conf/kbse/LinMZCZ15, DBLP:journals/tse/GouesHSBDFW15, DBLP:journals/tse/GouesHSBDFW15, DBLP:conf/msr/ZamanAH12, DBLP:conf/hotpar/JalbertPPS11, DBLP:conf/kbse/KuHCL07, DBLP:conf/icse/ChenYLCLWL21, DBLP:conf/se/SelakovicP17, DBLP:conf/cloud/GunawiHLPDAELLM14, DBLP:conf/icsm/HoMI0SKNR23, DBLP:journals/tosem/SantanaNAA24, DBLP:conf/msr/AzadIHR23, DBLP:conf/qsw/ZhaoWLLZ23, DBLP:conf/wcre/WangZRLJ23, DBLP:conf/kbse/WangWCCY22, DBLP:conf/sigsoft/ShenM0TCC21, DBLP:conf/sigsoft/WangLX0S21, DBLP:conf/kbse/EghbaliP20, DBLP:conf/kbse/FrancoGR17, DBLP:conf/kbse/WangDGGQYW17, DBLP:conf/icse/TanDGR18, DBLP:conf/icse/HumbatovaJBR0T20, DBLP:conf/kbse/RomanoLK021, DBLP:conf/msr/AmannNNNM16, DBLP:conf/sigsoft/GaoDQGW0HZW18, DBLP:conf/msr/BeyerGKRR24, DBLP:journals/ese/TambonNAKA24, DBLP:journals/pacmpl/DrososSAM024, DBLP:journals/tse/ZhaoXBCL23} & 68 \\
    \midrule
    Static Analysis & \cite{DBLP:conf/secrypt/SenanayakeKAP023, DBLP:conf/icse/LiangLSSFD22, DBLP:conf/icse/ZhengPLBEYLMS21, DBLP:conf/msr/KamienskiPBH21, DBLP:journals/jss/FerencGGTG20, DBLP:conf/icse/TanYYMR17, DBLP:journals/jss/WangCHZBZ23, DBLP:conf/fie/SandersWA24} & 8 \\
    \midrule
    Machine Learning & \cite{DBLP:conf/kbse/SongWCCLLWP23, DBLP:conf/kbse/AvulaVM23, DBLP:conf/dsa/XuYWA19} & 3 \\
    \midrule
    Others & \cite{DBLP:conf/kbse/RuanLZL24, DBLP:conf/raid/0001DACW23, DBLP:conf/msr/KeshavarzN22, DBLP:conf/msr/BuiSF22, DBLP:conf/prdc/PereiraAV22, DBLP:conf/promise/VieiraSRG19, DBLP:conf/icse/DmeiriTWBLDVR19, DBLP:conf/wcre/DelfimUMM19, DBLP:conf/msr/GkortzisMS18, DBLP:conf/sigsoft/Zhang0C0Z19, DBLP:conf/wcre/ReyesGSBM24} & 11 \\
    \bottomrule
  \end{tabularx}
\end{table}

To enhance the utility of their datasets, many dataset
creators opt to classify the defects in their datasets along different
axes, such as the root causes and symptoms of defects. Defect
categorization does not involve adding or removing entries from a
dataset, but simply adding metadata to defects already in a
dataset. \cref{tab:mine-categorization} shows the categorization
strategies for the datasets that use one, which include manual categorization, static analysis, and machine learning.

\paragraph{Manual Defect Categorization}
\label{sec-mine-categorize-manual}
By far the most common way for dataset creators to categorize the
defects in their datasets is through manual examination, with 68 (\ratio{68}{151})
datasets falling into this category. Similar to manual validation,
manual categorization allows researchers to operate reasonably
efficiently and flexibly.
Indeed, it is not uncommon for datasets to
employ both manual validation and manual categorization, (\eg
ECench~\cite{DBLP:conf/msr/KimKL22} and
ExcePy~\cite{DBLP:conf/wcre/ZhangYYCYZ22}), with 48 (\ratio{48}{151}) datasets in this group.

Defect categorization need not be entirely automatic or entirely
manual. For instance, \citet{DBLP:conf/msr/OhiraKYYMLFHIM15} label
``surprise'' and ``dormant'' bugs automatically, since the definitions
they use are easy to detect, while manually labeling ``blocking,''
``security,'' ``performance,'' and ``breakage'' bugs.

\paragraph{Categorization via Static Analysis}
\label{sec-mine-categorize-static-analysis}
Eight datasets in the survey use static analysis to
classify defects. For instance,
\citet{DBLP:conf/secrypt/SenanayakeKAP023} run two static analyzers,
Qark\webcite{qark_github} and MobSF\webcite{mobsf_github}, on the APKs and source files of Android applications and store
the analysis results in their dataset. Categorizing defects based on static
analysis has some of the same benefits and drawbacks of static
analysis-based collection, namely a fairly high precision, but a
limited scope of possible categories and the possibility of false
negatives (i.e., bugs with a missing classification).

\paragraph{Categorization via Machine Learning}
In total, three of the surveyed datasets classify their entries using
machine learning approaches. \citet{DBLP:conf/kbse/SongWCCLLWP23}
train a neural network to classify bugs given a bug report, fixing
commit message, bug-fix location, and bug-inducing location.
\citet{DBLP:conf/kbse/AvulaVM23} use an off-the-shelf LLM to generate
a description of what each bug fix in their dataset does.
\citet{DBLP:conf/dsa/XuYWA19} use latent Dirichlet allocation~\cite{DBLP:journals/jmlr/BleiNJ03} (a kind
of Bayesian network) to extract topic keywords from pull requests and
add them to each defect's metadata.

\begin{tcolorbox}[
  colback=white,
  colframe=gray!140,
  boxsep=2pt,
  left=2pt,
  right=2pt,
  top=1pt,
  bottom=1pt,
  title=\textbf{RQ2 Summary},
  fonttitle=\bfseries
]
\vspace{0.25em}
\begin{itemize}[leftmargin=10pt, topsep=2pt, itemsep=2pt]

    \item The majority of software defect datasets are sourced from software development processes, with issue reports and version control systems being most common.
    A smaller number rely on pre-existing datasets, vulnerability databases, developer Q\&A forums, coding assignments, and CI/CD systems.
    Some datasets employ defect synthesis techniques to generate synthetic defects for controlled experimentation.

    \item Dataset construction generally involves three stages: defect collection, validation, and categorization.
    Collection is most often performed through automated mining strategies such as keyword matching (\ratio{64}{151}) or mapping (\ratio{42}{151}) between issue reports and commits.
    Fewer datasets rely on manual collection, external oracles, the SZZ algorithm, static analysis, or machine learning techniques.
    Validation and categorization practices vary widely, though manual inspection remains most common in both stages.

\end{itemize}
\end{tcolorbox}

\section{Availability and Usability (RQ3)}
\label{sec-artifact}

\begin{figure}[t]
\centering
\small
\begin{minipage}{\textwidth}
    \centering
    \begin{minipage}[t]{0.5\textwidth}
        \centering
        \begin{subfigure}[t]{0.7\textwidth}
            \centering
            \includegraphics[width=\textwidth]{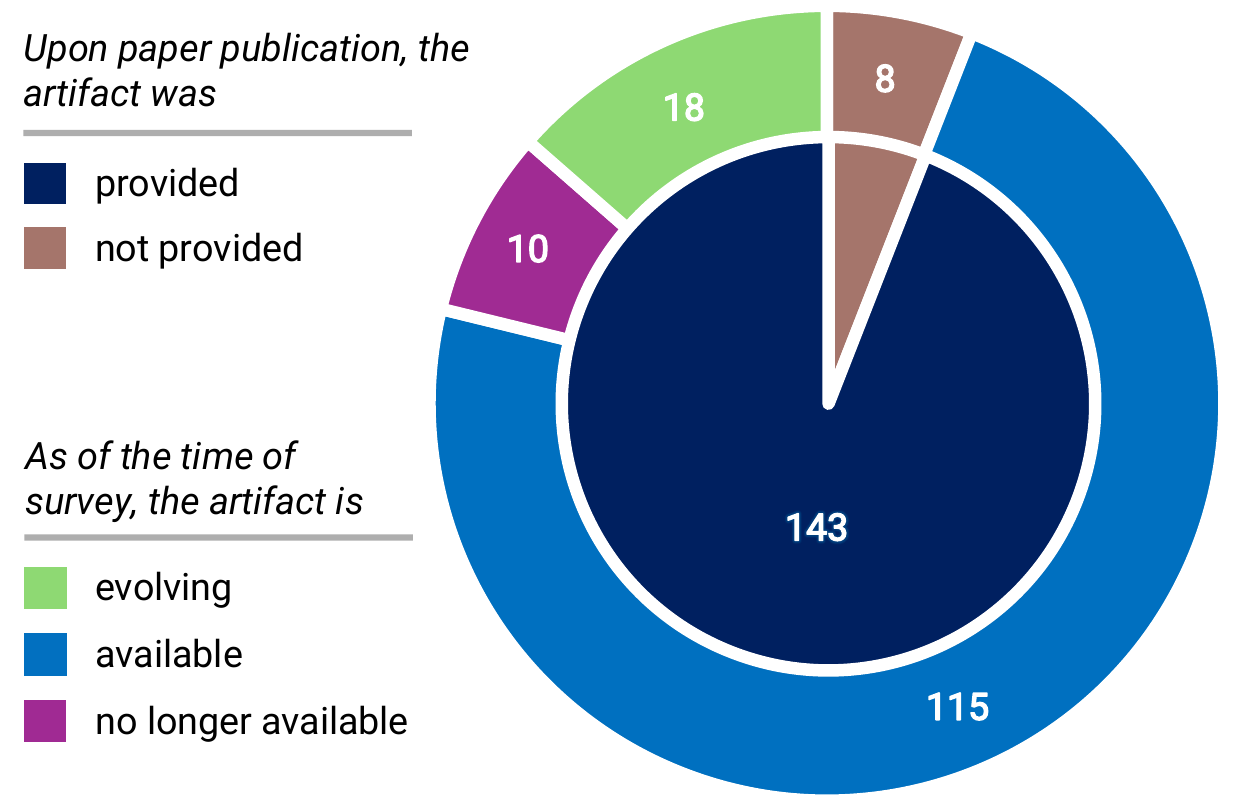}
            \caption{Availability of artifacts.}
            \label{fig-artifact-pie}
        \end{subfigure}
    \end{minipage}\hfill
    \begin{minipage}[t]{0.5\textwidth}
        \centering
        \begin{subfigure}[t]{0.7\textwidth}
            \centering
            \includegraphics[width=\textwidth]{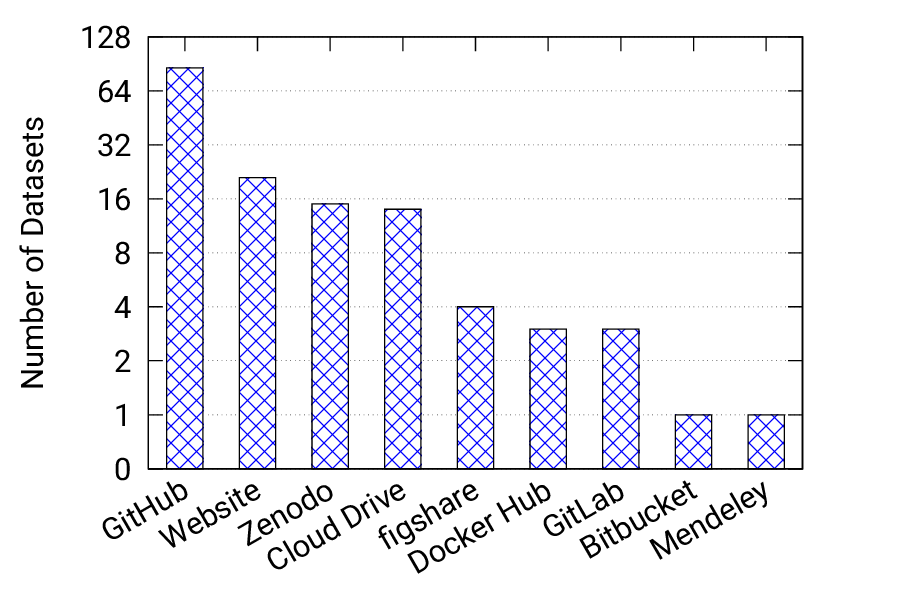}
            \caption{Hosting platforms. One dataset may be hosted on multiple platforms.}
            \label{fig-artifact-host}
        \end{subfigure}
    \end{minipage}
\end{minipage}

\caption{Availability and hosting platforms of software defect datasets.}
\end{figure}

The availability and usability of a dataset play a crucial role for its impact on other research and practical applications. Most of the dataset papers in this study provide accompanying artifacts that range from simple download links for datasets to more complex, executable frameworks designed to automate bug reproduction or testing. There are a few instances where the actual datasets are not provided at all, or where the artifacts are not easily accessible or usable. In this section, we examine and discuss the availability and usability of the \numdataset datasets we survey.

\subsection{Availability and Hosting Platforms}
\label{sec-availability}

\begin{table}
    \centering
    \caption{Availability of artifacts of surveyed datasets.}
    \tablefontsize
    \begin{tabularx}{\textwidth}{ll|X|r}
        \toprule
        \textbf{Availability} &                     & \textbf{Datasets}                                                                                                                                                                                                                                                                                                                                                                                                                                                                                                                                                                                                                                                                                                                                                                                                                                                                                                                                                                                                                                                                                                                                                                                                                                                                                                                                                                                                                                                                                                                                                                                                                                                                                                                                                                                                                                                                                                                                                                                                                                                                                                                                                                                                                                                                                                                                                                                                                                                                                                                                                                                                 & \textbf{Count} \\

        \midrule

        Provided     & \textit{Available}            & \cite{DBLP:journals/ese/TimperleyHSDW24, DBLP:conf/edcc/AndradeLV24, DBLP:conf/msr/LiuHLZCSHM24, DBLP:conf/kbse/RuanLZL24, DBLP:conf/msr/NiSYZW24, DBLP:journals/tosem/LiuMC24, DBLP:journals/tosem/ZhangCWCLMMHL24, DBLP:journals/tosem/JiangJWMLZ24, DBLP:conf/icse/CuiD0WSZWYXH0024, DBLP:conf/issta/YuXZ0LS24, DBLP:conf/issta/XuG024, DBLP:conf/ease/WaseemDA0M24, DBLP:conf/kbse/DasAM24, DBLP:conf/acl/TianYQCLPWHL0024, DBLP:conf/iclr/JimenezYWYPPN24, DBLP:conf/msr/PramodSTSW24, DBLP:journals/jss/CaiLZZS24, DBLP:conf/kbse/ZhongLZGWGL23, DBLP:conf/kbse/AvulaVM23, DBLP:conf/msr/MahbubSR23, DBLP:journals/jss/ZhaoMLZ23, DBLP:conf/icst/KimH23, DBLP:journals/ese/MorovatiNKJ23, DBLP:conf/asiaccs/LiangYDMHZ23, DBLP:conf/icse/SaavedraSM24, DBLP:conf/raid/0001DACW23, DBLP:conf/secrypt/SenanayakeKAP023, DBLP:journals/compsec/BrustSG23, DBLP:conf/aina/PonteRM23a, DBLP:conf/icse/GuanXLLB23, DBLP:conf/issta/XiongX0SWWP0023, DBLP:conf/msr/CsuvikV22, DBLP:journals/softx/PachoulyAK22, DBLP:conf/icse/LiangLSSFD22, DBLP:conf/msr/KeshavarzN22, DBLP:conf/msr/KimKL22, DBLP:conf/uss/ZhangP0W22, DBLP:conf/wcre/ZhangYYCYZ22, DBLP:conf/prdc/PereiraAV22, DBLP:journals/tse/ChenXLGLC22, DBLP:journals/pacmpl/PaltenghiP22, DBLP:conf/msr/WendlandSMMHMRF21, DBLP:conf/apsec/AkimovaBDKKMM21, DBLP:journals/stvr/GyimesiVSMBFM21, DBLP:conf/msr/KimKL21, DBLP:conf/icse/PaulTB21a, DBLP:conf/icse/ZhengPLBEYLMS21, DBLP:conf/sigsoft/NikitopoulosDLM21, DBLP:conf/splc/NgoNNV21, DBLP:journals/tse/AfroseXRMY23, DBLP:conf/msr/KamienskiPBH21, DBLP:conf/icse/Makhshari021, DBLP:journals/jss/FerencGGTG20, DBLP:journals/access/SanchezDMS20, DBLP:journals/jss/WuZCWM20, DBLP:conf/icsm/ZhangXL20, DBLP:conf/msr/KarampatsisS20, DBLP:conf/icst/BuresHA20, DBLP:conf/msr/FanL0N20, DBLP:journals/pacmpl/0001FSK20, DBLP:conf/msr/WangBJS20, DBLP:conf/icse/GarciaF0AXC20, DBLP:conf/promise/VieiraSRG19, DBLP:conf/icse/BentonGZ19, DBLP:conf/msr/RiganelliMMM19, DBLP:conf/msr/PontaPSBD19, DBLP:conf/dsa/XuYWA19, DBLP:conf/icics/LinXZX19, DBLP:conf/msr/SahaLLYP18, DBLP:conf/promise/FerencTLSG18, DBLP:conf/msr/GkortzisMS18, DBLP:conf/issta/ZhangCCXZ18, DBLP:conf/msr/GaoYJLYZ18, DBLP:conf/msr/MadeyskiK17, DBLP:conf/icse/TanYYMR17, DBLP:journals/ese/LiuWWXCWYZ19, DBLP:conf/kbse/LinMZCZ15, DBLP:journals/tse/GouesHSBDFW15, DBLP:journals/tse/GouesHSBDFW15, DBLP:conf/msr/LamkanfiPD13, DBLP:journals/ese/DAmbrosLR12, DBLP:conf/icse/ChenYLCLWL21, DBLP:conf/se/SelakovicP17, DBLP:conf/icsm/HoMI0SKNR23, DBLP:conf/uss/XuLDDLWPM23, DBLP:journals/tosem/SantanaNAA24, DBLP:journals/infsof/WangBWYGS23, DBLP:journals/jss/WangCHZBZ23, DBLP:conf/msr/AzadIHR23, DBLP:conf/qsw/ZhaoWLLZ23, DBLP:conf/wcre/WangZRLJ23, DBLP:conf/kbse/WangWCCY22, DBLP:conf/sigsoft/ShenM0TCC21, DBLP:conf/sigsoft/WangLX0S21, DBLP:conf/kbse/EghbaliP20, DBLP:conf/compsac/GuWL0019, DBLP:conf/kbse/FrancoGR17, DBLP:conf/oopsla/LinKCS17, DBLP:conf/kbse/YeCG23, DBLP:conf/icse/TanDGR18, DBLP:conf/fie/SandersWA24, DBLP:conf/icse/HumbatovaJBR0T20, DBLP:conf/icse/HuZLYH024, DBLP:conf/issta/WuJPLD0BS23, DBLP:conf/kbse/RomanoLK021, DBLP:conf/msr/OliverDAH24, DBLP:conf/wcre/ReyesGSBM24, DBLP:conf/msr/BeyerGKRR24, DBLP:journals/access/AntalVKMHF24, DBLP:journals/ese/TambonNAKA24, DBLP:journals/pacmpl/DrososSAM024, DBLP:journals/tosem/TufanoWBPWP19, DBLP:journals/tse/ZhaoXBCL23, DBLP:conf/kbse/XuZL23, DBLP:conf/sp/Dolan-GavittHKL16}     & 115    \\
                     & \textit{Evolving}           & \cite{DBLP:conf/msr/SilvaSM24, DBLP:conf/icse/WangHGWC024, DBLP:conf/icst/LeeKYY24, DBLP:conf/kbse/AnKCYY23, DBLP:conf/sigsoft/WuLZ024, DBLP:journals/tse/JiangLLZCNZHBZ23, DBLP:conf/msr/ApplisP23, DBLP:conf/icse/BhuiyanPVPS23, DBLP:conf/msr/BuiSF22, DBLP:conf/issta/Song0NW0DM22, DBLP:conf/cgo/YuanLLLLX21, DBLP:conf/sigsoft/WidyasariSLQPTT20, DBLP:conf/icse/DmeiriTWBLDVR19, DBLP:conf/msr/RaduN19, DBLP:conf/wcre/DelfimUMM19, DBLP:conf/promise/MitraR17, DBLP:conf/issta/JustJE14, DBLP:conf/msr/AmannNNNM16} & 18    \\
                     & \textit{No Longer Available} & \cite{DBLP:conf/kbse/SongWCCLLWP23, DBLP:conf/issre/Du0MZ21, DBLP:conf/hotsos/MurphyBSR20, DBLP:conf/msr/OhiraKYYMLFHIM15, DBLP:conf/kbse/KuHCL07, DBLP:conf/cloud/GunawiHLPDAELLM14, DBLP:conf/kbse/WangDGGQYW17, BegBunch, DBLP:conf/sigsoft/GaoDQGW0HZW18, DBLP:conf/sigsoft/Zhang0C0Z19}                                                                                                                                                                                                                                                                                                                                                                                                                                                                                                                                                                                                                                                                                                                                                                                                                                                                                                                                                                                                                                                                                                                                                                                                                                                                                                                                                                                                                                                                                                                                                                                                                                                                                                                                                                                                                                                                                                                                                                                                                                                                                                                                                                                                                                                       & 10     \\
                     \midrule
        Not Provided &                     & \cite{DBLP:conf/pst/TebibAAG24, DBLP:conf/qrs/YuW24, DBLP:conf/issre/LiuZDHDM022, DBLP:conf/msr/ZamanAH12, DBLP:conf/hotpar/JalbertPPS11, DBLP:conf/iwpc/RenL020, lu2005bugbench, DBLP:conf/icse-apr/YadavW24}                                                                                                                                                                                                                                                                                                                                                                                                                                                                                                                                                                                                                                                                                                                                                                                                                                                                                                                                                                                                                                                                                                                                                                                                                                                                                                                                                                                                                                                                                                                                                                                                                                                                                                                                                                                                                                                                                                                                                                                                                                                                                                                                                                                                                                                                                                                                                                                                         & 8     \\
    \bottomrule
    \end{tabularx}
    \label{tab-artifact-availability}
\end{table}

Ideally, all datasets should be publicly available to facilitate future studies. As shown in \cref{fig-artifact-pie} and \cref{tab-artifact-availability}, we found that all but eight datasets were made publicly available at the time of paper publication. In total, 143 (\ratio{143}{150}) dataset papers have released their datasets alongside the corresponding paper. For the 143 publicly available datasets, we further examined their availability at the time of our survey and categorized them into the following three groups:\\
\emph{(1) Available} means the dataset is still available at the time of our survey, as it was when the corresponding paper was published, but has not been updated. This group accounts for \ratio{115}{151} of the datasets we survey.\\
\emph{(2) Evolving} means the dataset is publicly available, and it has been updated or evolved since the time of paper publication. Such evolution could include adding new defects, deprecating stale defects, or updating the framework or tool to facilitate the use of the dataset. This group accounts for \ratio{18}{151} of the surveyed datasets.\\
\emph{(3) No Longer Available} means the dataset was available at some point in time, but is not available anymore as of the time of our survey. Further investigation shows the reasons include (a) institutional webpage used to host the dataset becoming no longer loadable~\cite{DBLP:conf/msr/OhiraKYYMLFHIM15,DBLP:conf/kbse/KuHCL07,DBLP:conf/kbse/WangDGGQYW17,BegBunch}, (b) datasets being removed from a hosting service~\cite{DBLP:conf/kbse/SongWCCLLWP23,DBLP:conf/issre/Du0MZ21}, (c) unresolvable shortened links~\cite{DBLP:conf/hotsos/MurphyBSR20}, and (d) datasets being lost by accident~\cite{DBLP:conf/cloud/GunawiHLPDAELLM14}.   This group accounts for \ratio{10}{151} of the surveyed datasets.

Hosting platforms play a significant role in ensuring the ease of access to datasets, as our analysis of the datasets that are no longer available shows that most of them are offline due to discontinued or unstable hosting services. \cref{fig-artifact-host} shows the distribution of hosting platforms for the available datasets we survey. Note that some datasets are hosted on multiple platforms. The most popular hosting platform is \github, which hosts 86 (\ratio{86}{143}) available datasets. The second most popular hosting approach is to self-host a dedicated website for the dataset. Other hosting platforms, such as Zenodo, cloud storage services (\eg \gdrive, OneDrive, Box), are also frequently used, while other platforms such as \figsshare, Docker Hub, GitLab, Bitbucket, and Mendeley are less common.

\subsection{Presentation}
\label{sec-presentation}

There are many ways to present the defects in a dataset, ranging from simply providing the link to the corresponding issue reports to creating complex executable frameworks. The way datasets present defects can impact their usability for both research and practical applications. We categorize the presentation of defects into three levels: (1) meta level, (2) code level, and (3) execution level. \cref{tab-dataset-presentation} shows the presentation levels of 133 surveyed datasets that are available or evolving.

\begin{table}
    \centering
    \caption{Presentation of datasets.}
    \tablefontsize
    \begin{tabularx}{\textwidth}{ll|X|r}
        \toprule
    \textbf{Presentation}   &           & \textbf{Datasets} & \textbf{Count}  \\
    \midrule
    Meta Level     & \textit{Metadata}  & \cite{DBLP:conf/edcc/AndradeLV24, DBLP:journals/tosem/LiuMC24, DBLP:journals/tosem/ZhangCWCLMMHL24, DBLP:journals/tosem/JiangJWMLZ24, DBLP:conf/icse/CuiD0WSZWYXH0024, DBLP:conf/issta/YuXZ0LS24, DBLP:conf/issta/XuG024, DBLP:conf/ease/WaseemDA0M24, DBLP:conf/kbse/DasAM24, DBLP:conf/kbse/AvulaVM23, DBLP:conf/msr/MahbubSR23, DBLP:conf/secrypt/SenanayakeKAP023, DBLP:conf/aina/PonteRM23a, DBLP:conf/icse/GuanXLLB23, DBLP:conf/issta/XiongX0SWWP0023, DBLP:journals/softx/PachoulyAK22, DBLP:conf/icse/LiangLSSFD22, DBLP:conf/msr/KeshavarzN22, DBLP:journals/tse/ChenXLGLC22, DBLP:journals/pacmpl/PaltenghiP22, DBLP:conf/msr/KimKL21, DBLP:conf/icse/PaulTB21a, DBLP:conf/sigsoft/NikitopoulosDLM21, DBLP:conf/msr/KamienskiPBH21, DBLP:conf/icse/Makhshari021, DBLP:journals/jss/FerencGGTG20, DBLP:journals/access/SanchezDMS20, DBLP:journals/jss/WuZCWM20, DBLP:conf/msr/FanL0N20, DBLP:journals/pacmpl/0001FSK20, DBLP:conf/msr/WangBJS20, DBLP:conf/icse/GarciaF0AXC20, DBLP:conf/promise/VieiraSRG19, DBLP:conf/msr/RaduN19, DBLP:conf/msr/PontaPSBD19, DBLP:conf/dsa/XuYWA19, DBLP:conf/msr/MadeyskiK17, DBLP:journals/ese/LiuWWXCWYZ19, DBLP:conf/kbse/LinMZCZ15, DBLP:conf/msr/LamkanfiPD13, DBLP:journals/ese/DAmbrosLR12, DBLP:conf/icse/ChenYLCLWL21, DBLP:conf/icsm/HoMI0SKNR23, DBLP:conf/uss/XuLDDLWPM23, DBLP:journals/tosem/SantanaNAA24, DBLP:journals/infsof/WangBWYGS23, DBLP:journals/jss/WangCHZBZ23, DBLP:conf/msr/AzadIHR23, DBLP:conf/qsw/ZhaoWLLZ23, DBLP:conf/wcre/WangZRLJ23, DBLP:conf/kbse/WangWCCY22, DBLP:conf/sigsoft/ShenM0TCC21, DBLP:conf/sigsoft/WangLX0S21, DBLP:conf/kbse/EghbaliP20, DBLP:conf/compsac/GuWL0019, DBLP:conf/kbse/FrancoGR17, DBLP:conf/fie/SandersWA24, DBLP:conf/icse/HumbatovaJBR0T20, DBLP:conf/icse/HuZLYH024, DBLP:conf/kbse/RomanoLK021, DBLP:conf/msr/OliverDAH24, DBLP:journals/ese/TambonNAKA24, DBLP:journals/pacmpl/DrososSAM024, DBLP:journals/tosem/TufanoWBPWP19, DBLP:journals/tse/ZhaoXBCL23}        & 65      \\
    \midrule
    Code Level     & \textit{Snippet}   & \cite{DBLP:journals/jss/CaiLZZS24, DBLP:conf/icse/WangHGWC024, DBLP:conf/kbse/ZhongLZGWGL23, DBLP:conf/raid/0001DACW23, DBLP:conf/apsec/AkimovaBDKKMM21, DBLP:conf/icse/ZhengPLBEYLMS21, DBLP:conf/icics/LinXZX19, DBLP:conf/promise/FerencTLSG18}        & 8      \\
                   & \textit{Diff}      & \cite{DBLP:conf/msr/KimKL22, DBLP:conf/uss/ZhangP0W22, DBLP:conf/msr/KarampatsisS20, DBLP:conf/kbse/YeCG23}        & 4      \\
                   & \textit{File}      &   \cite{DBLP:journals/jss/ZhaoMLZ23, DBLP:conf/msr/CsuvikV22, DBLP:conf/prdc/PereiraAV22, DBLP:conf/splc/NgoNNV21, DBLP:journals/tse/AfroseXRMY23, DBLP:conf/icsm/ZhangXL20, DBLP:conf/icst/BuresHA20, DBLP:conf/msr/GaoYJLYZ18, DBLP:conf/icse/TanYYMR17, DBLP:journals/tse/GouesHSBDFW15, DBLP:conf/se/SelakovicP17}      & 11      \\
    \midrule
    Execution Level & \textit{Project}   & \cite{DBLP:journals/ese/MorovatiNKJ23, DBLP:conf/issta/ZhangCCXZ18, DBLP:conf/promise/MitraR17, DBLP:journals/tse/GouesHSBDFW15}        & 4      \\
                   & \textit{Binary}    & \cite{DBLP:journals/compsec/BrustSG23, DBLP:conf/msr/WendlandSMMHMRF21, DBLP:conf/icse/TanDGR18}        & 3      \\
                   & \textit{Framework} & \cite{DBLP:journals/ese/TimperleyHSDW24, DBLP:conf/msr/LiuHLZCSHM24, DBLP:conf/kbse/RuanLZL24, DBLP:conf/msr/NiSYZW24, DBLP:conf/acl/TianYQCLPWHL0024, DBLP:conf/iclr/JimenezYWYPPN24, DBLP:conf/msr/PramodSTSW24, DBLP:conf/msr/SilvaSM24, DBLP:conf/icst/LeeKYY24, DBLP:conf/kbse/AnKCYY23, DBLP:conf/sigsoft/WuLZ024, DBLP:conf/icst/KimH23, DBLP:conf/asiaccs/LiangYDMHZ23, DBLP:conf/icse/SaavedraSM24, DBLP:journals/tse/JiangLLZCNZHBZ23, DBLP:conf/msr/ApplisP23, DBLP:conf/icse/BhuiyanPVPS23, DBLP:conf/wcre/ZhangYYCYZ22, DBLP:conf/msr/BuiSF22, DBLP:conf/issta/Song0NW0DM22, DBLP:journals/stvr/GyimesiVSMBFM21, DBLP:conf/cgo/YuanLLLLX21, DBLP:conf/sigsoft/WidyasariSLQPTT20, DBLP:conf/icse/BentonGZ19, DBLP:conf/icse/DmeiriTWBLDVR19, DBLP:conf/msr/RiganelliMMM19, DBLP:conf/wcre/DelfimUMM19, DBLP:conf/msr/SahaLLYP18, DBLP:conf/msr/GkortzisMS18, DBLP:conf/issta/JustJE14, DBLP:conf/oopsla/LinKCS17, DBLP:conf/issta/WuJPLD0BS23, DBLP:conf/msr/AmannNNNM16, DBLP:conf/wcre/ReyesGSBM24, DBLP:conf/msr/BeyerGKRR24, DBLP:journals/access/AntalVKMHF24, DBLP:conf/kbse/XuZL23, DBLP:conf/sp/Dolan-GavittHKL16}        & 38     \\
    \bottomrule
    \end{tabularx}
    \label{tab-dataset-presentation}
    \end{table}

\emph{(1) Meta Level.} Meta-level datasets provide the details about defects only in textual form. For example, a table listing all defects with links to corresponding issue reports is considered meta-level. Such metadata may also include the date of creation for an issue report, a severity label, or other annotations.
Meta-level datasets are helpful as a reference or resource for analysis but do not include direct means for code analysis or defect reproduction. From the datasets that are still available at the time of this survey, 65 (\ratio{65}{133}) are meta-level.

\emph{(2) Code Level.} Code-level datasets provide isolated code snippets or source files from the defective program, as well as code diffs that show the changes made to the program to fix the defects. For example, a dataset that provides both buggy and fixed versions of source files without providing the entire project is considered code-level. Code-level datasets are helpful for source code analysis, but do not include direct means to run or reproduce the defects. It takes extra effort to integrate (\eg via \texttt{git apply}) the provided code into a complete project to reproduce the defects.
From the datasets that are still available at the time of this survey, 23 (\ratio{23}{133}) are code-level.

\emph{(3) Execution Level.} Execution-level datasets offer more hands-on resources by providing executable artifacts to help reproduce the bugs. For example, a dataset that provides a framework to automatically check out the defective program and reproduce the building and testing process is considered execution-level. Other forms of execution-level datasets include those that provide complete projects with build configurations or datasets that provide executable binaries. Execution-level datasets are particularly useful for practical applications, such as the evaluation of fault localization, test generation, or automated program repair techniques. From the datasets that are still available at the time of this survey, 45 (\ratio{45}{133}) are execution-level.

\begin{tcolorbox}[
  colback=white,
  colframe=gray!140,
  boxsep=2pt,
  left=2pt,
  right=2pt,
  top=1pt,
  bottom=1pt,
  title=\textbf{RQ3 Summary},
  fonttitle=\bfseries
]
\vspace{0.25em}
\begin{itemize}[leftmargin=10pt, topsep=2pt, itemsep=2pt]

    \item Nearly all datasets (\ratio{143}{151}) were released publicly at publication.
    Among them, \ratio{115}{143} remain available, \ratio{18}{143} have evolved with updates, and \ratio{10}{143} are no longer available due to discontinued hosting or link decay.
    Long-term persistence remains a concern.

    \item GitHub dominates hosting, with \ratio{86}{133} of available datasets.
    Dedicated websites, Zenodo, and cloud storage are also common.
    Datasets that disappeared were typically hosted on institutional or self-managed servers, indicating that unstable hosting contributes to data loss.

    \item Dataset usability varies by presentation level.
    About \ratio{65}{133} provide only meta-level information (issue links, descriptions), \ratio{23}{133} include code-level artifacts (buggy files, diffs), and \ratio{45}{133} offer execution-level resources (complete projects, reproduction frameworks).
    Execution-level datasets enable direct evaluation but require additional maintenance.

\end{itemize}
\end{tcolorbox}

\section{Uses of Software Defect Datasets (RQ4)}
\label{sec-usage}

Understanding how software defect datasets are used in practice is essential not only for assessing their impact on research and practice, but also for revealing current needs.
To this end, we conduct a systematic analysis of research work citing the top-five most-cited\footnote{As of January 1, 2025, on Google Scholar.} software defect dataset papers among the ones considered in this survey, which are \dfj~\cite{DBLP:conf/issta/JustJE14}, Bug Prediction Dataset~\cite{DBLP:journals/ese/DAmbrosLR12}, Big-Vul~\cite{DBLP:conf/msr/FanL0N20}, the ManyBugs and IntroClass benchmarks~\cite{DBLP:journals/tse/GouesHSBDFW15}, and SWE-Bench~\cite{DBLP:conf/iclr/JimenezYWYPPN24}.
All five datasets are general-purpose and not limited to any specific application domain.
Four datasets primarily focus on functional defects, while Big-Vul specifically targets security vulnerabilities.
In terms of programming languages, \dfj and the Bug Prediction Dataset collect defects from \java projects; Big-Vul and the ManyBugs and IntroClass benchmarks focus on \cncpp defects; and SWE-Bench, the most recent among them, is in \python.

\paragraph{Methodology} We categorize the uses of software defect datasets by conducting a systematic analysis of papers citing these datasets. Initially, we manually review a subset of citing papers to develop a preliminary categorization. Using an open-coding approach, we continuously refine the categories as we review additional papers. This taxonomy construction process involves over 150 randomly selected citing papers, leading to the following categorization scheme:
\begin{itemize}[left=0pt]
    \item \textbf{Empirical Research}: Research on existing software defects and existing software engineering techniques, which can be further divided into empirical analyses and empirical evaluations.
    \begin{itemize} [left=1em,label=$\circ$]
        \item \ul{Empirical Analyses}: Papers that analyze existing software defects, such as defect characteristics, root causes, and fix patterns.
        \item \ul{Empirical Evaluations}: Papers that perform retrospective evaluations solely on existing software engineering techniques using software defect datasets.
    \end{itemize}
    \item \textbf{Technical Evaluations}: Evaluations of existing or new software engineering techniques using software defect datasets, which can be further divided into detection, testing, and repair.
    \begin{itemize}[left=1em,label=$\circ$]
        \item \ul{Detection}: Papers that use software defect datasets to evaluate bug detection and localization techniques.
        \item \ul{Testing}: Papers that use software defect datasets to evaluate testing techniques, such as test generation and test prioritization.
        \item \ul{Repair}: Papers that use software defect datasets to evaluate automated program repair techniques, such as code repair and issue solving.
    \end{itemize}
    \item \textbf{New Benchmark}: Papers that introduce new benchmarks.
    \item \textbf{Others}: Papers that cite software defect datasets only as related work but without directly using them.
\end{itemize}

The ``empirical research'' and ``technical evaluations'' categories are not mutually exclusive, as a paper may use a dataset for both purposes.
For instance, a paper that empirically evaluates \textit{existing} program repair tools using established software defect datasets is classified into both empirical research and the repair category of technical evaluations, while a paper that introduces a \textit{new} repair technique and evaluates it falls solely under technical evaluation.

Based on the above scheme, we use the \texttt{o3-mini} model from OpenAI to automatically categorize papers using their titles and abstracts.
The model is instructed with a structured prompt\footnote{The full prompt is provided in the supplementary materials~\cite{defect_datasets_2025_17402613}.} that enumerates our taxonomy and labeling rules, including that a paper may have multiple labels, that \textit{new benchmark} is mutually exclusive with other categories, and that ambiguous cases should be marked as \textit{others}.
Two authors jointly validate a random sample of 100 LLM-labeled papers, independent of the open-coding process, since open coding did not record per-paper labels.
All 100 sampled papers have satisfactory classifications: 93 exactly match the human labels, and only 7 show minor discrepancies (\eg non-contradictory extra categories). For instance, paper~\cite{DBLP:conf/issta/Xia024} is labeled as both ``Empirical Study'' and ``Repair'' by the model, whereas the human label includes only ``Repair'', because the abstract highlights substantial empirical evaluation. Likewise, paper~\cite{DBLP:conf/icse/DingFISCAWR025} is labeled as both ``Empirical Study'' and ``Detection'' for the same reason, while the human label includes only ``Detection''. We have also inspected the model-provided reasoning summaries, which generally (i) propose candidate categories based on cues in the title and abstract, (ii) apply the taxonomy rules to include or exclude labels, and (iii) resolve overlaps by prioritizing the primary contribution over supporting components.

\begin{figure}[t]
    \centering
    \small
    \begin{minipage}{\textwidth}
        \centering
        \begin{minipage}[b]{0.5\textwidth}
            \centering
            \begin{subfigure}[b]{\linewidth}
                \centering
                \includegraphics[width=\linewidth]{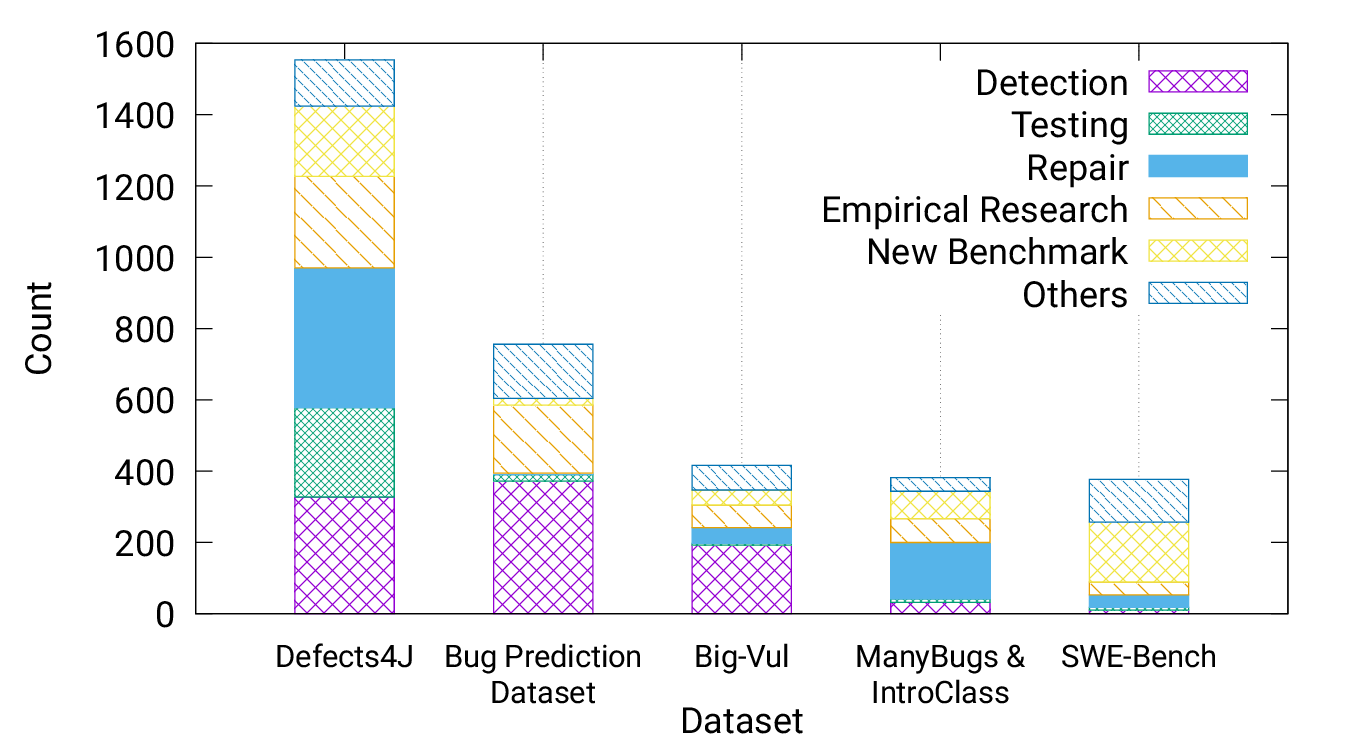}
                \caption{Breakdown of uses.}
                \label{fig-citation-breakdown}
            \end{subfigure}
        \end{minipage}\hfill
        \begin{minipage}[b]{0.5\textwidth}
            \centering
            \begin{subfigure}[b]{0.76\linewidth}
                \centering
                \includegraphics[width=\linewidth]{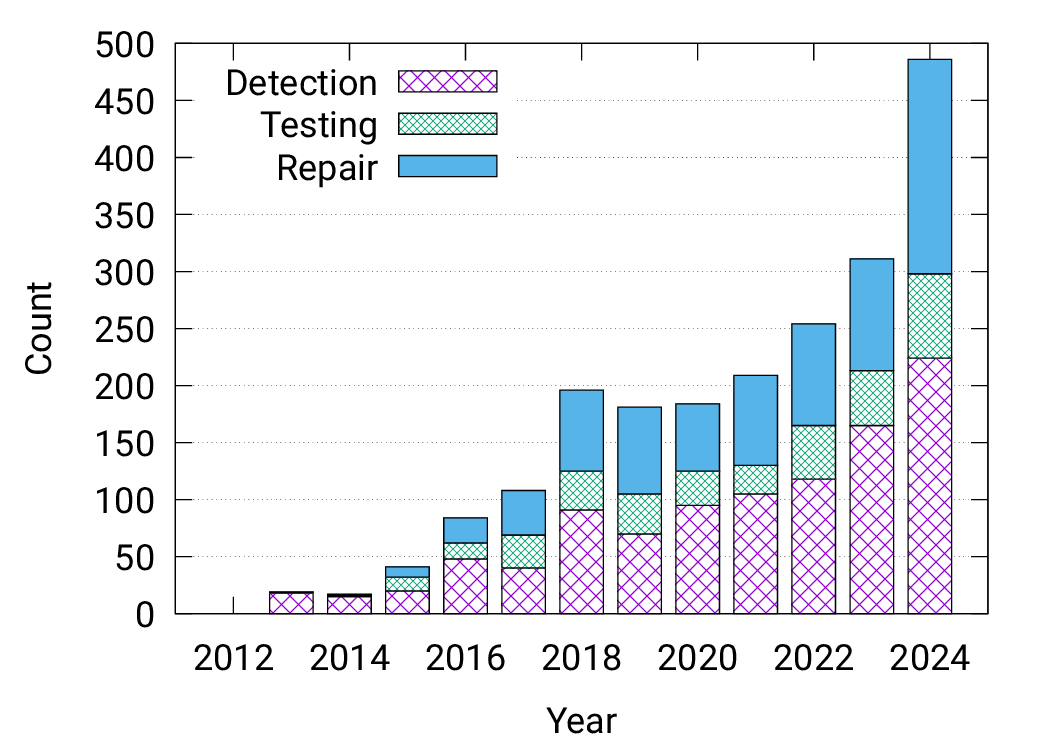}
                \caption{Trend of techniques evaluated.}
                \label{fig-citation-trend}
            \end{subfigure}
        \end{minipage}
    \end{minipage}
    \caption{Uses of software defect datasets.}
    \label{fig-citation-usage}
\end{figure}

\cref{fig-citation-breakdown} illustrates that empirical research and technical evaluations are the primary use categories for the citing papers analyzed across the five most-cited datasets. Technical evaluations, which can be further divided into evaluations of detection, testing, and repair techniques, demonstrate a growing demand for software defect datasets, as shown in \cref{fig-citation-trend}. This section focuses on these two primary categories to provide a comprehensive overview of the uses of the most-cited software defect datasets. As our analysis specifically focuses on direct dataset utilization, we exclude papers that introduce new benchmarks and the ``others'' category from the detailed discussion.

\subsection{Empirical Research}
Software defect datasets can serve as valuable resources for empirical research to understand various aspects of software defects. Such empirical research can be broadly categorized into two groups. On one hand, studies may focus directly on analyzing software defects themselves, typically with regard to their manifestations, symptoms, root causes, and fixes. For example, \citet{DBLP:conf/wcre/SobreiraDDMM18} present an in-depth analysis of 395 patches from \dfj. By analyzing the patch size, location, and the underlying repair patterns, the study provides insights and future direction for automated program repair, such as highlighting the necessity for repair strategies capable of effectively handling method calls. On the other hand, empirical research can also explore interactions between software defects and existing software engineering techniques. For example, \citet{DBLP:conf/sigsoft/DurieuxDMA19} conduct an empirical evaluation on eleven automated program repair tools using five software defect datasets. Their study shows that the effectiveness of automated program repair tools varies significantly across different datasets and concludes various causes of failures for the automated program repair. Both types of empirical research significantly benefit from pre-collected, well-structured software defect datasets, as these datasets offer diverse, realistic, and accessible instances of software defects, thus enabling rigorous analysis and reliable comparisons across different studies.

\subsection{Technical Evaluations}
Another primary use of the most-cited software defect datasets is to evaluate existing or new software engineering techniques, such as fault detection/localization (\eg \cite{DBLP:conf/icst/SarhanGB22,DBLP:conf/issta/AnY22,DBLP:conf/se/SchlichtigSNB23,DBLP:conf/icse/ZengWY0Z022,DBLP:conf/compsac/YanCBL22,DBLP:journals/tse/WenCTWHHC21,DBLP:journals/tse/WangLYSLDZL21,DBLP:conf/setta/YanJZH21,DBLP:conf/scam/SarhanVB21,DBLP:journals/tse/PerezAD21,DBLP:conf/issre/KimAFY21,DBLP:journals/fac/NilizadehLPN24,DBLP:journals/spe/SunJWFWC24}), test generation/prioritization (\eg \cite{DBLP:journals/fcsc/ZhangXYM24,DBLP:conf/qrs/JabbarHF23,DBLP:conf/kbse/RotaruV23,DBLP:journals/stvr/ChaimBONA23,DBLP:journals/tosem/ChenCWZWCZW23,DBLP:journals/ese/AlmullaG22,DBLP:journals/saem/BiswasRDMM22,DBLP:conf/icst/GazzolaMOPT22}), and automated program repair (\eg \cite{DBLP:conf/dsc/YeXFZYG23,DBLP:journals/tosem/LiuZLKKGLKB23,DBLP:journals/tosem/TianLLKKHLWKB23,DBLP:conf/icse/FanGMRT23,DBLP:journals/tse/ChakrabortyDAR22}). Fault detection/localization aims to identify defects in software in an effective, efficient, and automated manner. Test generation/prioritization is about generating or prioritizing test cases to maximize the likelihood of detecting defects, while minimizing the cost of testing. Both detection and testing techniques have long attracted significant interest from the software engineering, programming languages, and other communities, as such techniques are essential for improving software quality and reliability. The goal of automated program repair~\cite{DBLP:journals/cacm/GouesPR19} is to fix defective software without human intervention, which is a challenging task that has gained increasing attention in recent years. The evaluation of these techniques is crucial for understanding their effectiveness, efficiency, and limitations, and software defect datasets play a critical role in this process. By providing standardized, realistic, and reproducible instances of software defects, software defect datasets enable researchers and practitioners to evaluate their techniques in a controlled environment.

\cref{fig-citation-breakdown} illustrates that detection and repair techniques are the most frequently evaluated. \dfj is used for evaluating all three types of techniques, and it is also the predominant dataset used for evaluating testing techniques.
The Bug Prediction Dataset~\cite{DBLP:journals/ese/DAmbrosLR12} and Big-Vul~\cite{DBLP:conf/msr/FanL0N20} are primarily used for evaluating detection techniques, while both the ManyBugs and IntroClass benchmarks~\cite{DBLP:journals/tse/GouesHSBDFW15} and SWE-Bench~\cite{DBLP:conf/iclr/JimenezYWYPPN24} are used more extensively for evaluating repair techniques.
\cref{fig-citation-trend} illustrates the unprecedented growth in the number of proposed software engineering techniques, particularly in the area of automated program repair. This surge in recent years underscores the increasing importance and demand for high-quality software defect datasets.

\begin{tcolorbox}[
  colback=white,
  colframe=gray!140,
  boxsep=2pt,
  left=2pt,
  right=2pt,
  top=1pt,
  bottom=1pt,
  title=\textbf{RQ4 Summary},
  fonttitle=\bfseries
]
\vspace{0.25em}
\begin{itemize}[leftmargin=10pt, topsep=2pt, itemsep=2pt]

    \item Software defect datasets are primarily utilized in empirical studies and technical evaluations.
    Empirical work investigates defect characteristics, root causes, and fixes, while technical evaluations emphasize testing, detection, and automated repair.

    \item The most-cited datasets are frequently reused in software engineering research.
    Their adoption indicates a shift toward evaluating automated techniques like automated repair, test generation, and defect detection, highlighting the increasing demand for executable and large-scale datasets suitable for data-driven and LLM-based methods.

\end{itemize}
\end{tcolorbox}
\section{Discussion and Opportunities for Future Work}
\label{sec-discussion}

In this section, we discuss the challenges and opportunities in creating and maintaining software defect datasets, and provide future directions for enhancing the utility of such datasets for the research community.
We focus on dataset creation (\cref{sec-discussion-creating}) and maintenance (\cref{sec-discussion-maintaining}).

\subsection{Creation of Software Defect Datasets}
\label{sec-discussion-creating}

\paragraph{Scale, Diversity, Quality, and Cost}

Though various methodologies for constructing defect datasets have been proposed (\cref{sec-construction}), the balance between scale, diversity, quality, and cost remains a challenge. Intuitively, datasets of high quality tend to be smaller and more costly to create, such as the hand-crafted \dfj~\cite{DBLP:conf/issta/JustJE14}. In this sense, further efforts are still needed for fully automated defect mining. However, an important bottleneck for automated mining is the lack of a ground truth that indicates the actual defects present in the codebase. For example, keyword-based mining, the most commonly used technique, shows a relatively low precision of 51.9\% and recall of 58.8\%~\cite{DBLP:conf/icse/TianLL12}. It would be desirable to establish a ground truth without manual labelling or testing. BugSwarm~\cite{DBLP:conf/icse/DmeiriTWBLDVR19} and Bears~\cite{DBLP:conf/wcre/DelfimUMM19} have shown that by leveraging the fail-pass pattern in \cicd pipelines, one can automatically mine defects with high precision. Such techniques may be adopted into the mining processes for other sources, without narrowing to \cicd build histories. Similarly, ReposVul~\cite{DBLP:conf/icse/WangHGWC024} have shown that the combination of LLMs and static analysis can achieve up to 95\% precision in defect mining. Such approaches can be further explored to improve the quality of defect datasets while keeping the manual effort low.

\paragraph{Scope Selection}

Software defect datasets vary in scope, ranging from general-purpose datasets to domain- or type-specific datasets. Datasets with specialized scopes are desirable for more effective benchmarking of targeted techniques, or for fine-tuning LLMs for specific purposes~\cite{DBLP:conf/asianhost/FuYDGQ23}. However, as highlighted in \cref{sec-scope-focus}, there is a lack of datasets focusing on specific domains, non-functional defects, and some fast-growing programming languages. For example, despite the widespread adoption of multi-threading capabilities in \python, there is currently no available software defect dataset dedicated to concurrency defects in this language.
Future research efforts could prioritize creating datasets that cover more specialized application domains, defect types, or programming languages, to better support the development of targeted software engineering techniques.

\paragraph{Precise Defect Isolation and Annotation}
Though tremendous efforts have been made in defect mining, the granularity of defect datasets is usually at the repository-, file-, or method-level, without fully precise defect location information that pinpoints defects, e.g., at the line-level.
Ideally, datasets should isolate defects, e.g., by removing any unrelated code changes from a bug-fixing commit. For example, to isolate actual defects from newly added features, refactorings, and other code changes, \dfj~\cite{DBLP:conf/issta/JustJE14} has manually reviewed and filtered the code diffs that fix a given bug. Such a manual review and filtering process is costly and time-consuming, though; thus, it is preferable to have automated techniques to precisely locate defects. Such isolation techniques could be based on refactoring detection~\cite{DBLP:conf/icse/JiangLNZH21} or probabilistic delta debugging~\cite{DBLP:conf/kbse/SongWCCLLWP23}.
Another related problem is the annotation of defects, e.g., with the defect's category. Though some datasets  provide such annotations, they are typically manually labeled or inferred from other metadata such as exceptions~\cite{DBLP:conf/icse/DmeiriTWBLDVR19}. Recently, machine learning and natural language processing techniques have shown promising results for automated defect annotation~\cite{DBLP:conf/kbse/SongWCCLLWP23,DBLP:conf/dsa/XuYWA19,DBLP:journals/corr/abs-2408-05534}. More investigation is needed on the effectiveness and efficiency of such techniques, and on how they can be integrated into the process of creating defect datasets.

\paragraph{Dataset Organization}
Effectively organizing a dataset is crucial for maximizing its availability and usability. Currently, there is no standardized format for organizing software defect datasets, which can lead to difficulties in dataset access and usage. GitHub is the most popular platform for hosting datasets and tool chains, but it does not ensure long-term dataset availability and is susceptible to data loss or deletion. It is desirable to have datasets archived in a more stable and reliable repository, such as Zenodo or figshare, in addition to \github~\cite{ACMArtifactBadging2025}. Furthermore, the organization of datasets should facilitate future evaluation of new techniques. \cref{sec-presentation} shows that half of the datasets provide only textual metadata (\ie meta-level), which is insufficient for running software defects. Our analysis indicates that, after normalizing citation counts by dataset age (\ie citations per year), execution-level datasets exhibit higher uptake (mean 53.65, median 12.00) than code-level (mean 21.01, median 12.75) and meta-level datasets (mean 18.07, median 8.50). Future efforts should focus on providing more comprehensive and reproducible software artifacts for defects. At a minimum, we recommend the following artifact reproducibility checklist: (1) a metadata index that records each defect and links it to the corresponding buggy and fixed code versions; (2) an environment descriptor, such as a container image or environment specification; (3) clear build and test instructions with expected outcomes; and (4) a runnable reproduction driver that demonstrates the defect and confirms the fix.

\subsection{Maintenance of Software Defect Datasets}
\label{sec-discussion-maintaining}

\paragraph{Reproducibility} Ensuring the reproducibility of defects in software defect datasets is crucial for researchers to validate their findings and advance existing work. However, achieving reproducibility poses a significant challenge in software defect datasets, regardless of the construction methodology or dataset organization~\cite{DBLP:conf/icse/ZhuR23}. Recent work shows that the major reason for defects not being reproducible is inaccessible dependencies, which datasets can mitigate by caching dependencies~\cite{DBLP:conf/icse/ZhuR23,DBLP:conf/icse/SaavedraSM24,DBLP:conf/msr/SilvaSM24}. In this direction, more techniques are required to improve the long-term reproducibility of datasets while maintaining the balance between dataset size and usability.

\paragraph{Dataset Evolution}

Software defect datasets can benefit from further evolution after initial publication, but most are not updated after release, leading to outdated information.
Datasets that do not evolve may cause data leakage when evaluating machine learning-based approaches, especially LLMs, as defects may be in the training corpus~\cite{DBLP:journals/corr/abs-2304-11938}.
Future datasets should support continuous updates by adding new defects while maintaining historical data for longitudinal studies.
Continuous mining with automated pipelines~\cite{DBLP:conf/icse/DmeiriTWBLDVR19,DBLP:conf/icse/ZhuGFR23} and community-driven efforts~\cite{DBLP:conf/icse/McGuireSAR23} show promise for dataset evolution.
However, frequent updates complicate comparisons with prior work, as different studies may use distinct dataset versions.
Clear versioning, documentation, and stable baselines are crucial for fair and reproducible comparisons.
Future research should explore best practices for balancing new defects with historical data maintenance and incentivizing dataset creators to evolve published datasets.

\section{Threats to Validity}
\label{sec-threats}

\paragraph{Selection and implementation bias}
Despite our comprehensive search across four major bibliographic databases, we may miss work published outside these databases, or in venues not covered by our selection criteria.
Also, excluding preprints and non-peer-reviewed works ensures quality but may miss emerging datasets.
Automated scripts in data collection may contain unnoticed errors, and human judgment in screening may introduce subjectivity.

\paragraph{Annotation and interpretation bias}
Manual annotation of dataset attributes involves subjective interpretation.
Although we measure inter-rater agreement across multiple independent annotations, minor biases may still persist.
The LLM-based analysis in \cref{sec-usage} is manually validated, but some errors may remain due to model limitations.

\paragraph{Construct validity}
Our analytical dimensions may omit factors such as community engagement, and differing classification granularity within each dimension may affect cross-study comparability. Our usage analysis prioritizes the most-cited datasets and may not generalize; dataset evolution after publication may also affect the stability of findings.

\section{Conclusion}
\label{sec-conclusion}

This survey provides a comprehensive overview of the landscape of software defect datasets, including their scope, construction, availability, usability, and practical uses. As fundamental resources to the software engineering, programming languages, computer security, computer systems, and machine learning communities, software defect datasets have seen significant growth in both supply and demand in recent years. This article enables users to identify datasets that best align with their research and practical needs, and it gives future dataset creators guidance on best practices for dataset development and maintenance. Based on the discussions of existing datasets, we also propose several future directions to improve the quality and accessibility of software defect datasets. To facilitate exploration and selection of software defect datasets, we make our artifact publicly available at~\cite{defect_datasets_2025_17402613} and provide an interactive website where users can browse and filter datasets based on their specific needs: \url{https://defect-datasets.github.io/}.

{\small
\bibliographystyle{ACM-Reference-Format}
\bibliography{reference, survey-papers}
}

\end{document}